\documentclass[onehalfspace]{easychithesis}

\usepackage{amsmath}
\usepackage{xspace}
\usepackage{epsfig}
\usepackage{graphics}
\usepackage{graphicx}
\usepackage{natbib}
\usepackage{math-cmds,math-envs}
\usepackage{proof}
\usepackage{prooftree}
\usepackage{code}
\usepackage{listings}
\usepackage{localcode}
\usepackage{mathpartir}
\usepackage{lscape}
\usepackage{verbatim}
\usepackage{url}
\bibpunct();A{},
\let\cite=\citep

\renewcommand{\paragraph}[1]{{\bf {#1}}}

\newcommand{\rname}[1]{{\scshape{#1}}}
\newcommand{\rlabel}[1]{\hspace*{-1mm}\mbox{(\rname{#1})}}

\newcommand{\out}[1] {}

\newcounter{codeLineCntr}

\newenvironment{codeListingNormal}
 {\setcounter{codeLineCntr}{0}
  \vspace{-.1in}
  \ttfamily
  \begin{tabbing}}
 {\end{tabbing}
 \vspace{-.1in}
}

\reversemarginpar
\newif\ifnotes
\notestrue

\newcommand{\punt}[1]{}

\newcommand{\chref}[1]{Chapter~\ref{ch:#1}}

\newcommand{\secref}[1]{Section~\ref{sec:#1}}

\newcommand{\figref}[1]{Figure~\ref{fig:#1}}

\newcommand{\thmref}[1]{Theorem~\ref{thm:#1}}

\newcommand{\lemref}[1]{Lemma~\ref{lem:#1}}

\newcommand{\defref}[1]{Definition~\ref{def:#1}}

\newcommand{\proc}[1]{\ifmmode\mbox{\textsc{#1}}\else\textsc{#1}\fi}

  \newcommand{\func}[1]{\ifmmode\mathrm{#1}\else\textrm{#1}fi} %

\newcommand{\irr}[3]{\inferrule*[right={\scriptsize{#1}},leftskip=3mm,rightskip=3mm]{#2}{#3}}

\newcommand{\systemf}{System~\textsf{F}\xspace}

\newcommand{\lambdarec}{\textsf{LRec}\xspace}

\newcommand{\taub}{\ensuremath{\bar{\tau}}}
\newcommand{\rhob}{\ensuremath{\bar{\rho}}\xspace}

\newcommand{\Gammab}{\ensuremath{\bar{\Gamma}}}
\newcommand{\Deltab}{\ensuremath{\bar{\Delta}}}
\newcommand{\thetab}{\ensuremath{\bar{\theta}}}
\newcommand{\vb}{\ensuremath{\bar{v}}}
\newcommand{\eb}{\ensuremath{\bar{e}}}
\newcommand{\cb}{\ensuremath{\bar{c}}}

\newcommand{\mlpolyr}{{\bf MLPolyR}}

\newcommand{\Caption}[2]{\caption{{#1} {\small\advance\baselineskip-3pt #2}}}

\newcommand{\pos}[2]{\ensuremath{\mathsf{pos}({#1},{#2})}}

\newcommand{\labelsOf}[1]{\ensuremath{\mathsf{labels}({#1})}}

\newcommand{\boolty}{\ensuremath{\mathsf{bool}}}
\newcommand{\stringty}{\ensuremath{\mathsf{string}}}
\newcommand{\intty}{\ensuremath{\mathsf{int}}}

\newcommand{\val}{\ensuremath{\mathsf{val}}}
\newcommand{\eMod}[2]{\ensuremath{\mathbf{module~}{#1}\mathsf{~=~}{#2}}}
\newcommand{\eVal}[2]{\ensuremath{\mathbf{val~}{#1}\mathsf{~=~}{#2}}}
\newcommand{\eTemp}[3]{\ensuremath{\mathbf{template~}{#1}{~(#2)~}=~{#3}}}

\newcommand{\cc}[1]{\mbox{\textcircled{\small{#1}}}}

\newcommand{\itemsubcase}[1]{\item{\mbox{\it Subcase \small{\bf {#1}}}:}}
\newcommand{\kh}[2]{\ensuremath{\lambda k:\cont{#1}.\lambda h:\hdlr{#2}}}

\newcommand{\LAM}[1]{\ensuremath{\Lambda{#1}}}

\newcommand{\ra}{\ensuremath{\rightarrow}}
\newcommand{\dra}{\ensuremath{\Rightarrow}}
\newcommand{\hra}{\ensuremath{\hookrightarrow}}
\newcommand{\rowArrow}{\ensuremath{\rightarrowtail}}

\newcommand{\era}[1]{\ensuremath{\stackrel{#1}{\rightarrow}}}
\newcommand{\ehra}[1]{\ensuremath{\stackrel{#1}{\hookrightarrow}}}
\newcommand{\etCase}[3]{\ensuremath{{#1} \ehra{#3} {#2}}}

\newcommand{\tmaps}{\ensuremath{\blacktriangleright}}

\newcommand{\fmaps}{\ensuremath{\vartriangleright}}

\newcommand{\Ans}{\ensuremath{\mathsf{ans}}\xspace}
\newcommand{\Int}{\ensuremath{\mathsf{int}}\xspace}
\newcommand{\tRowEmp}{\ensuremath{\centerdot}\xspace}
\newcommand{\tRec}[1]{\ensuremath{\left\{ {#1} \right\}}}
\newcommand{\tRecN}[3]{\ensuremath{\left\{\:{#3}\:\right\}_{#1}^{#2}}}
\newcommand{\tSum}[1]{\ensuremath{\left< {#1} \right>}}
\newcommand{\tSumN}[3]{\ensuremath{\langle\:{#3}\:\rangle_{#1}^{#2}}}
\newcommand{\tBind}[2]{\ensuremath{{#1}~\mathbf{as}~{#2}}}
\newcommand{\tCase}[2]{\ensuremath{{#1} \hra {#2}}}

\newcommand{\eMat}[2]{\ensuremath{\mathbf{match}~{#1}~{\mathbf{with}}~{#2}}}
\newcommand{\eLet}[3]{\mathbf{let}~{#1}={#2}~\mathbf{in}~{#3}}
\newcommand{\eLetrec}[3]{\mathbf{letrec}~{#1}={#2}~\mathbf{in}~{#3}}
\newcommand{\eFun}[3]{\ensuremath{\mathbf{fun}~{#1}~{#2}={#3}}}
\newcommand{\eAbs}[2]{\ensuremath{\lambda{#1}.{#2}}}

\newcommand{\eTabs}[2]{\ensuremath{\Lambda{#1}.{#2}}}

\newcommand{\eTapp}[2]{\ensuremath{{#1}[{#2}]}}

\newcommand{\eIfz}[3]{\ensuremath{\mathbf{if0}~({#1},{#2},{#3})}}

\newcommand{\recPlus}{\ensuremath{~\otimes~}}
\newcommand{\recMinus}{\ensuremath{~\oslash~}}
\newcommand{\eRec}[1]{\ensuremath{\mathbf{\{}{#1}\mathbf{\}}}}
\newcommand{\eRecN}[3]{\ensuremath{\left\{\:{#3}\:\right\}_{#1}^{#2}}}

\newcommand{\eRecExt}[2]{\ensuremath{{#1} \recPlus {#2}}}
\newcommand{\eRecSub}[2]{\ensuremath{{#1} \recMinus {#2}}}

\newcommand{\matPlus}{\ensuremath{~\oplus~}}
\newcommand{\matMinus}{\ensuremath{~\ominus~}}
\newcommand{\eCase}[1]{\ensuremath{\{\:{#1}\:\}}}
\newcommand{\eCaseN}[3]{\ensuremath{\{\:{#3}\:\}_{#1}^{#2}}}

\newcommand{\eCaseExt}[2]{\ensuremath{{#1} \matPlus {#2}}}
\newcommand{\eCaseSub}[2]{\ensuremath{{#1} \matMinus {#2}}}

\newcommand{\EXN}{\ensuremath{E_{\mathbf{exn}}}}
\newcommand{\eRaise}[1]{\mathbf{raise}~{#1}}
\newcommand{\eHandle}[2]{\ensuremath{{#1}~\mathbf{handle}~{#2}}}
\newcommand{\eUnhandle}[2]{\ensuremath{{#1}~\mathbf{unhandle}~{#2}}}
\newcommand{\eRehandle}[2]{\ensuremath{{#1}~\mathbf{rehandle}~{#2}}}
\newcommand{\eRestore}[2]{\ensuremath{\mathbf{restore}_{~#1}~{#2}}}

\newcommand{\elRec}[1]{\ensuremath{\mathbf{\langle}{#1}\mathbf{\rangle}}}
\newcommand{\elRecN}[3]{\ensuremath{\left\langle\:{#3}\:\right\rangle_{#1}^{#2}}}
\newcommand{\indexOf}[3]{\ensuremath{\mathsf{indexOf}({#1},{#2},{#3})}}
\newcommand{\lengthOf}[2]{\ensuremath{\mathsf{lengthOf}({#1},{#2})}}

\newcommand{\projT}[3]{\ensuremath{\mathsf{proj}_{t}({#1},{#2},{#3})}}
\newcommand{\projL}[2]{\ensuremath{\mathsf{proj}_{l}({#1},{#2})}}
\newcommand{\Eu}{\ensuremath{\underline{E}}}
\newcommand{\eeu}{\ensuremath{\underline{e}}}
\newcommand{\vvu}{\ensuremath{\underline{v}}}
\newcommand{\ssu}{\ensuremath{\underline{s}}}

\newcommand{\ok}{~\mbox{\bf ok}~}

\newcommand{\lacks}{\ensuremath{\setminus}}

\newcommand{\prog}[1]{\ensuremath{{#1}~\mathsf{program}}}

\newcommand{\FTV}{\ensuremath{\mathrm{FTV}}}

\newcommand{\eLength}[1]{\ensuremath{\mathbf{len}({#1})}}
\newcommand{\eIfzero}[3]{\ensuremath{\mathbf{ifzero~}({#1},{#2},{#3})}}

\newcommand{\tx}[1]{\mathrm{#1}}
\newcommand{\reorder}{\ensuremath{\approx}}
\newcommand{\Eb}{\ensuremath{\bar{E}}}
\newcommand{\EL}{\ensuremath{\mathsf{EL}}}
\newcommand{\IL}{\ensuremath{\mathsf{IL}}}
\newcommand{\istyp}[2]{\ensuremath{{#1} \ts {#2} : \star}}
\newcommand{\isrow}[2]{\ensuremath{{#1} \ts {#2} : \varnothing}}

\newcommand{\cont}[1]{\ensuremath{{#1}~\mathsf{cont}}}
\newcommand{\hdlr}[1]{\ensuremath{{#1}~\mathsf{hdlr}}}
\newcommand{\comp}[2]{\ensuremath{({#1},{#2})~\mathsf{comp}}}
\newcommand{\cpsRa}[3]{\ensuremath{{#1} \era{#2} {#3}}}
\newcommand{\cpsCa}[3]{\ensuremath{{#1} \ehra{#2} {#3}}}
\newcommand{\cpsSum}[1]{\ensuremath{\tSum{\!\mid\!{#1}\!\mid\!}}}
\newcommand{\reify}[3]{\ensuremath{\mathbf{reify}[{#1}][{#2}]~{#3}}}

\newcommand{\tsv}{\ensuremath{~{\ts}_{\mathsf{v}}~}}
\newcommand{\tsp}{\ensuremath{~{\ts}_{\mathsf{\rho}}~}}

\newcommand{\wf}{\ensuremath{\mathsf{wf}}}
\newcommand{\wfc}[2]{\ensuremath{\ts ({#1},{#2})~\wf}}

\begin{document}
\title{Type Safe Extensible Programming}
\author{Wonseok Chae}
\date{August 2009}
\department{Computer Science}
\division{Physical Sciences}
\degree{Doctor of Philosophy in Computer Science}

\maketitle
\maketitlewithsignature
\dedication
\begin{center}
To my wife Namhee, and to my parents.
\end{center}

\topmatter{Abstract}



Software products evolve over time. Sometimes they evolve by adding
new features, and sometimes by either fixing bugs or replacing
outdated implementations with new ones. When software engineers fail
to anticipate such evolution during development, they will eventually
be forced to re-architect or re-build from scratch. Therefore, it has
been common practice to prepare for changes so that software products
are extensible over their lifetimes. However, making software
extensible is challenging because it is difficult to anticipate
successive changes and to provide adequate abstraction mechanisms over
potential changes. Such extensibility mechanisms, furthermore, should
not compromise any existing functionality during extension. Software
engineers would benefit from a tool that provides a way to add
extensions in a reliable way. It is natural to expect programming
languages to serve this role. Extensible programming is one effort to
address these issues.

In this thesis, we present type safe extensible programming using the
MLPolyR language. MLPolyR is an ML-like functional language whose type
system provides type-safe extensibility mechanisms at several
levels. 
After presenting the language, we will show how these
extensibility mechanisms can be put to good use in the context of
product line engineering. Product line engineering is an emerging
software engineering paradigm that aims to manage variations, which
originate from successive changes in software.

\topmatter{Acknowledgments}
This thesis would not have been possible without the support and
encouragement from: my dear advisor Matthias Blume for his guidance
during my graduate study; my dissertation committee, Umut Acar and
David MacQueen for their valuable feedbacks; John Reppy, Derek Dreyer,
Robby Findler, Amal Ahmed, Matthew Fluet, and Xinyu Feng for sharing
their good insights on programming languages; fellow colleagues
Matthew Hammer, George Kuan, Mike Rainey, Adam Shaw, Lars Bergstrom,
Jacob Matthews, and Jon Riehl for keeping me motivated for this whole
process; Sakichi Toyoda and his son Kiichiro Toyoda for their
philanthropy; my classmates Hoang Trinh and Jacob Abernethy for their
bravery to become first TTI-C students; President Mitsuru Nagasawa and
Dean Stuart Rice for sharing their experienced wisdom; Chief academic
officer David McAllester and Academic advisor Nathan Srebro who always
said yes when I asked for their help; Frank Inagaki, Makoto Iwai,
Motohisa Noguchi, Gary Hamburg, Adam Bohlander, Carole Flemming, Julia
MacGlashan, Christina Novak, Dawn Gardner, Don Coleman, Hiroyuki Oda,
and Seiichi Mita for their continued support; Kyo C. Kang for his
inspiration; Hyunsik Choi and Kiju Kim for being friends of long
standing; Father Mario, Sister Lina, Deacon Paul Kim, Karen Kim, and
Myoung Keller who offered their prayer for me; the Eunetown villagers
at St. Andrew Kim Parish for sharing their delicious food and love;
Denise Swanson and J. K. Rowling whose works gave me both rest and
energy for further study; Jewoo and Ahjung for teaching me the joy of
being a father; Younghak Chae and Jongwon Lee for being the most
loving and supportive parents in the world; Jungsook Jeong for being
the best mother-in-law in the universe; and, finally, my wife, Namhee
Kim, for her unflagging belief in my talent.







%
%

\tableofcontents

\listoffigures

\listoftables

\mainmatter
\doublespacing

\chapter{Introduction}
\label{ch:intro}
Software products evolve over time. Sometimes they evolve by adding
new features, and sometimes by fixing bugs that a previous release
introduced. In other cases, they evolve by replacing outdated
implementations with better ones. Unless software engineers anticipate
such evolution during development, they will eventually be forced to
re-implement them again from scratch.  Therefore, it has become common
practice to prepare for extensibility when we design a software system
so that it can evolve over its lifetime.  For example, look at the
recent release history of the SML/NJ compiler:
\begin{itemize}
\item 1/13/09. v110.69. \underline{Add} new concurrency instructions
  to MLRISC. \underline{Fix} problem with CM tools.
\item 9/17/08. v110.68. \underline{Improve} type checking and type
  error messages. \underline{Re-implement} the RegExp
  library. \underline{Fix} bugs in ml-ulex. \underline{Update}
  documentation. \underline{Add} NLFFI support in Microsoft Windows.
\item 11/15/07. v110.67. \underline{Fix} performance
  bugs. \underline{Support} Mac OS X 10.5 (Leopard) on both Intel and
  PPC Macs. \underline{Drop} support for Windows 95 and 98.
\end{itemize} 
The SML/NJ compiler has evolved by means of adding and replacing
functionality since its birth around the early 1990s. Interestingly,
its evolution is sequential in that all its changes have been
integrated together into a new release~\cite{Buckley05}. In this
scenario, we are interested in easily adding extensions to an existing
system, and therefore extensibility mechanisms become our major
concern. Furthermore, we would like to have extensibility mechanisms
which do not compromise any functions in the base system.  Hence,
software engineers need a tool that provides a way to add extensions
in a reliable way, and it is natural to expect programming languages
to function in this way. Functional languages such as SML and Haskell
have already improved safety in the sense that ``well-typed programs
do not go wrong.''~\cite{milner:78:wrong}.  Beyond this, we would like
to have a language safe enough to guarantee that nothing bad happens
{\em during extensions.}  This approach will work well for sequential
evolution since extensible languages make it easy to extend one
version into another in a reliable way.

There are many cases, however, where software changes can not be
integrated into the original product, and as a result, different
versions begin to coexist. Moreover, there are even situations where
such divergence is planned from the beginning. A marketing plan may
introduce a product lineup with multiple editions. Windows Vista,
which ships in six editions, is such an example. These editions are
roughly divided into two target markets, consumer and business, with
editions varying to meet the specific needs of a large spectrum of
customers~\cite{microsoft}.  Then, each edition may evolve {\em
  independently} over time.  Unless we carefully manage each change in
different editions, multiple versions that originate from one source
start to coexist separately. They quickly become so incompatible that
they require separate maintenance, even though much of their code is
duplicated. This quickly leads to a maintenance nightmare. In such a
case, the role of programming languages become limited and, instead,
we need a way of managing variability in a product lineup.

\begin{figure}
\centering
\includegraphics[scale=1]{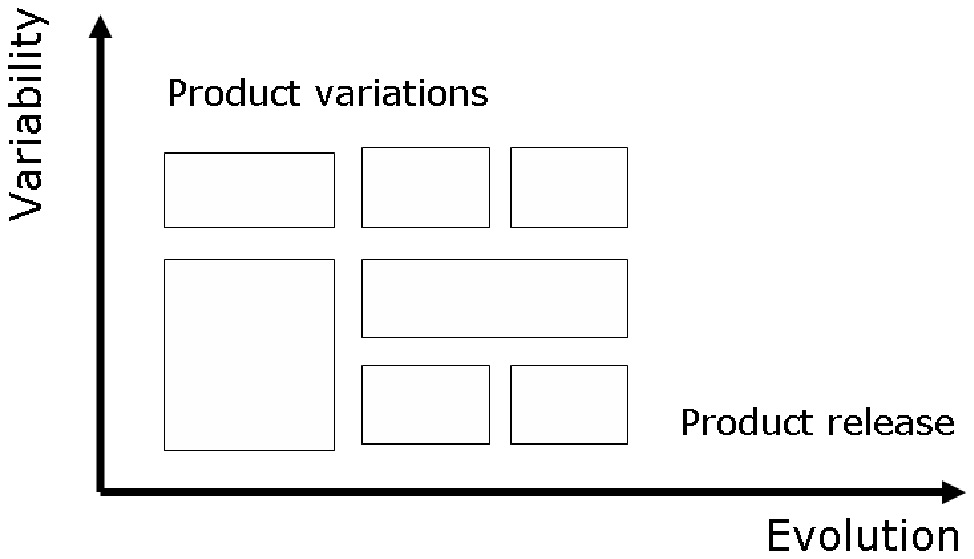}
\caption{Variability and evolution~\citep{Mikael00}.}
\label{fig:evolution}
\end{figure} 

Svahnberg studies the relationship between variability and evolution,
as shown in~\figref{evolution}, where product variations and product
release span two dimensions. As his figure suggests, a set of products
evolve over time just as one product does. Any extensibility mechanism
which does not take these two dimensions into consideration can not
fully provide satisfactory solutions.

In this thesis, we propose type safe extensible programming which
takes two dimensions into consideration.  In particular, our language
provides extensibility mechanisms at multiple levels of granularity,
from the fine degree (at the core expression level) to the coarse
degree (at the module level).  At the same time, in order to manage
variability, we adopt product line engineering as a developing
paradigm, and then provide a development process which guides how to
apply this paradigm to our extensibility mechanisms:
\begin{itemize}
\item A core language that supports polymorphic extensible records,
first-class cases and type safe exception handling (Section \ref{ext-prog});
\item A module system that supports separate compilation 
in the presence of the above features (Section \ref{large-prog});
\item A development process that supports the construction of
a family of systems (Section \ref{pl-prog}).
\end{itemize}

\chapter{Related work}

Extensible programming is a programming style that focuses on
mechanisms to extend a base system with additional functionality. The
main idea of extensible programming is to use the existing artifacts
(e.g., code, documents, or binary executables) but extend them to fit
new requirements and extensibility mechanisms take an important role
in simplifying such activities.  Building extensible systems has
received attention because it is seen as a way to reduce the
development cost by reusing the existing code base, not by developing
them from scratch. Furthermore, nowadays software products need to
support extensibility from the beginning since the current computing
environment demands a high level of adaptability by software products.
Extensible programming provides language features designed for
extensibility in oder to simplify the construction of extensible
systems. In the remainder of this section, we will study similar
works that take extensibility and adaptability in software into
consideration.


\section{The extensible language approach}

Software evolves by means of adding and/or replacing its functionality
over time. Such extensibility has been studied extensively in the
context of compilers and programming languages. Previous work on
extensible compilers has proposed new techniques on how to easily add
extensions to existing programming languages and their compilers.
For example, JaCo is an extensible compiler for Java based on
extensible algebraic types~\cite{Zenger-Odersky01,Zenger-Odersky05}.
The Polyglot framework implements an extensible compiler where even
changes of compilation phases and manipulation of internal abstract
syntax trees are possible ~\cite{nystrom03polyglot}.  Aspect-oriented
concepts are also applied to extensible compiler
construction~\cite{Xiaoqing+:05}.

However, most of these existing solutions do not attempt to pay
special attention to the {\em set of extensions} they produce.
Extensions are best accomplished if the original code base was
designed for extensibility. Even worse, successive extensions can make
the code base difficult to learn and hard to change substantially.
For example, the GNU Compiler Collection~\cite{gcc} started as an
efficient C compiler but has evolved to officially support more than
seven programming languages and a large number of target
architectures. However, a variety of source languages and target
architectures have resulted in a complexity that makes it difficult to
do GCC development~\cite{Vichare}. This effect apparently even led to
some rifts within the GCC developer community~\cite{Matzan}.

\section{The design patterns approach}

In software engineering, extensibility is one kind of design principle
where the goal is to minimize the impact of future changes on existing
system functions. Therefore, it has become common practice to prepare
for future changes when we design systems.  The concept of {\em Design
  patterns} takes an important role in this context
~\cite{Gamma:1995:DPE}. Each pattern provides design alternatives
which take changes into consideration so that the system is robust
enough to accommodate such changes. For example, the {\em visitor
  pattern} makes it easy to define a new operation without changing
the classes of the members on which it is performed. It is
particularly useful when the classes defining the object structure
rarely change.  By clearly defining intent, applicability and
consequences of their application, patterns will help programmers
manage changes.

However, design patterns are not generally applicable to
non-object-oriented languages. Even worse, Norvig shows how it is
trivial to implement various design patterns in dynamic
languages~\cite{Norvig98}. Some criticize that design patterns are
just workarounds for missing language features~\cite{Monteiro06}.

\section{The feature-oriented programming approach}

Product line engineering is an emerging paradigm 
for construction of a family of products~\cite{Kyo+:02,Kwanwoo+:02,SEI}.
This paradigm encourages developers to focus on developing 
a set of products rather than on developing one particular product. 
Therefore, mechanisms for managing variability through 
the design and implementation phases are essential. 
While most efforts in product line engineering have focused on 
principles and guidelines, only a few have suggested concrete
mechanisms of implementing variations.
Consequently, their process-centric approach is too
abstract to provide a working solution in a particular language.
For example, the Feature-Oriented Reuse Method (FORM) often suggests
parameterization techniques, but implementation details are
left to developers \cite{Kang98,Lee2000}.
Therefore, preprocessors, e.g., macro systems, have been used in many
examples in the literature as the feature delivery
method~\cite{Kang98,KangKLK05}.  For example, the macro language in
FORM determines inclusion or exclusion of some code segments based on
the feature selection.
Macro languages have some advantage in that they can be mixed
easily with any target programming languages, however, feature
specific segments are scattered across multiple classes, so code
can easily become complicated. Even worse, since 
general purpose compilers do not understand the macro language,
any error appearing in feature code segments cannot be detected
until all feature sets are selected and the corresponding code
segments are compiled.

In order to take advantage of the current compiler technology
including static typing and separate compilation, we need native
language support. Therefore, feature-oriented programming emerges as
an attempt to provide better support for feature
modularity~\cite{Lopez05}.  FeatureC++~\cite{Apel05featurec++:on} and
AHEAD~\cite{Batory2004} are such language extensions to C++ and Java,
respectively.  In these approaches, features are implemented as
distinct units and then they are combined to become a product.
However, there still is no formal type system, so these languages do
not guarantee the absence of type errors during feature
composition~\cite{Thaker07}. Recently, such a formal type system has
been proposed for a simple, experimental feature-oriented
language~\cite{Apel08}.

\section{The generic programming approach}

The idea of generic programming is to implement the common part
once and parameterize variations so that different products can be
instantiated by assigning distinct values as parameters. Higher-order
modules, also known as {\it functors} -- e.g., in the Standard ML
programming language (SML), are a typical example in that they can be
parameterized on values, types and even other modules, possibly
including higher-order ones~\cite{smlnj}.  The SML module system has
been demonstrated to be powerful enough to manage variations in the
context of product lines~\cite{Chae08}.

However, its type system sometimes impose restrictions which require
code duplication between functions on data types.  Many proposals to
overcome this restriction have been presented. For example, MLPolyR
proposes extensible cases \cite{blume+:06:mlpolyr}, and OCaml proposes
polymorphic variants \cite{Garrigue00}.

Similarly, templates in C++ provide parameterization over types and
have been extensively studied in the context of programming families
~\cite{Krzyszttof}. Recently, an improvement that would provide
better support of generic programming has been proposed~\cite{Reis06}.
Originally, Java and C\# did not support parameterized types but now
both support similar concepts~\cite{Torgersen04,Garcia07}.

Sometimes the generic programming approach is criticized for its
difficulty in identifying variation points and defining parameters
\cite{Gacek2001}. However, systematic reasoning (e.g., product line
analysis done by product line architects) can ease this burden by
providing essential information for product line implementation
~\cite{Chae08}.

\section{The generative programming approach}
Generative programming is a style of programming that utilizes code
generation techniques which make it possible to generate code from
generic artifacts such as specifications, diagrams, and
templates~\cite{Czarnecki04}. This approach is similar to the generic
programming approach in that a specialized program can be obtained
from a generic one, but the generative programming approach focuses on
the usage of domain specific languages and their code generators while
the generic programming approach focuses on the usage of the built-in
language features such as templates and functors.

\section{The open programming approach}

Extensions can be added generally by modifying source code.  In this
compile-time form of extensions, a program needs to be compiled for
extensions to become available. However, in some cases, a software
product need to modify its behavior dynamically during its execution.
Non-stop applications are such examples. Sometimes, a certain type of
change can be arranged to be picked up by a linker during load-time.
Open programming is an attempt at addressing these issues in the
context of programming languages. For instances, Java can dynamically
load (class-) libraries for this sort of thing. Rossberg proposes the
Alice ML programming language which reconciles open programming
concepts with strong typing~\cite{Rossberg07}. 

Similarly, there have been attempts to upgrade software while it is
running. Appel illustrated the usage of ``applicative'' module linking
to demonstrate how to replace a software module without having the
downtime~\cite{Princeton94hot-slidingin}. However, it was Erlang that
made this ``hot-sliding'' or ``hot code swapping'' idea
popular~\cite{Erlang}.  In Erlang, old code can be phased out and
replaced by new code, which makes it easier to fix bugs and upgrade a
running system.

Unlike these approaches, we focus on compile-time extensions by
modifying source code with minimal efforts.

\chapter{Type safe extensible programming}
\label{ext-prog}
\section{Introduction}

The \mlpolyr{} language has been specifically designed to support
type-safe extensible programming at a relatively fine degree of
granularity. Its records are polymorphic and extensible unlike in most
programming languages where records must be explicitly declared and
are not extensible. As their duals, polymorphic sums with extensible
cases make composable extensions possible.  Moreover, by taking
advantage of representing exceptions as sums and assigning exception
handlers polymorphic, extensible row types, we can provide type-safe
exception handling, which suggests ``well-typed programs do not have
uncaught exceptions.''

To understand the underlying mechanism, it is instructive to first
look at an example. The following sections informally provide such
examples that highlight the extensible aspect of the \mlpolyr{}
language.  Then, we show how these constructors provide a solution to
the expression problem which is considered one of the most fundamental
problems in the study of extensibility (\secref{exp-prob}).

Theoretical aspects of this language (derived from the previously
published conference papers ~\citep{blume+:06:mlpolyr,
  blume+:08:exception}) are presented in the following sections.
First, we consider an implicitly typed {\em external language} \EL{}
that extends $\lambda$-calculus with polymorphic extensible records,
extensible cases and exceptions.  Our implementation rests on a
deterministic, type sensitive semantics for \EL{} based on elaboration
(i.e., translation) into an explicitly typed {\em internal language}
\IL{}. The elaboration process involves type inference for \EL{}. Our
compiler for \mlpolyr{} provides efficient type reconstruction of
principal types by using a variant of the well-known algorithm
W~\citep{Milner78}.  Finally, \IL{} is translated into a variant of an
untyped language, called \lambdarec{}, which is closer to machine
code. Therefore, our compiler is structured as follows:
\[
\underbrace{\stackrel{\mathsf{EL}}{~}}_{\stackrel{\mathsf{Implcitly~typed}}{\stackrel{\vdots}{\secref{el}}}} 
\stackrel{\mathsf{CPS~and~Dual~transformation}}{\vector(1,0){100}} 
\underbrace{\stackrel{\mathsf{IL}}{~}}_{\stackrel{\mathsf{Explicitly~typed}}{\stackrel{\vdots}{\secref{il}}}}
\stackrel{\mathsf{Index~Passing}}{\vector(1,0){80}}
\underbrace{\stackrel{\mathsf{LRec}}{~}}_{\stackrel{\mathsf{Untyped~\lambda-calculus}}{\stackrel{\vdots}{\secref{lrec}}}}
\]\\

\subsection{Polymorphic extensible records}
\label{sec:recs}

\mlpolyr{}~supports polymorphic extensible records.  One of its record
expressions has the form {\tt \{ a = e, ... = $r$ \}}. This creates
a new record which extends record {\tt r} with a new field {\tt a}.
Table~\ref{table:rec} shows more such record operations.  Record
update and renaming operations can be derived by combining extension
and subtraction operations.

\begin{table}
\[
\begin{array}{|@{~~}l@{~~~~}||@{~~~~}l|l|}
\hline
\mathsf{Selection}   & \val~sel\_a~:~\forall \beta : \tRec{a}. 
                       \tRec{a:\tau, \beta} \ra \tRec{\tau} \\ 
                     & \eFun{sel\_a}{r}{r.a} \\
\hline
\mathsf{Extension}   
                     & \val~add\_a~:~\forall \beta : \tRec{a}. 
                       \tRec{\beta} \ra \tRec{a:\intty, \beta} \\ 
                     & \eFun{add\_a}{r}{\eRec{a=1,\ldots=r}} \\  
\hline
\mathsf{Subtraction} 
                     & \val~sub\_a~:~\forall \beta : \tRec{a}.\forall \alpha. 
                       \eRec{a:\alpha, \beta} \ra \tRec{\beta} \\ 
                     & \eFun{sub\_a}{\eRec{a=\_,\dots=r}}{r} \\  
\hline
\mathsf{Update}      
                     & \val~upd\_a~:~\forall \beta : \tRec{a}.\forall \alpha. 
                       \eRec{a:\alpha, \beta} \ra \tRec{a:\intty,\beta} \\ 
                     & \eFun{upd\_a}{r}{\eLet{\eRec{a=\_,\dots=rest}}{r}{\eRec{a=1,\ldots=rest}}} \\
\hline
\mathsf{Rename}      
                     & \val~ren\_a~:~\forall \beta : \tRec{a,b}.\forall \alpha. 
                       \eRec{a:\alpha, \beta} \ra \tRec{b:\alpha,\beta} \\ 
                     & \eFun{ren\_a}{r}{\eLet{\eRec{a=a',\dots=rest}}{r}{\eRec{b=a',\ldots=rest}}} \\
\hline
\end{array}
\]
\caption{Basic operations on records in \mlpolyr.}
\label{table:rec}
\end{table}

To understand the extension mechanism, let us first look at an example.
Since records are first-class values, we can abstract over the record being
extended and obtain a function {\tt add\_a} that extends any argument
record (as long as it does not already contain {\tt a}) with a
field {\tt a}.  Such a function can be thought of as the
``difference'' between its result and its argument:

\begin{lstlisting}[style=mlpolyr]
   fun add_a r = { a = 1, ... = r }
\end{lstlisting}

\noindent Here the difference consists of a field labeled {\tt a} of type {\tt
  int} and value {\tt 1}.  The type of function {\tt add\_a} is inferred
as $\forall\beta:\tRec{a}.\tRec{\beta}\ra\tRec{a:\Int,\beta}$
where $\beta:\tRec{a}$ represents a constraint that a row variable $\beta$
must not contain a label $a$.
We can write similar functions {\tt add\_b} and {\tt add\_c} 
which add fields {\tt b} of type $\boolty$ and {\tt c} of type $\stringty$ respectively:

\begin{lstlisting}[style=mlpolyr]
   fun add_b r = { b = true, ... = r } 
   fun add_c r = { c = "hello", ... = r }
\end{lstlisting}

\noindent We can then ``add up'' record differences represented
by {\tt add\_a}, {\tt add\_b}, {\tt add\_c} by composing these functions:

\begin{lstlisting}[style=mlpolyr]
   fun add_ab r = add_a (add_b r)
   fun add_bc r = add_b (add_c r)
\end{lstlisting}
where the inferred types are respectively:
\[
\begin{array}{ll}
\hspace{-3cm}\mathsf{val~add\_ab:~} & \forall\beta:\tRec{a,b}.\tRec{\beta}\ra\tRec{a:\Int,b:\boolty,\beta} \\
\hspace{-3cm}\mathsf{val~add\_bc:~} & \forall\beta:\tRec{b,c}.\tRec{\beta}\ra\tRec{b:\boolty,c:\stringty,\beta}
\end{array}
\]

\noindent Finally, we can create actual records by ``adding'' differences to the empty
record:
\begin{lstlisting}[style=mlpolyr]
   val a = add_a {}
   val ab = add_ab {}
   val bc = add_bc {}
\end{lstlisting}

\subsection*{Records as classes}

Extensible records continue to receive attention 
since they can also be used
as a type-theoretical basis for object-oriented programming~\cite{Remy98}. 
For example, assuming polymorphic records and references in place,
we can define a base class, and then create sub-classes 
with additional methods
in order to obtain the same effect of code reuse via inheritance.

As a demonstration of records as classes (followed by 
Pierce's encoding~\cite{Pierce02}),
we first define a $\mathsf{counter}$ class which provides two methods: 
1) $\mathsf{get}$ returns the current value of a field $i$ by dereferencing  
and 2) $\mathsf{inc}$ increments its value by first reading and then 
assigning its incremental) as follows:
\begin{lstlisting}[style=mlpolyr]
   val counterClass = fn x =>
       {get = fn _ => x!i, 
        inc = fn _ => x!i := x!i + 1
       }
\end{lstlisting}
where $!$ is a dereferencing operator and $:=$ is an assignment operator.
Then, individual $\mathsf{counter}$ objects can be obtained 
by a counter generator $\mathsf{newCounter}$ which applies 
$\mathsf{counterClass}$ to a record with a reference field $i$:
\begin{lstlisting}[style=mlpolyr]
   val newCounter = fn _ => let
       val x = {| i = 0 |}
       in counterClass x
       end
\end{lstlisting}
where $\{|\ldots|\}$ denotes a mutable record.
Furthermore, by taking advantage of extensible records, 
we can implement a subclass $\mathsf{resetCounterClass}$ 
which extends the base class $\mathsf{counterClass}$ with a new method $\mathsf{reset}$ 
like this:
\begin{lstlisting}[style=mlpolyr]
   val resetCounterClass = fn x =>
       {...=counterClass x,
        reset = fn _ => x!i := 0
       }
\end{lstlisting}
where $\ldots$ refers to the same fields that the base class contains,
so the returned value contains one more field named $\mathsf{reset}$.
Similarly, individual $\mathsf{resetCounter}$ objects can be obtained
by a generator $\mathsf{newResetCounter}$:
\begin{lstlisting}[style=mlpolyr]
   val newResetCounter = fn _ => let
       val x = {| i= 0 |} 
       in resetCounterClass x 
       end
\end{lstlisting}

\subsection{Extensible programming with first-class cases}
\label{sec:recs-sums}
Variants are dual of records in the same manner as logical $\vee$
is dual to $\wedge$:
\[
\begin{array}{c@{~~}c@{~~}c}
\neg{\eRec{a \wedge b}}    & = & \left<\neg a \vee \neg b\right> \\
\neg \left<a \vee b\right> & = & \eRec{\neg a \wedge \neg b} \\
\end{array}
\]



\noindent Then, as in any dual construction, 
the introduction form of the primal
corresponds to the elimination form of the dual.  Thus, elimination
forms of sums (e.g., {\bf match}) correspond to
introduction forms of records.  In particular, record extension (an
introduction form) corresponds to the extension of {\em cases} (an
elimination form).  This duality motivates making cases first-class
values as opposed to a mere syntactic form. 
With cases being first-class and extensible, one can
use the usual mechanisms of functional abstraction in a style of
programming that facilitates composable extensions.  

Here is a function representing the difference
between two code fragments, one of which can handle case {\tt `A}
while the other, represented by the argument $c$, cannot:

\begin{lstlisting}[style=mlpolyr]
   fun add_A c = cases `A () => print "A" 
                 default: c
\end{lstlisting}

\noindent where data type constructors ($`A~()$) are represented by
prefixing their names with a backquote character `.  Note that
function {\tt add\_A} corresponds to {\tt add\_a} of the dual (in
Section~\ref{sec:recs}).  The type inferred for {\tt add\_A} is
$\forall\beta:\tRec{`A}.(\tCase{\tSum{\beta}}{()}) \ra
(\tCase{\tSum{`A:(),\beta}}{()})$ where a type
$\tCase{\tSum{\rho}}{\tau}$ denotes the type of first-class cases,
$\tSum{\rho}$ is the sum type that is being handled, and $\tau$ is the
result. We also assume that $()$ denotes a unit type.

Examples for functions {\tt add\_B} and {\tt add\_C}
(corresponding to {\tt add\_b} and {\tt add\_c} in the dual) are:

\begin{lstlisting}[style=mlpolyr]
   fun add_B c = cases `B () => print "B"
                 default: c 
   fun add_C c = cases `C () => print "C"
                 default: c
\end{lstlisting}

\noindent As in the dual, we can now compose difference functions to
obtain larger differences:

\begin{lstlisting}[style=mlpolyr]
   fun add_AB c = add_A (add_B c)
   fun add_BC c = add_B (add_C c)
\end{lstlisting}

\noindent By applying a difference to the empty case {\bf nocases} we
obtain case values:

\begin{lstlisting}[style=mlpolyr]
   val case_A = add_A nocases
   val case_AB = add_AB nocases
   val case_BC = add_BC nocases
\end{lstlisting}

\noindent These values can be used in a {\bf match} form.
The {\bf match} construct is the elimination form for the case arrow
$\hra$.  The following expression will cause {\tt "B"} to be printed:

\begin{lstlisting}[style=mlpolyr]
   match `B () with case_BC
\end{lstlisting}

The previous examples demonstrate how functional record extension in
the primal corresponds to code extension in the dual.  The latter
feature gives rise to a simple programming pattern facilitating {\em
  composable extensions}.  Composable extensions can be used as a
principled approach to solving the well-known {\em expression problem}
described by Wadler~\citep{Wadler98a}. We will show how our composable
extensions provide a solution to the expression problem in the
following section (\secref{exp-prob}).

\subsection{Exception handlers as extensible cases}
\label{sec:exception}

Exceptions are an indispensable part of modern programming
languages.  They are, however, handled poorly, especially by
higher-order languages such as ML and Haskell: in both
languages a well-typed program can unexpectedly fail due to an
uncaught exception.  \mlpolyr{} enriches the type system
with type-safe exception handling by relying on representing 
exceptions as sums and assigning exception handlers 
polymorphic, extensible row types.
Our syntax distinguishes between the act of
establishing a new exception handler ($\mathbf{handle}$) and that of
overriding an existing one ($\mathbf{rehandle}$). The latter can be
viewed as a combination of $\mathbf{unhandle}$ (which removes an
existing handler) and $\mathbf{handle}$. This design choice makes 
it possible to represent exception types as row types 
without need for additional complexity. 
From a usability perspective, the design makes
overriding a handler explicit, reducing the likelihood of this
happening by mistake.

We will now visit a short sequence of simple program fragments,
roughly ordered by increasing complexity.  None of the examples
exhibits uncaught exceptions.  The rejection of any one of them by a
compiler would constitute a false positive.  The type system and the
compiler that we describe accept them all.

Of course, baseline functionality consists of being able to match a
manifest occurrence of a raised exception with a manifestly matching
handler:
\begin{lstlisting}[style=mlpolyr]
   (... raise `Neg 10 ...) handle `Neg i => ...
\end{lstlisting}

\noindent The next example moves the site where the exception is
raised into a separate function.  To handle this in the type system,
the function type constructor $\rightarrow$ acquires an additional
argument $\rho$ representing the set of exceptions that may be raised by an
application, i.e., function types have the form $\tau_1 \era{\rho} \tau_2$.
This is about as far as existing static
exception trackers that are built into programming languages (e.g.,
Java's {\bf throws} declaration) go.
\begin{lstlisting}[style=mlpolyr]
   fun foo x = if x < 0 then raise `Neg x else ...
   (... foo y ...) handle `Neg i => x ...
\end{lstlisting}

\noindent But we also want to be able to track exceptions through
calls of higher-order functions such as {\tt map}, which themselves do
not raise exceptions while their functional arguments might:
\begin{lstlisting}[style=mlpolyr]
   fun map f [] = []
     | map f (x::xs) = f x :: map f xs
   (... map f l ...) handle `Neg i => ...
\end{lstlisting}

\noindent Moreover, in the case of curried functions and partial
applications, we want to be able to distinguish stages that do not
raise exceptions from those that might.  In the example of {\tt map},
there is no possibility of any exception being raised when {\tt map}
is partially applied to the function argument; all exceptions are
confined to the second stage when the list argument is supplied:
\begin{lstlisting}[style=mlpolyr]
   val mfoo = map foo
   (... mfoo l ...) handle `Neg i => ...
\end{lstlisting}

\noindent Here, the result {\tt mfoo} of the partial application acts as
a data structure that carries a latent exception.  In the general
case, exception values can occur in any data structure.  For example,
the SML/NJ Library~\citep{smlnj-lib} provides a constructor
function for hash tables which accepts a programmer-specified exception
value which becomes part of the table's representation from where it
can be raised, for example when an attempt is made at
looking up a non-existing key.

The following example shows a similar but simpler situation.  Function
{\tt check} finds the first pair in the given list whose left
component does not satisfy the predicate {\tt ok}.  If such a pair
exists, its right component, which must be an exception value, is
raised.  To guarantee exception safety, the caller of {\tt check} must be
prepared to handle any exception that might be passed along in the
argument of the call:
\begin{lstlisting}[style=mlpolyr]
   fun check ((x, e)::rest) = if ok x then check rest else raise e
     | check [] = ()
   (... check [(3, `A 10), (4, `B true)] ...) handle `A i => ... 
                                                   | `B b => ...
\end{lstlisting}

\noindent Finally, exception values can participate in complex data
flow patterns.  The following example illustrates this by showing an
exception {\tt `A} that carries another exception {\tt `B} as its
payload.  The payload {\tt `B 10} itself gets raised by the exception
handler for {\tt `A} in function {\tt f2}, so a handler for {\tt `B}
on the call of {\tt f2} suffices to make this fragment exception-safe:
\begin{lstlisting}[style=mlpolyr]
   fun f1 () = ... raise `A (`B 10) ...
   fun f2 () = f1 () handle `A x => raise x
   (... f2 () ...) handle `B i => ...
\end{lstlisting}

\section{Case study: A two-way extensible interpreter}
\label{sec:exp-prob}

There are two axes along which we can extend a system: functionality
and variety of data. For the first axis, we can add more functionality
on the basic set of data. For the second axis, we can add to the
variety of data on which the basic functions perform.  Ideally, two
dimensional extensions should be orthogonal.  However, depending on
the context, extensions along one axis can be more difficult than
along the other. Simultaneous two-way extensions can be even more
difficult. This phenomenon can be easily explained in terms of
expressions (data) and evaluators (functions), which the reason Wadler
called it {\em the expression problem}~\citep{Wadler98a}.  This
section discusses a two-way extensible interpreter that precisely
captures this phenomenon. Our intention with this case study is to
define a real yet simple example that extends its functionality in an
interesting way.

\subsection*{Base language}

Let us consider a Simple Arithmetic Language (SAL) that contains terms
such as numbers, variables, additions, and a let-binding form.  Not
all expressions that conform to the grammar are actually ``good''
expressions. We want to reject expressions that have ``dangling''
references to variables which are not in scope.  The judgment $\Gamma
\ts e \ok$ expresses that $e$ is an acceptable expression if it
appears in a context described by $\Gamma$.  In this simple case,
$\Gamma$ keeps track of which variables are currently in scope, so we
take it to be a set of variables.  An expression is acceptable as a
{\em program} if it is an expression that makes no demands on its
context, i.e., $\varnothing \ts e \ok$.  When discussing the dynamic
semantics of a language, we need to define its values, i.e., the
results of a computation.  In SAL, values are simply natural numbers.
Then, our evaluation semantics describes the entire evaluation process
as one ``big step''. We write $(E, e) \Downarrow n$ to say that $e$
evaluates to $n$ under environment $E$. The environment is a finite
mapping from variables to values.

\begin{figure}
$\mathsf{Values} ~~~~ n \in N$\\
$\mathsf{Variables} ~ x \in Var$\\
$\mathsf{Terms}  ~~~~~ e \bnfdef n \bnfalt x \bnfalt e + e \bnfalt \eLet{x}{e}{e}$\\
\hrule
\vspace{3mm}
\framebox{$\Gamma \ts e \ok$}\\ \\
$\mathsf{Typing~env.} ~ \Gamma \bnfdef \varnothing \bnfalt \Gamma,x$
\begin{mathpar}
\irr{}
{ }
{\Gamma \ts n \ok}
\and
\irr{}
{x \in \Gamma}
{\Gamma \ts x \ok}
\and
\irr{}
{
\Gamma \ts e_1 \ok \+
\Gamma \ts e_2 \ok
}
{\Gamma \ts e_1 + e_2 \ok}
\and
\irr{}
{
\Gamma \ts e_1 \ok \+
\Gamma,x \ts e_2 \ok
}
{\Gamma \ts \eLet{x}{e_1}{e_2} \ok}
\end{mathpar}
\hrule
\vspace{3mm}
\framebox{$(E,e) \Downarrow n$}\\ \\
$\mathsf{Environment} ~ E \in \mathsf{Var} \rightarrow N$
\begin{mathpar}
\irr{}
{E(x) = n}
{(E,x) \Downarrow n}
\and
\irr{}
{
(E,e_1) \Downarrow n_1 \+
(E,e_2) \Downarrow n_2 \+
n_1 + n_2 = n
}
{(E, e_1 + e_2) \Downarrow n}
\and
\irr{}
{ }
{(E,n) \Downarrow n}
\and
\irr{}
{
(E,e_1) \Downarrow n_1 \+
(E[x \rightarrow n_1],e_2) \Downarrow n_2
}
{(E, \eLet{x}{e_1}{e_2}) \Downarrow n_2}
\end{mathpar}
\caption{The Simple Arithmetic Languages (SAL): syntax (top), 
the static semantics (2nd) and the evaluation semantics (bottom).}
\label{fig:pl-base}
\end{figure}

\figref{base-int} shows a simple implementation for the base
interpreter which is the composition of the function $\mathsf{check}$
(realizing the static semantics) and $\mathsf{eval}$ (realizing the
evaluation semantics).  As explained in \secref{recs-sums}, our
language \mlpolyr{} has polymorphic sum types. The type system is
based on R\'{e}my-style {\em row polymorphism}, handles equi-recursive
types, and can infer principal types for all language constructs.  For
function $\mathsf{eval}$ in \figref{base-int}, the compiler calculates
the following type. 

\parbox{5in}{\begin{codeListingNormal}
xxxxxxx\=(($\alpha$ {\bf as} <\= \kill
{\bf val} eval: \\
$\forall \beta:\varnothing$. 
\> (($\alpha$ {\bf as} <`Let {\bf of} (string, $\alpha$, $\alpha$), \\
\>\>                   `Num {\bf of} int, \\
\>\>                   `Plus {\bf of} ($\alpha$, $\alpha$),\\
\>\>                   `Var {\bf of} string>), string $\era{\beta}$ int) $\era{\beta}$ int \\
\end{codeListingNormal}}

Here $\alpha$ is a recursive sum type,
indicated by keyword {\bf as} and a type row closed in {\tt < $\ldots$
  >}. $\beta$ is a row type variable constrained to a particular {\it
  kind} representing a set of labels that must be absent in any
instantiation.

\begin{figure}
\begin{lstlisting}[style=mlpolyr]
   (* environment *)
   fun bind (a, x, env) y = 
       if String.compare (x, y) == 0 then a else env y

   fun empty x = 
       raise `Fail (String.concat [``unbound variable: ``, x, ``\n''])

   (* the static semantics *)
   (* check returns () or fails with `Fail *) 
   fun check (e, env) = match e with
       cases `Var x => env x
           | `Num n => ()
           | `Plus (e1, e2) => (check (e1, env); check (e2, env))
           | `Let (x, e1, e2) => (check (e1, env); 
                                  check (e2, bind ((), x, env))

   (* the evaluation semantics *)
   fun eval (e, env) = match e with
       cases `Var x => env x
           | `Num n => n
           | `Plus (e1, e2) => eval (e1, env) + eval (e2, env)
           | `Let (x, e1, e2) => 
              eval (e2, bind (eval (e1, env), x, env))

   (* the interpreter obtained by composing two functions *)
   fun interp e = 
       try r = (check (e, empty); eval (e, empty))
       in r
       handling `Fail msg => (String.output msg; -1)
       end
\end{lstlisting}
\caption{A simple implementation for the base interpreter.}
\label{fig:base-int}
\end{figure}

\subsection*{Preparation for extensions}

Because it is desirable to extend the base language by new language
features, we had better prepare for language extensions.  In
\mlpolyr{}, first-class extensible cases can be helpful to make code
extensible. Case expressions have an elimination form,
$\eMat{e_1}{e_2}$ where $e_1$ is a scrutinee and $e_2$ is a case
expression.  First, we separate cases from the scrutinee in the
$\mathbf{match}$ expression. Then, we parameterize them by closing
over their free variables.  One of these free variables is the
recursive instance of the current function itself. This design
achieves open-recursion. With this setting, it becomes easy to add a
new variant (i.e., new cases).  For example, \figref{prepare} shows
the old function $\mathsf{check}$ becomes a pair of
$\mathsf{check}$\_$\mathsf{case}$ and $\mathsf{check}$. The new
version of $\mathsf{eval}$ follows the same pattern. For
$\mathsf{eval\_case}$, the compiler calculates the following type and
here it shows that its return type is the case type denoted by
$\tSum{\rho} \ehra{} \tau$:

\parbox{5in}{\begin{codeListingNormal}
xxxxxxx\=(<\= \kill
{\bf val} eval\_case: \\
$\forall \beta:\varnothing$. 
\> (($\alpha$, string $\era{\beta}$ int) $\era{\beta}$ int, string $\era{\beta}$ int) $\ra$ \\
\> (<`Let {\bf of} (string, $\alpha$, $\alpha$), \\
\>\>  `Num {\bf of} int, \\
\>\>  `Plus {\bf of} ($\alpha$, $\alpha$),\\
\>\>  `Var {\bf of} string>) $\ehra{\beta}$ int) \\
\end{codeListingNormal}}

\begin{figure}
\begin{lstlisting}[style=mlpolyr]
   (* extensible cases for the static semantics *)
   fun check_case (check, env) =
       cases `Var x => env x
           | `Num n => ()
           | `Plus (e1, e2) => (check (e1, env); check (e2, env))
           | `Let (x, e1, e2) => (check (e1, env); 
                                  check (e2, bind ((), x, env))

   (* close open recursion for the static semantics *)
   fun check (e, env) = match e with check_case (check, env)

   (* extensible cases for the evaluation semantics *)
   fun eval_case (eval, env) = 
       cases `Var x => env x
           | `Num n => n
           | `Plus (e1, e2) => eval (e1, env) + eval (e2, env)
           | `Let (x, e1, e2) => 
              eval (e2, bind (eval (e1, env), x, env))

   (* close open recursion for the evaluation semantics *)
   fun eval (e, env) = match e with eval_case (eval, env)
\end{lstlisting}
\caption{Preparation for extensions.}
\label{fig:prepare}
\end{figure}

\begin{figure}
\begin{center}
$\mathsf{Terms}  ~~~~~ e \bnfdef \ldots \bnfalt \eIfz{e}{e}{e}$\\
\end{center}
\hrule
\vspace{3mm}
\framebox{$\Gamma \ts e \ok$}\\ \\
\begin{mathpar}
\irr{}
{
\Gamma \ts e_1 \ok \+
\Gamma \ts e_2 \ok \+
\Gamma \ts e_3 \ok
}
{\Gamma \ts \eIfz{e_1}{e_2}{e_3} \ok}
\end{mathpar}
\hrule
\vspace{3mm}
\framebox{$(E,e) \Downarrow n$}\\ \\
\begin{mathpar}
\irr{}
{
(E, e_1) \Downarrow 0 \+
(E, e_2) \Downarrow n_2
}
{
(E, \eIfz{e_1}{e_2}{e_3}) \Downarrow n_2
}
\and
\irr{}
{
(E, e_1) \Downarrow n_1 \+
n_1 \neq 0 \+
(E, e_3) \Downarrow n_3
}
{
(E, \eIfz{e_1}{e_2}{e_3}) \Downarrow n_3}
\end{mathpar}
\hrule
\vspace{3mm}
\framebox{$(k,E,e) \Rightarrow (k',E',e') ~~~ 
           (k,E,e) \Rightarrow (n, k') ~~~ 
           (n,k) \Rightarrow (n',k') ~~~
           (n,k) \Rightarrow (k', E, e)$
         }\\ \\
$\mathsf{Frame}  ~~~ f \bnfdef \tSum{[]+e,E} \bnfalt \tSum{n+[]} \bnfalt
                               \tSum{\eLet{x}{[]}{e},E} \bnfalt 
                               \tSum{\eIfz{[]}{e}{e}}$\\
$\mathsf{Stack}  ~~~~ k \bnfdef \cdot \bnfalt f \triangleright k$\\
\begin{mathpar}
(k,E,x) \Rightarrow (E(x),k)
\and
(k,E,n) \Rightarrow (n,k)
\and\and
(k,E,e_1+e_2) \Rightarrow (\tSum{[]+e_2,E}\triangleright k,E,e_1)
\and
(k,E,\eLet{x}{e_1}{e_2}) \Rightarrow (\tSum{\eLet{x}{[]}{e_2},E}\triangleright k,E,e_1)
\and
(k,E,\eIfz{e_1}{e_2}{e_3}) \Rightarrow (\tSum{\eIfz{[]}{e_2}{e_3},E}\triangleright k,E,e_1)
\and
(n, \tSum{[]+e,E} \triangleright k) \Rightarrow (\tSum{n+[]}\triangleright k,E,e)
\and
(n, \tSum{n'+[]}) \triangleright k) \Rightarrow (n' + n, k)
\and
(n, \tSum{\eLet{x}{[]}{e_2},E} \triangleright k) \Rightarrow (k, E[x\rightarrow n],e_2)
\and
(0, \tSum{\eIfz{[]}{e_2}{e_3},E} \triangleright k) \Rightarrow (k, E, e_2)
\and
(n, \tSum{\eIfz{[]}{e_2}{e_3},E} \triangleright k) \Rightarrow (k, E, e_3) ~~~\mathsf{where}~~~ n \neq 0
\end{mathpar}
\hrule
\vspace{3mm}
\framebox{$e \gg e'$}\\
\begin{mathpar}
{n_1 + n_2 \gg n; n = n_1 + n_2}
\and
{\eIfz{0}{e_2}{e_3} \gg e_2}
\and
{\eIfz{n}{e_2}{e_3} \gg e_3; n \neq 0}
\end{mathpar}
\caption{Language extensions: syntax (top), the static semantics (2nd), 
the evaluation semantics (3rd), the machine semantics (4th) and
optimization rules (bottom).}
\label{fig:pl-ext}
\end{figure}

\subsection*{Language extensions}

\figref{pl-ext} shows how the base language grows. As a conditional
term $\mathsf{If0}$ is introduced, the corresponding rule sets for
both the static semantics ($\mathsf{check}$) and the evaluation
semantics ($\mathsf{eval}$) are changed.  Instead of the evaluation
semantics, alternatively, we can define the machine semantics
($\mathsf{eval}_{\mathsf{m}}$) which makes control explicit by
representing computation stages as stacks of frames.  Each frame $f$
corresponds to a piece of work that has been postponed until a
sub-computation is complete.  Our machine semantics follows the
conventional single-step transition rules between states
\citep{Harper05}.  It consists of expression states $(k, E, e)$, value
states $(n, k)$ and a transition relation between states where $k$ is
a stack and $e$ is the current expression. The empty stack is $\cdot$
and a frame $f$ on top of stack $k$ is written $f \triangleright k$.
The machine semantics is given as a set of single-step transition
rules $(k, E, e) \Rightarrow (k', E', e')$ and $(n, k) \Rightarrow
(n', k')$ between states.  Additionally, optimization rules may be
introduced.  We write $e \gg e'$ to say that $e$ is translated into
$e'$ by performing some simple optimization. In our running example,
we consider constant folding and short-circuiting techniques.

\subsection*{Implementation of extensions}

With our preparation for extensions in place, we only have to focus on
a single new case ($\mathsf{`If0}$) by letting the original set of
other cases be handled by $\mathsf{check}$\_$\mathsf{case}$.
\figref{ext-int} shows how an extended checker $\mathsf{echeck}$, now
handling five cases including $\mathsf{`If0}$, is obtained by closing
the recursion through applying $\mathsf{echeck}$\_$\mathsf{case}$ to
$\mathsf{echeck}$~(Line 8).  The extension of $\mathsf{eval}$, called
$\mathsf{eeval}$, is constructed analogously by applying
$\mathsf{eeval\_case}$ whose types is computed as follows:

\parbox{5in}{\begin{codeListingNormal}
xxxxxxx\=(<\= \kill
{\bf val} eeval\_case: \\
$\forall \beta:\varnothing$. 
\> (($\alpha$, string $\era{\beta}$ int) $\era{\beta}$ int, string $\era{\beta}$ int) $\ra$ \\
\> (<`If0 {\bf of} ($\alpha$, $\alpha$, $\alpha$), \\
\>\>  `Let {\bf of} (string, $\alpha$, $\alpha$), \\
\>\>  `Num {\bf of} int, \\
\>\>  `Plus {\bf of} ($\alpha$, $\alpha$),\\
\>\>  `Var {\bf of} string>) $\ehra{\beta}$ int) \\
\end{codeListingNormal}}

Finally, the extended
interpreter can be obtained by applying $\mathsf{eeval}$ and
$\mathsf{echeck}$, instead of $\mathsf{eval}$ and $\mathsf{check}$
(Line 22).

\begin{figure}
\begin{lstlisting}[style=mlpolyr]
   (* extends check_case with a new case (`If0) *)
   fun echeck_case (check, env) =
       cases `If0 (e1, e2, e3) => 
             (check (e1, env); check (e2, env); check (e3, env))
       default: check_case (check, env)
     
   (* close open recursion with the extension *)
   fun echeck (e, env) = match e with echeck_case (echeck, env)

   (* extends eval_case with a new case (`If0) *)
   fun eeval_case (eval, env) = 
       cases `If0 (e1, e2, e3) =>
             if eval (e1, env) == 0 
             then eval (e2, env) else eval (e3, env)
       default: eval_case (eval, env)

   (* close open recursion with the extension *)
   fun eeval (e, env) = match e with eeval_case (eeval, env)

   (* the extended interpreter by composing extended functions *)
   fun einterp e = 
       try r = (echeck (e, empty); eeval (e, empty))
       in r
       handling `Fail msg => (String.output msg; -1)
       end
\end{lstlisting}
\caption{Implementation for extensions.}
\label{fig:ext-int}
\end{figure}

Adding new kinds of functions such as a new optimizer ($\mathsf{opt}$) 
does not require any preparation in \mlpolyr{}. 
For example, the combinator $\mathsf{opt}$ which performs
constant folding may be inserted to build an optimized one:
\begin{lstlisting}[style=mlpolyr]
   (* the helper function for handling three cases *)
   fun nope f = cases `VAR x => f (`VAR x)
	            | `PLUS (e1, e2) => f (`PLUS (e1, e2))
	            | `LET (x, e1, e2) => f (`LET (x, e1, e2))

   (* `PLUS (`NUM n1, `NUM n2) >> `NUM (n1+n2)  *)
   (*  otherwise, return arguments as received. *)
   fun chkPLUS (e1, e2) = match e1 with 
       cases `NUM n1 => (match e2 with 
              cases `NUM n2 => `NUM (n1+n2) 
              default: nope (fn _ => `PLUS (e1, e2)))
       default: nope (fn _ => `PLUS (e1, e2))

   (* the optimization rules *)
   fun opt e = match e with
       cases `Var x => `Var x
           | `Num n => `Num n
           | `Plus (e1, e2) => chkPlus (opt e1, opt e2)
           | `Let (x, e1, e2) => `Let (x, e1, e2)

   (* the optimized interpreter by composing three functions *)
   fun optimizedInterp e = 
       try r = (check (e, empty); eval (opt e, empty))
       in r
       handling `Fail msg => (String.output msg; -1)
       end
\end{lstlisting}

\noindent where we define a function $\mathsf{chkPlus}$ which returns
$\mathsf{`Num (n_1+n_2)}$ if two arguments are recursively optimized
to $\mathsf{`Num (n_1)}$ and $\mathsf{`Num (n_2)}$, respectively.
Otherwise, it returns $\mathsf{`Plus (opt~e_1, opt~e_2)}$.  Even
though adding functions does not impose any trouble, $\mathsf{opt}$
itself should also be prepared for extension because $\mathsf{opt}$
itself may be extended to support a conditional term:
\begin{lstlisting}[style=mlpolyr]
   (* extensible cases for the optimization rules *)
   fun opt_case opt =
       cases `Var x => `Var x
           | `Num n => `Num n
           | `Plus (e1, e2) => chkPlus (opt e1, opt e2)
           | `Let (x, e1, e2) => `Let (x, e1, e2)

   (* close open recursion for the optimization rules *)
   fun opt e = match e with opt_case opt
\end{lstlisting}
 
\subsection*{Related work}

By using the well-known {\em expression problem}, we have demonstrated
the \mlpolyr{} language features make it possible to easily extend
existing code with new cases. Such extensions do not require any
changes to code in a style of composable extensions. These language
mechanisms play an important role in providing a solution to the
expression problem.  Since Wadler described the difficult of the
two-way extensions, there have been many attempts at solving the
expression problem.

Most of them have been studied in an object-oriented
context~\citep{Odersky97, Bourdoncle97, Findler-Flatt99, Flatt99,
   Bruce03}.  Some tried to adopt functional style using
the {\em Visitor} design pattern to achieve easy extensions to adding
new operations~\cite{Gamma:1995:DPE}.  However, this approach made it
difficult to add new data. To obtain extensibility in both dimensions,
variants were proposed such as the {\em Extensible Visitor} pattern
and extensible algebraic datatypes with defaults
~\cite{Krishnamurthi98,Zenger-Odersky01} but they did not guarantee
static type safety.  Torgersen provided his solution using generics
and a simple trick (in order to overcome typing problems) in
Java~\cite{Torgersen04}. His insight was to use genericity to allow
member functions to extend without modifying the type of parent's
class but his approach required rather complex programming protocols
to be observed.

As the functional approach, Garrigue presented his solution based on
polymorphic variants in OCaml~\cite{Garrigue98,Garrigue00}.  As Zenger
and Odersky point out~\cite{Zenger-Odersky01}, variant dispatching
requires explicit forwarding of function calls.  This is a consequence
of the fact that in Garrigue's system, extensions need to know what
they are extending.  As a result, his solution is similar to our
two-way extensible interpreter example but somewhat less general.

Because extensions along one direction can be more difficult than
along the other depending on implementation mechanisms, the expression
problem is often said to reveal ``tension in language
design''\cite{Wadler98a}. Naturally, there have been attempts to live
in the ``best of both worlds'' in order to design languages powerful
enough to provide better solutions.  For example, the Scala language
integrates features of object-oriented and functional languages and
provides type-safe solutions by using its abstract types and mixin
composition~\cite{Zenger-Odersky05}.  OCaml also presents the similar
solutions due to the benefits of its integration of object-oriented
features to ML~\cite{Remy98,Remy04}.  As a smooth way of integration,
OML and Extensible ML (EML) generalize ML constructs to support
extensibility instead of directly providing class and method
definitions as in OCaml~\cite{Reppy96,MillCoCraig02}. Especially, EML
supports extensible functions as well as extensible
datatypes. However, a function's extensibility in EML is second-class
and EML requires explicitly type annotations due to difficulty of
polymorphic type inference in the presence of subtyping while
extensible cases in \mlpolyr{} are first-class values and fully
general type inference is provided by a variant of the classic
algorithm $W$~\cite{Milner78} only extended to handle R\'emy-style row
polymorphism and equi-recursive types.

\section{The External Language (\EL)}
\label{sec:el}
In this section, we explore theoretical aspects of the \mlpolyr{}
language that we have seen informally. First, we start by describing
\EL{}, our implicitly typed {\em external} language that provides
sums, cases, and mechanisms for raising as well as handling
exceptions.

%
%
\subsection{Syntax}
%
%

\begin{figure}
\[
\begin{array}{lrrl}
\mathsf{Terms} &
e & \bnfdef & n \bnfalt x \bnfalt l~e \bnfalt e_1~e_2 \bnfalt
	      \eFun{f}{x}{e} \bnfalt \eLet{x}{e_1}{e_2} \\
& & \bnfalt & \eRecN{i=1}{n}{l_i = e_i} \bnfalt \eRecExt{e_1}{\eRec{l=e_2}} 
              \bnfalt \eRecSub{e}{l} \bnfalt e.l \\
& & \bnfalt & \eCaseN{i=1}{n}{l_i~x_i \dra e_i} \bnfalt
	      \eMat{e_1}{e_2} \bnfalt \eCaseSub{e}{l} \bnfalt
              \eCaseExt{e_1}{\eCase{l~x \dra e_2}} \\
& & \bnfalt & \eRaise{e} \bnfalt \eHandle{e_1}{\eCase{l~x \dra e_2}} \\
& & \bnfalt & \eRehandle{e_1}{\eCase{l~x \dra e_2}} \bnfalt
	      \eHandle{e_1}{\eCase{x \dra e_2}} \bnfalt
              \eUnhandle{e}{l} \\[1mm]

\mathsf{Values} &
v & \bnfdef & n \bnfalt \eFun{f}{x}{e} \bnfalt  l~v \bnfalt 
              \eRecN{i=1}{n}{l_i = v_i} \bnfalt 
              \eCaseN{i=1}{n}{l_i~x_i \dra e_i} \\
\mathsf{Kinds} &
\kappa & \bnfdef & \star \bnfalt L  \\

\mathsf{Label~sets} &
L      &\bnfdef  & \{l_1, \ldots, l_n\} \bnfalt \varnothing \\

\mathsf{Types} &
\tau & \bnfdef & \alpha \bnfalt \Int \bnfalt \tau_1 \era{\rho} \tau_2 \bnfalt 
                 \etCase{\tSum{\rho_1}}{\tau}{\rho_2} \bnfalt
                 \tRec{\rho} \bnfalt \tSum{\rho} \bnfalt
                 \tBind{\alpha}{\tSum{\rho}} \\
& \rho & \bnfdef & \alpha \bnfalt \tRowEmp \bnfalt l:\tau, \rho \\
& \theta & \bnfdef & \tau \bnfalt \rho \\

\mathsf{Schemas} &
\sigma & \bnfdef & \tau \bnfalt \forall \alpha:\kappa.\sigma \\

\mathsf{Typenv} &
\Gamma & \bnfdef & \varnothing \bnfalt \Gamma, x \mapsto \sigma \\

\mathsf{Kindenv} &
\Delta & \bnfdef & \varnothing \bnfalt \Delta, \alpha \mapsto \kappa
\end{array}
\]
\caption{External language (\EL) syntax.}
\label{fig:el-syntax}
\end{figure}

 \figref{el-syntax} shows the definitions of
expressions $e$ and values $v$. We have integer constants $n$,
variables $x$, injection into sum types $l~e$, applications $e_1~e_2$,
recursive functions $\eFun{f}{x}{e}$, {\it let}-bindings
$\eLet{x}{e_1}{e_2}$.  For record expressions, we have record
constructors $\eRec{l_1 = e_1, \ldots, l_n = e_n}$ (which we will
often abbreviate as $\eRecN{i=1}{n}{l_i = e_i}$), record extensions
$\eRecExt{e_1}{\eRec{l=e_2}}$, record subtractions $\eRecSub{e}{l}$
and record selections $e.l$.  For case expressions, we have case
constructors $\eCase{l_1~x_1 \dra e_1, \ldots, l_n~x_n \dra e_n}$
(abbreviated as $\eCaseN{i=1}{n}{l_i~x_i\dra e_i}$), case extensions
$\eCaseExt{e_1}{\eCase{l~x \dra e_2}}$, case subtractions
$\eCaseSub{e}{l}$ and match expressions $\eMat{e_1}{e_2}$ which
matches $e_1$ to the expression $e_2$ whose value must be a case.
There are also $\eRaise{e}$ for raising exceptions and several forms
for managing exception handlers: The form $\eHandle{e_1}{\eCase{l~x
    \dra e_2}}$ establishes a handler for the exception constructor
$l$.  The new exception context is used for evaluating $e_1$, while
the old context is used for $e_2$ in case $e_1$ raises $l$. The old
context cannot already have a handler for $l$.  The form
$\eRehandle{e_1}{\eCase{l~x \dra e_2}}$, on the other hand, overrides
an existing handler for $l$.  Again, the original exception context is
restored before executing $e_2$.  The form $\eHandle{e_1}{\eCase{x
    \dra e_2}}$ establishes a new context with handlers for {\em all}
exceptions that $e_1$ might raise.  As before, $e_2$ is evaluated in
the original context.  The form $\eUnhandle{e}{l}$ evaluates $e$ in a
context from which the handler for $l$ has been removed.  The original
context must have a handler for $l$.

The type language for \EL{} is also given in \figref{el-syntax}.  It
contains type variables ($\alpha,\beta,\ldots$), base types (e.g.,
$\Int$), constructors for function- and case types ($\ra$ and $\hra$),
record types ($\tRec{\rho}$), sum types ($\tSum{\rho}$),
recursive sum types ($\tBind{\alpha}{\tSum{\rho}}$), 
the empty row type ($\tRowEmp$), and row
types with at least one typed label ($l:\tau,\rho$).  Notice that
function- and case arrows take {\em three} type arguments: the domain,
the co-domain, and a row type describing the exceptions that could be
raised during an invocation.  A type $\theta$ is either an ordinary type 
$\tau$ or a row type $\rho$.  
Kinding judgments of the form $\Delta \ts \tau : \kappa$
(stating that in the current kinding context $\Delta$ type $\tau$ has
kind $\kappa$) are used to distinguish between these cases and to
establish that types are well-formed.  As a convention, wherever
possible we will use meta-variables such as $\rho$ for row types and
$\tau$ for ordinary types.  Where this distinction is not needed, for
example for polymorphic instantiation (\rname{var} in 
~\figref{el-typing-a}), we will use the letter $\theta$.

%
%

\begin{figure}
\[
\begin{array}{c}









\infer
{\Delta \ts \alpha : \star}
{\Delta(\alpha) = \star}
\quad

\infer
{\Delta \vdash \Int : \star}
{\strut}
\quad

\infer
{\istyp{\Delta}{\tau \era{\rho} \tau'}}
{\istyp{\Delta}{\tau} & \istyp{\Delta}{\tau'} & \isrow{\Delta}{\rho}}
\quad

\infer
{\Delta \ts \tRowEmp : L}
{\strut}
\quad

\infer
{\istyp{\Delta}{\tSum{\rho}}}
{\isrow{\Delta}{\rho}}

\\[2mm]

\infer
{\Delta \ts \alpha : L}
{L \subseteq \Delta(\alpha)}

\quad

\infer
{\istyp{\Delta}{\etCase{\tSum{\rho'}}{\tau}{\rho}}}
{\isrow{\Delta}{\rho'} & \istyp{\Delta}{\tau} & \isrow{\Delta}{\rho}}
\quad

\infer
{\Delta \ts (l:\tau, \rho) : L}
{
\istyp{\Delta}{\tau} &
\Delta \ts \rho : L \cup \tRec{l} &
l \not\in L
}
\end{array}
\]
\caption{Well-formedness for types in \EL{}.}
\label{fig:el-wf}
\end{figure}


Ordinary types have kind $\star$.  A row type $\rho$ has kind $L$
where $L$ is a set of labels which are known not to occur in
$\rho$.  An unconstrained row variable has kind $\varnothing$.
Inference rules are given in \figref{el-wf}.
The use of a kinding judgment in a typing rule
constrains $\Delta$ and ultimately propagates kinding information back
to the \rname{let/val} rule in ~\figref{el-typing-a} 
where type variables are bound and kinding
information is used to form type schemas denoted by $\sigma$.

%
%

\subsection{Operational semantics}
%
%

\begin{figure}[t]
\[
\begin{array}{lcl}
e    & \bnfdef & \ldots \bnfalt \eRestore{}{} \\[2mm]

E & \bnfdef & [] \bnfalt l~E \bnfalt 
               E~e \bnfalt v~E \bnfalt
               \eLet{x}{E}{e} \\
  & \bnfalt & \eRec{\ldots,l_{i-1}=v_{i-1},l_i=E,l_{i+1}=e_{i+1},\ldots} \bnfalt 
              E.l \bnfalt \\
  & \bnfalt & \eRecExt{E}{\eRec{l=e}} \bnfalt 
              \eRecExt{v}{\eRec{l=E}} \bnfalt 
              \eRecSub{E}{l} \\
  & \bnfalt & \eCaseExt{E}{\eCase{l~x \dra e}} \bnfalt
              \eCaseSub{E}{l} \\
  & \bnfalt & \eMat{E}{e} \bnfalt \eMat{v}{E} \\
  & \bnfalt & \eRaise{E} \bnfalt \eRestore{\EXN}{E} \\[2mm]

r & \bnfdef & (\eFun{f}{x}{e})~v \\
  & \bnfalt & \eLet{x}{v}{e} \\
  & \bnfalt & v.l \\
  & \bnfalt & \eRecExt{v_1}{\eRec{l = v_2}} \\
  & \bnfalt & \eRecSub{v}{l} \\
  & \bnfalt & \eCaseExt{v}{\eCase{l~x \dra e}} \\
  & \bnfalt & \eCaseSub{v}{l} \\
  & \bnfalt & \eMat{v_1}{v_2} \\
  & \bnfalt & \eRaise{l~v} \\
  & \bnfalt & \eHandle{e_1}{\eCase{l~x \dra e_2}} \\
  & \bnfalt & \eRehandle{e_1}{\eCase{l~x \dra e_2}} \\
  & \bnfalt & \eUnhandle{e}{l} \\
  & \bnfalt & \eHandle{e_1}{\eCase{x \dra e_2}} \\
  & \bnfalt & \eRestore{\EXN}{v} \\[2mm]



\EXN & \bnfdef & \eRec{l_1 = E_1, \ldots, l_n = E_n}
\end{array}
\]
\caption{Evaluation contexts $E$, redexes $r$ and exception contexts \EXN.}
\label{fig:el-contexts}
\end{figure}


We give an operational small-step semantics for \EL{} as a
context-sensitive rewrite system in a style inspired by Felleisen and
Hieb~\cite{felleisen-hieb:92:eval-contexts}.  An {\em evaluation
  context} $E$ is essentially a term with one sub-term replaced by a
hole (see \figref{el-contexts}). Any closed expression $e$ that is not
a value has a unique decomposition $E[r]$ into an evaluation context
$E$ and a redex $r$ that is placed into the hole within $E$.
Evaluation contexts in this style of semantics represent continuations.  
The rule for handling an
exception could be written simply as
\begin{math}
    E[\eHandle{(E'[\eRaise{l~v}])}{\eCase{l\,x \dra e}}] \mapsto E[e[v/x]],
\end{math}
but this requires an awkward side-condition stating that $E'$ must not
also contain a handler for $l$.  We avoid this difficulty by
maintaining the exception context separately and explicitly on a
per-constructor basis.  This choice makes it clear that exception
contexts can be seen as extensible records of continuations.  However,
we now also need to be explicit about where a computation re-enters
the scope of a previous context.  This is the purpose of
{restore}-frames of the form \eRestore{\EXN}{E} that we added to the
language, but which are assumed not to occur in source
expressions. There are real-world implementations of languages
with exception handlers where restore-frames have a concrete
manifestation.  For example, SML/NJ~\cite{smlnj} represents the
exception handler as a global variable storing a continuation.  When
leaving the scope of a handler, this variable gets assigned the
previous exception continuation.

An {\em exception context} $\EXN$ is a record
$\eRec{l_1=E_1,\ldots,l_n=E_n}$ of evaluation contexts
$E_1,\ldots,E_n$ labeled $l_1,\ldots,l_n$.  A {\em reducible
  configuration} $(E[r],\EXN)$ pairs a redex $r$ in context $E$ with a
corresponding exception context $\EXN$ that represents all
exception handlers that are available when reducing $r$.  A {\em final
  configuration} is a pair $(v,\eRec{})$ where $v$ is a value. Given a
reducible configuration $(E[r], \EXN)$, we call the pair $(E,\EXN)$
the {\em full context} of $r$.

The semantics is given as a set of single-step transition rules from
reducible configurations to configurations:
$(E[r],\EXN) \mapsto (E[e],\EXN')$. That is, a pair of an evaluation context
with a redex $E[r]$ and an exception context $\EXN$ evaluates to
a pair of an evaluation context with an evaluated expression $E[e]$ and
a new exception context $\EXN'$ in a single step. A program (i.e., a closed
expression) $e$ evaluates to a value $v$ if $(e,\eRec{})$ can be
reduced in the transitive closure of our step relation to a final
configuration $(v,\eRec{})$.  Rules unrelated to exceptions are
standard and leave the exception context unchanged.  The rule for
$\eRaise{l~v}$ selects field $l$ of the exception context and places
$v$ into its hole. The result, paired with the empty exception
context, is the new configuration which, by construction, will have
the form $(E'[\eRestore{\EXN'}{v}],\eRec{})$ so that the next step
will restore exception context $\EXN'$.
The rules for \eHandle{e_1}{\eCase{l~x\dra e_2}} 
and \eRehandle{e_1}{\eCase{l~x\dra e_2}} 
as well as \eUnhandle{e}{l} are very similar to each other: one
adds a new field to the exception context, another replaces an
existing field, and the third drops a field.  All exception-handling
constructs augment the current evaluation context with a {\bf
  restore}-form so that the original context is re-established if and
when $e_1$ reduces to a value.

%
%

%
%

\begin{landscape}
\begin{figure}
\[
\begin{array}{rcll}

(E[(\eFun{f}{x}{e})~v], \EXN) & \longmapsto &  
(E[e[\eFun{f}{x}{e}/f,v/x]], \EXN) 
& \rlabel{app} \\[1mm]

(E[\eLet{x}{v}{e}], \EXN) & \longmapsto &  
(E[e[v/x]], \EXN) 
& \rlabel{let} \\[1mm]

(E[\eRec{\ldots,l_i = v_i,\ldots}.l_i], \EXN) & \longmapsto & 
(E[v_i], \EXN)
& \rlabel{r/sel} \\[1mm]

(E[\eRecExt{\eRecN{i=1}{n}{l_i = v_i}}{\eRec{l = v}}], \EXN) & \longmapsto & 
(E[\eRec{l_1 = v_1,\ldots,l_n = v_n, l = v}], \EXN)
& \rlabel{r/ext} \\[1mm]

(E[\eRecSub{\eRecN{i=1}{n}{l_i = v_i}}{l_j}], \EXN) & \longmapsto & 
(E[\eRecN{i=1,i \neq j}{n}{l_i = v_i}], \EXN)
& \rlabel{r/sub} \\[1mm]

(E[\eCaseExt{\eCaseN{i=1}{n}{l_i~x_i \dra e_i'}}{\eCase{l~x \dra e}}], \EXN) & \longmapsto & 
(E[\eCase{l_1~x_1 \dra e_1',\ldots,l_n~x_n \dra e_n', l~x \dra e}], \EXN)
& \rlabel{c/ext} \\[1mm]

(E[\eCaseSub{\eCaseN{i=1}{n}{l_i~x_i \dra e_i'}}{l}], \EXN) & \longmapsto & 
(E[\eCaseN{i=1,i\neg j}{n}{l_i~x_i \dra e_i'}], \EXN)
& \rlabel{c/sub} \\[1mm]

(E[\eMat{l_i~v}{\eCase{...,l_i~x_i \dra e_i,...}}], \EXN) & \longmapsto & 
(E[e_i[v/x_i]], \EXN)
& \rlabel{match} \\[1mm]

(E[\eRaise{l_i~v}], \eRec{\ldots,l_i=E_i,\ldots}) & \longmapsto &  
(E_i[v], \eRec{}) 
& \rlabel{raise} \\[1mm]

(E[\eHandle{e_1}{\eCase{l~x \dra e_2}}], \EXN) & \longmapsto &  
(E[\eRestore{\EXN}{e_1}], \EXN') 
& \rlabel{handle} \\[1mm]
& & \hspace{-50mm}\tx{where} ~~~\EXN = \eRecN{i=1}{n}{l_i=E_i} \\
& & \hspace{-50mm}\tx{and} ~~~~~~\EXN' = \eRec{l_1=E_1,\ldots,l_n=E_n,l=E[\eLet{x}{\eRestore{\EXN}{[]}}{e_2}]}\\

(E[\eRehandle{e_1}{\eCase{l_j~x \dra e_2}}], \EXN) & \longmapsto & 
(E[\eRestore{\EXN}{e_1}], \EXN') 
& \rlabel{rehandle} \\[1mm]
& & \hspace{-50mm}\tx{where} ~~~ \EXN = \eRecN{i=1}{n}{l_i=E_i} \tx{~and~}
                  \EXN' = \eRecN{i=1}{n}{l_i=E'_i} \tx{~and~}
                  \forall i \neq j . E'_i = E_i \\
& & \hspace{-50mm}\tx{and}  ~~~~~~
		  E'_j = E[\eLet{x}{\eRestore{\EXN}{[]}}{e_2}] \\
(E[\eUnhandle{e}{l_j}], \EXN) & \longmapsto &  
(E[\eRestore{\EXN}{e}], \EXN') 
& \rlabel{unhandle} \\[1mm]
& &\hspace{-50mm}\tx{where} ~~~ \EXN = \eRecN{i=1}{n}{l_i=E_i} \tx{~and~}
		                \EXN' = \eRecN{i=1,i\neq j}{n}{l_i=E_i} \\
(E[\eHandle{e_1}{\eCase{x \dra e_2}}], \EXN) & \longmapsto &  
(E[\eRestore{\EXN}{e_1}], \EXN') 
& \rlabel{handle~all} \\[1mm]
& & \hspace{-50mm}\tx{where}~~~
	 \EXN' = \eRecN{i=1}{n}{l_i=E[\eLet{x}{l_i(\eRestore{\EXN}{[]})}{e_2}]}
         \tx{~(for~some~n)}\\
(E[\eRestore{\EXN'}{v}], \EXN) & \longmapsto &  
(E[v], \EXN') 
& \rlabel{restore}
\end{array}
\]
\caption{Operational semantics for \EL.}
\label{fig:el-opsem}
\end{figure}
\end{landscape}


%
%
\subsection{Static semantics}
%
%

\begin{figure}
\begin{mathpar}
\irr{\rlabel{var}}
{
\Gamma(x) = \forall \alpha_1:\kappa_1 \ldots \forall \alpha_n:\kappa_n.\tau \+
\forall i_{\in 1..n}. \Delta \ts \theta_i : \kappa_i
}
{
\Delta;\Gamma  \tsv  x : \tau[\theta_1 / \alpha_1,\ldots,\theta_n / \alpha_n]
}
\and
\irr{\rlabel{int}}
{ }
{
\Delta;\Gamma  \tsv  n : \Int
}
\and
\irr{\rlabel{c}}
{
\forall i_{\in 1..n}. \Delta;\Gamma,x_i\mapsto\tau_i \ts e_i : \tau;\rho \+
\isrow{\Delta}{(l_1:\tau_1,\ldots,l_n:\tau_n,\tRowEmp)}
}
{
\Delta;\Gamma \tsv \eCaseN{i=1}{n}{l_i~x_i \dra e_i} 
: \etCase{\tSumN{i=1}{n}{l_i:\tau_i}}{\tau}{\rho}
}
\and
\irr{\rlabel{fun/val}}
{
\Delta;\Gamma, f \mapsto (\forall \alpha:\varnothing.\tau_2 \era{\alpha} \tau),
        x \mapsto\tau_2  \tsv  e : \tau \+
\istyp{\Delta}{\tau_2} \+
\isrow{\Delta}{\rho}
}
{
\Delta;\Gamma \tsv \eFun{f}{x}{e} : \tau_2 \era{\rho} \tau
}
\and
\irr{\rlabel{fun/non-val}}
{
\Delta;\Gamma, f \mapsto \tau_2 \era{\rho} \tau,  x \mapsto \tau_2  
  \ts  e : \tau ; \rho \+
\istyp{\Delta}{\tau_2} \+
\isrow{\Delta}{\rho}
}
{
\Delta;\Gamma \tsv  \eFun{f}{x}{e} : \tau_2 \era{\rho} \tau
}
\end{mathpar}
\hrule
\begin{mathpar}
\irr{\rlabel{teq/v}}
{
\Delta:\Gamma \tsv e : \tau ~~~
\tau \reorder \tau'
}
{
\Delta;\Gamma \tsv e : \tau'
}
\and
\irr{\rlabel{teq}}
{
\Delta;\Gamma \ts e: \tau ; \rho ~~~
\tau \reorder \tau' ~~~
\rho \reorder \rho'
}
{
\Delta;\Gamma  \ts  e : \tau' ; \rho'
}
\and
\irr{\rlabel{value}}
{
\Delta;\Gamma \tsv e : \tau ~~~
\isrow{\Delta}{\rho}
}
{
\Delta;\Gamma \ts e : \tau ; \rho
}
\end{mathpar}
\hrule
\begin{mathpar}
\irr{\rlabel{app}}
{
\Delta; \Gamma  \ts  e_1 : \tau_2 \era{\rho} \tau ; \rho \+
\Delta; \Gamma  \ts  e_2 : \tau_2 ; \rho
}
{
\Delta; \Gamma  \ts  e_1 \ e_2 : \tau ; \rho
}
\and
\irr{\rlabel{let/val}}
{
\alpha_1,\ldots,\alpha_n = \FTV(\tau_1) \setminus \FTV(\Gamma) ~~~
\Delta,\alpha_1\mapsto\kappa_1,\ldots,\alpha_n\mapsto\kappa_n; \Gamma
   \tsv e_1 : \tau_1 \+\+
\Delta;\Gamma,x\mapsto
    \forall \alpha_1:\kappa_1.\ldots\forall\alpha_n:\kappa_n.\tau_1
        \ts e_2 : \tau_2 ; \rho
}
{
\Delta; \Gamma \ts \eLet{x}{e_1}{e_2}: \tau_2 ; \rho
}
\and
\irr{\rlabel{let/non-val}}
{
\Delta; \Gamma \ts e_1 : \tau_1 ; \rho \+
\Delta; \Gamma, x \mapsto \tau_1 \ts e_2 : \tau_2 ; \rho 
}
{
\Delta; \Gamma \ts \eLet{x}{e_1}{e_2}: \tau_2 ; \rho
}
\and
\irr{\rlabel{dcon}}
{
\Delta; \Gamma \ts e : \tau ; \rho' \+
\isrow{\Delta}{(l:\tau, \rho)}
}
{
\Delta; \Gamma  \ts l \ e : \tSum{l:\tau, \rho} ; \rho'
}
\and
\irr{\rlabel{roll}}
{
\Delta; \Gamma \ts e : \tSum{\rho[\tBind{\alpha}{\tSum{\rho}}/\alpha]};\rho'
}
{
\Delta; \Gamma \ts e : \tBind{\alpha}{\tSum{\rho}};\rho'
}
\and
\irr{\rlabel{unroll}}
{
\Delta; \Gamma \ts e : \tBind{\alpha}{\tSum{\rho}};\rho'
}
{
\Delta; \Gamma \ts e : \tSum{\rho[\tBind{\alpha}{\tSum{\rho}}/\alpha]};\rho'
}
\end{mathpar}
\caption{Typing rules for \EL{} for syntactic values (top),
  type equivalence and lifting (2nd) and
  basic computations (bottom).}
\label{fig:el-typing-a}
\end{figure}

\begin{figure}
\begin{mathpar}
%
%
\irr{\rlabel{r}}
{
\forall i_{\in 1..n}.\Delta; \Gamma \ts e_i : \tau_i;\rho \+
\Delta \ts (l_1:\tau_1,\ldots,l_n:\tau_n,\tRowEmp):\varnothing
}
{
\Delta; \Gamma \ts \eRecN{i=1}{n}{l_i=e_i}: \tRecN{i=1}{n}{l_i:\tau_i}; \rho
}
\and
\irr{\rlabel{r/ext}}
{
\Delta; \Gamma \ts e_1 : \tRec{\rho};\rho' \+
\Delta \ts (l:\tau_2,\rho):\varnothing \+
\Delta; \Gamma \ts e_2 : \tau_2;\rho'
}
{
\Delta; \Gamma \ts \eRecExt{e_1}{\tRec{l=e_2}} : \tRec{l:\tau_2,\rho};\rho'
}
\and
\irr{\rlabel{r/sub}}
{
\Delta; \Gamma \ts e : \tRec{l:\tau, \rho}; \rho'
}
{
\Delta; \Gamma \ts \eRecSub{e}{l}:\tRec{\rho};\rho'
}
\and
\irr{\rlabel{select}}
{
\Delta; \Gamma \ts e : \tRec{l:\tau,\rho};\rho'
}
{
\Delta; \Gamma \ts e.l : \tau;\rho'
}
\end{mathpar}
\hrule
\begin{mathpar}
%
%
\irr{\rlabel{c/ext}}
{
\Delta;\Gamma \ts e_1 : \etCase{\tSum{\rho_1}}{\tau}{\rho} ; \rho' \+
\isrow{\Delta}{(l:\tau_1,\rho_1)} \+
\Delta;\Gamma, x\mapsto\tau_1 \ts e_2 : \tau ; \rho
}
{
\Delta;\Gamma \ts \eCaseExt{e_1}{\eCase{l~x \dra e_2}} : \etCase{\tSum{l:\tau_1, \rho_1
}}{\tau}{\rho} ; \rho'
}
\and
\irr{\rlabel{c/sub}}
{
\Delta;\Gamma \ts e_1 : \etCase{\tSum{l:\tau',\rho_1}}{\tau}{\rho} ; \rho'
}
{
\Delta;\Gamma \ts \eCaseSub{e_1}{l} : \etCase{\tSum{\rho_1}}{\tau}{\rho} ; \rho'
}
\and
\irr{\rlabel{match}}
{
\Delta; \Gamma \ts e_1 : \tSum{\rho} ; \rho' \+
\Delta; \Gamma \ts e_2 : \etCase{\tSum{\rho}}{\tau}{\rho'} ; \rho'
}
{
\Delta; \Gamma \ts \eMat{e_1}{e_2}: \tau ; \rho'
}
\end{mathpar}
\hrule
\begin{mathpar}
%
%
\irr{\rlabel{raise}}
{
\Delta; \Gamma \ts e : \tSum{\rho} ; \rho \+
\istyp{\Delta}{\tau}
}
{
\Delta; \Gamma \ts \eRaise{e} : \tau ; \rho
}
\and
\irr{\rlabel{handle}}
{
\Delta; \Gamma \ts e_1 : \tau ; l:\tau', \rho ~~~
\Delta; \Gamma, x\mapsto\tau' \ts e_2 : \tau ; \rho
}
{
\Delta; \Gamma \ts \eHandle{e_1}{\eCase{l~x \dra e_2}} : \tau ; \rho
}
\and
\irr{\rlabel{unhandle}}
{
\Delta; \Gamma \ts e : \tau ; \rho ~~~
\isrow{\Delta}{(l:\tau',\rho)}
}
{
\Delta; \Gamma \ts \eUnhandle{e}{l} : \tau ; l:\tau', \rho
}
\and
\irr{\rlabel{rehandle}}
{
\Delta; \Gamma \ts e_1 : \tau ; l:\tau', \rho ~~~
\Delta; \Gamma, x\mapsto\tau' \ts e_2 : \tau ; l:\tau'', \rho
}
{
\Delta; \Gamma \ts \eRehandle{e_1}{\eCase{l~x \dra e_2}} : \tau ; l:\tau'', \rho
}
\and
\irr{\rlabel{handle-all}}
{
\Delta; \Gamma \ts e_1 : \tau ; \rho' ~~~
\Delta; \Gamma, x\mapsto\tSum{\rho'} \ts e_2 : \tau ; \rho
}
{
\Delta; \Gamma \ts \eHandle{e_1}{\eCase{x \dra e_2}} : \tau ; \rho
}
\and
\fbox{\hspace{3mm}$
\irr{\rlabel{program}}
{
\varnothing;\Gamma_0 \ts e : \Int ; \tRowEmp
}
{
\Gamma_0 \ts \prog{e}
}
$\hspace{3mm}}
\end{mathpar}
\caption{Typing rules for \EL{} for for computations
  involving records (top), cases (2nd) and exceptions (bottom).
  The judgment for whole programs is shown in the framed box.}
\label{fig:el-typing-c}
\end{figure}


The type $\tau$ of a closed expression $e$ characterizes the values
that $e$ can evaluate to. From a dual point of view it describes the
values that the evaluation context $E$ must be able to receive.  In
our operational semantics $E$ is extended to a full context
$(E,\EXN)$, so the goal is to develop a type system with judgments
that describe the full context of a given expression.  Our typing
judgments have an additional component $\rho$ that describes $\EXN$ by
individually characterizing its constituent labels and evaluation
contexts.

General typing judgments have the form $\Delta;\Gamma \ts e : \tau ;
\rho$, expressing that $e$ has type $\tau$ and exception type $\rho$.
The typing environment $\Gamma$ is a finite map assigning types to the
free variables of $e$.  Similarly, the kinding environment $\Delta$
maps the free type variables of $\tau$, $\rho$, and $\Gamma$ to their
kinds.

The {\bf typing rules} for \EL{} are given in \figref{el-typing-a}
and \figref{el-typing-c}.
Typing is syntax-directed; for most syntactic constructs there is
precisely one rule, the only exceptions being the rules for
\textbf{fun} and \textbf{let} which rely on the notion of {\em
  syntactic values} to distinguish between two sub-cases.
As usual, in rules that introduce polymorphism we impose the {\em
  value restriction} by requiring certain expressions to be {\em
  valuable}.  Valuable expressions do not have effects and, in
particular, do not raise exceptions.  We use a separate typing
judgment of the form $\Delta;\Gamma \tsv e : \tau$ for syntactic
values (\rname{var}, \rname{int}, \rname{c}, \rname{fun/val}, and
\rname{fun/non-val}). Judgments for syntactic values
are lifted to the level of judgments for general expressions by the
\rname{value} rule.  The \rname{value} rule leaves the exception
type $\rho$ unconstrained.
Administrative rules \rname{teq} and \rname{teq/v} deal with type
equivalences $\tau \reorder \tau'$, which expresses the relationship
between two (row-) types where they are considered equal
up to permutation of their fields. Rules for $\tau \reorder \tau'$ are 
described in \figref{reorder}.

\begin{figure}
\begin{mathpar}
\irr{}
{
}
{\alpha \approx \alpha}
\and
\irr{}
{
}
{\Int \approx \Int
}
\and
\irr{}
{
\tau_1 \approx \tau_1' \+
\tau_2 \approx \tau_2' \+
\rho \approx \rho'
}
{\tau_1 \era{\rho} \tau_2 \approx \tau_1' \era{\rho'} \tau_2'}
\and
\irr{}
{\rho \approx \rho'}
{\tRec{\rho} \approx \tRec{\rho'}}
\and
\irr{}
{\rho \approx \rho'}
{\tSum{\rho} \approx \tSum{\rho'}}
\and
\irr{}
{\rho \approx \rho'}
{\tBind{\alpha}{\tSum{\rho}} \approx \tBind{\alpha}{\tSum{\rho'}}}
\and
\irr{}
{
\rho_1 \approx \rho_1' \+
\tau \approx \tau' \+
\rho_2 \approx \rho'2
}
{\etCase{\tSum{\rho_1}}{\tau}{\rho_2} \approx \etCase{\tSum{\rho_1'}}{\tau'}{\rho'_2}}
\and
\irr{}
{
}
{\beta \approx \beta}
\and
\irr{}
{
}
{\tRowEmp \approx \tRowEmp}
\and
\irr{}
{
}
{l: \tau, \beta \approx l: \tau, \beta}
\and
\irr{}
{
}
{l: \tau, \tRowEmp \approx l: \tau, \tRowEmp}
{\strut}
\and
\irr{}
{\#~ \mbox{is a permutation of {1,\ldots,k }}}
{
l_1:\tau_1, \ldots, l_k : \tau_k, \rho ~\approx~
l_{\#(1)}:\tau_{\#(1)}, \ldots, l_{\#(k)} : \tau_{\#(k)}, \rho
}
\end{mathpar}
\caption{The reordering judgment $\approx$.}
\label{fig:reorder}
\end{figure}

Rules unrelated to exceptions simply propagate a single exception type
without change.  This is true even for expressions that have more than
one sub-term, matching our intuition that the exception type
characterizes the exception context.  For example, consider function
application $e~e'$: The rules do not use any form of sub-typing to
express that the set of exceptions is the union of the three sets
corresponding to $e$, $e'$, and the actual application.
We rely on polymorphism to collect exception information
across multiple sub-terms.  As usual, polymorphism is introduced by
the \rname{let/val} rule for expressions $\eLet{x}{e_1}{e_2}$ where
$e_1$ is a syntactic value.

The rules for handling and raising exceptions establish bridges
between ordinary types and handler types
(i.e., types of exception handler contexts).  Exceptions
themselves are simply values of sum type; the {\bf raise} expression
passes such values to an appropriate handler.  Notice that the
corresponding rule equates the row type of the sum with the row type
of the exception context; there is no implicit subsumption here.
Instead, subsumption takes place where the exception payload is
injected into the corresponding sum type (\rname{dcon}).

Rule \rname{handle-all} is the inverse of \rname{raise}.  The form
$\eHandle{e_1}{\eCase{x \dra e_2}}$ establishes a handler that catches {\em
  any} exception emanating from $e_1$.  The exception is made
available to $e_2$ as a value of sum type bound to variable $x$.
Operationally this corresponds to replacing the current exception
handler context with a brand-new one, tailor-made to fit the needs of
$e_1$.  The other three constructs do not replace the exception
handler context wholesale but adjust it incrementally: {\bf handle}
adds a new field to the context while retaining all other fields; {\bf
  rehandle} replaces an existing handler at a specific label $l$ with
a new (potentially differently typed) handler at the same $l$; {\bf
  unhandle} removes an existing handler.  There are strong parallels
between \rname{c/ext} (case extension) and \rname{handle}, although
there are also some significant differences due to the fact that exception
handlers constitute a hidden part of the context while cases are
first-class values.  

Whole programs are closed up to some initial basis
environment $\Gamma_0$, raise no exceptions, and evaluate to \Int.
This is expressed by a judgment $\Gamma_0 \ts \prog{e}$.

\subsection{Properties of \EL{}}
The rule for the ``handle-all'' construct $\eHandle{e_1}{\eCase{x \dra e_2}}$
stands out because it is non-deterministic.  Since we represent each
handled exception constructor separately, the rule must \emph{guess} the
relevant set of constructors $\{l_1,\ldots,l_n\}$.  Introducing
non-determinism here might seem worrisome, but we can justify it by
observing that different guesses never lead to different outcomes:

\begin{lemma}[Uniqueness]
  If $(e,\eRec{}) \mapsto^{*} (v,\eRec{})$ and $(e,\eRec{})
  \mapsto^{*} (v',\eRec{})$, then $v = v'$.
\end{lemma}
\begin{proof}
By a bi-simulation between configurations, 
where two configurations are related 
if they are identical up to records.
Records may have different sets of labels, but common fields
must themselves be related.  It is easy to see that each step of the
operational semantics preserves this relation.
\end{proof}

However, guessing too few or too many labels can get the program
stuck.  Fortunately, for well-typed programs there always exists a
good choice.  The correct choice can be made deterministically by
taking the result of type inference into account, giving rise to a
type soundness theorem for \EL.  Type soundness is expressed in terms
of a well-formedness condition $\ts (E[e],\EXN)~\wf$ on
configurations. Along with the well-formedness of a configuration,
we define typing rules for a full context $(E,\EXN)$ of $r$ 
given a reducible configuration $(E[r], \EXN)$ 
in \figref{pf-el-fc}.
%
%

\begin{figure}
\[
\begin{array}{c}
\hspace{-8.3cm}
\framebox{$\ts (E, \EXN) : \tau;\rho$}
\\
\\
\infer
{\ts ([], \eRec{}) : \Int;\tRowEmp}
{\strut}
\qquad

\infer
{\ts (E[l~[]], \EXN) : \tau;\rho}
{
\ts (E, \EXN) : \tSum{l:\tau,\rho'};\rho \qquad
\varnothing \ts \rho' : \varnothing
}
\\[2mm]

\infer
{\ts (E[[]~e], \EXN) : \tau' \era{\rho} \tau;\rho}
{
\ts (E, \EXN) : \tau;\rho \qquad
\varnothing;\Gamma_0 \ts e : \tau';\rho
}
\qquad

\infer
{\ts (E[v~[]], \EXN) : \tau;\rho}
{
\begin{array}{c}
\varnothing;\Gamma_0 \tsv v : \tau \era{\rho'} \tau' \qquad
\varnothing \ts \rho' : \varnothing \\
\ts (E, \EXN) : \tau';\rho
\end{array}
}
\\[2mm]

\infer
{\ts (E[\eLet{x}{[]}{e}], \EXN) : \tau;\rho}
{
\ts (E, \EXN) : \tau';\rho \qquad
\varnothing;\Gamma_0,x:\tau \ts e : \tau';\rho
}
\qquad

\infer
{
\ts (E[[].l], \EXN) : \tRec{l:\tau,\rho'};\rho
}
{
\ts (E, \EXN) : \tau;\rho
}
\\[2mm]

\infer
{
\ts (E[\eRec{\ldots,l_{i-1}=v_{i-1},l_i=[],l_{i+1}=e_{i+1},\ldots}], \EXN) : \tau_i;\rho 
}
{
\ts (E, \EXN) : \tRec{l_i:\tau_i,\rho'};\rho \qquad
\varnothing;\Gamma_0 \ts  \eRec{\ldots,l_{i-1}=v_{i-1},l_{i+1}=e_{i+1},\ldots}: \tRec{\rho'};\rho
}
\\[2mm]

\infer
{
\ts (E[\eRecExt{[]}{\eRec{l=e}}], \EXN) : \tRec{\rho'};\rho
}
{
\ts (E, \EXN) : \tRec{l:\tau,\rho'};\rho \qquad
\varnothing;\Gamma_0 \ts e : \tau;\rho
}
\qquad

\infer
{
\ts (E[\eRecSub{[]}{l}], \EXN) : \tRec{l:\tau,\rho'};\rho
}
{
\ts (E, \EXN) : \tRec{\rho'};\rho
}
\\[2mm]

\infer
{
\ts (E[\eRecExt{v}{\eRec{l=[]}}], \EXN) : \tau;\rho 
}
{
\ts (E, \EXN) : \tRec{l:\tau;\rho'};\rho \qquad
\varnothing;\Gamma_0 \tsv v : \tRec{\rho'}
}
\qquad

\infer
{
\ts (E[\eCaseSub{[]}{l}], \EXN) : \etCase{\tSum{l:\tau',\rho_l}}{\tau}{\rho'};\rho
}
{
\ts (E, \EXN) : \etCase{\tSum{\rho_l}}{\tau}{\rho'};\rho
}
\\[2mm]

\infer
{
\ts (E[\eCaseExt{[]}{\eCase{l~x \dra e}}], \EXN)
: \etCase{\tSum{\rho_1}}{\tau}{\rho};\rho'
}
{
\ts (E, \EXN) : \etCase{\tSum{l:\tau_1,\rho_1}}{\tau}{\rho};\rho' \qquad
\varnothing;\Gamma_0,x:\tau_1 \ts e : \tau; \rho
}
\\[2mm]

\infer
{\ts (E[\eMat{[]}{e}], \EXN) : \tSum{\rho};\rho'}
{
\ts (E, \EXN) : \tau';\rho' \qquad
\varnothing;\Gamma_0 \ts e : \etCase{\tSum{\rho}}{\tau'}{\rho'}; \rho'
}
\quad

\infer
{\ts (E[\eRaise~[]], \EXN) : \tSum{\rho};\rho}
{\tsp \EXN : \rho}
\\[4mm]

\infer
{
\ts (E[\eMat{v}{[]}], \EXN) : \etCase{\tSum{\rho}}{\tau}{\rho'};\rho'
}
{
\ts (E, \EXN) : \tau;\rho' \qquad
\varnothing;\Gamma_0 \tsv v : \tSum{\rho}
}
\qquad

\infer
{\ts (E[\eRestore{\EXN'}{[]}], \EXN) : \tau;\rho}
{
\ts (E, \EXN') : \tau;\rho' \qquad
\tsp \EXN : \rho
}
\\[5mm]

\hspace{-9cm}
\framebox{$\tsp \EXN : \rho$}
\\

\infer
{\tsp \eRec{} : \tRowEmp}
{\strut}
\qquad

\infer
{
\tsp \eRec{l_i=E_i}_{i=1\ldots n} : l_1:\tau_1,\ldots,l_n:\tau_n
}
{\forall i. \ts (E_i, \eRec{}) : \tau_i; \tRowEmp}

\end{array}
\]
\caption{Given a reducible configuration $(E[r], \EXN)$, 
Typing rules for a full context of $r$.}
\label{fig:pf-el-fc}
\end{figure}


\begin{definition}[Well-formedness of a configuration]
\label{def:wf}
\begin{mathpar}
\irr{}
{
}
{
\wfc{v}{\eRec{}}
}
\and
\irr{}
{
\varnothing;\Gamma_0 \ts e : \tau; \rho \qquad
                     \ts (E, \EXN) : \tau; \rho
}
{
\wfc{E[e]}{\EXN}
}
\end{mathpar}
\end{definition}

Then, we can prove type soundness using the standard technique of
preservation and progress. Before we can proceed to establishing them,
we need a few technical lemmas. Some of them are standard: inversion,
cannonical forms, substitution and weakening.

\begin{lemma}[Cannonical forms]
\begin{enumerate}
 \item if $v$ is a value of type $\Int$, then $v = n$.
 \item if $v$ is a value of type $\tau_1 \era{\rho} \tau_2$, then $v = \eFun{f}{x}{e}$.
 \item if $v$ is a value of type $\tRecN{i=1}{n}{l_i:\tau_i}$, then $v = \eRecN{i=1}{n}{l_i=v_i}$.
 \item if $v$ is a value of type $\tSum{\rho}$, then $v = l~v'$.
 \item if $v$ is a value of type $\etCase{\tSum{\rho_1}}{\tau}{\rho_2}$,
   then $v = \eCaseN{i=1}{n}{l_i~x_i \dra e_i}$ for some $n$.
\end{enumerate}
\label{lemma:cannonical}
\end{lemma}

\begin{proof}
By induction of $\tau$ with the inversion lemma.
\end{proof}

\begin{lemma}[Substitution]
If $\varnothing;\Gamma_0,x:\forall\alpha:\kappa.\tau' \ts e:\tau;\rho$ and
$\varnothing,\alpha:\kappa;\Gamma_0 \ts v : \tau';\rho$,
then $\varnothing;\Gamma_0 \ts e[v/x] : \tau;\rho$
\end{lemma}

\begin{proof}
By induction on $e$.
\end{proof}

\begin{lemma}[Weakening]
\begin{enumerate}
\item If $\varnothing;\Gamma_0 \ts e:\tau;\rho$ and $x \notin Dom(\Gamma_0)$,
      then $\varnothing;\Gamma_0,x:\tau' \ts e : \tau;\rho$
\item If $\varnothing;\Gamma_0 \ts e:\tau;\rho$,
      then $\alpha_1:\kappa_1,\ldots,\alpha_n:\kappa_n;\Gamma_0 \ts e : \tau;\rho$
\end{enumerate}
\end{lemma}

\begin{proof}
By induction on $e$.
\end{proof}

In addition to the standard lemmas, we establish two special lemmas
to simplify the main lemma:

\begin{lemma}[Restore]
\label{lem:restore}
\begin{enumerate}
\item If $\ts (E, \EXN) : \tau'; \rho $
      and $\varnothing;\Gamma_0,x:\tau \ts e : \tau'; \rho$, \\
      then $\tsp \eRec{l=E[\eLet{x}{\eRestore{\EXN}{[]}}{e}]} : ~ l:\tau$.

\item If $\ts (E, \EXN) : \tau'; \rho $
      and $\varnothing;\Gamma_0,x:\tSumN{i=1}{n}{l_i:\tau_i} \ts e : \tau'; \rho$, \\
      then $\tsp \eRecN{i=1}{n}{l_i=E[\eLet{x}{l_i~(\eRestore{\EXN}{[]}}{e})]}
           : ~ l_1:\tau_1,\ldots,l_n:\tau_n$.
\end{enumerate}
\end{lemma}
\begin{proof}
By typing rules for a full context.
\end{proof}

\begin{lemma}[Exception context]
\label{lem:econ}
If $\ts (E, \EXN) : \tau;\rho$, then $\tsp \EXN:\rho$.
\end{lemma}

\begin{proof}
By induction on $E$.
\end{proof}

Given these we can show preservation:

\begin{lemma}[Preservation]
If $\wfc{E[e]}{\EXN]}$ and $(E[e], \EXN) \mapsto (E'[e'], \EXN')$,
then $\wfc{E'[e']}{\EXN'}$
\end{lemma}

\begin{proof}
The proof proceeds by case analysis according to the derivation of
$(E[e], \EXN) \mapsto (E'[e'], \EXN')$.
The cases are entirely standard except that some cases
use \lemref{econ} and \lemref{restore}. We present such a case
for example.
\begin{itemize}
\item {\em Case} \rname{handle}: $(E[\eHandle{e_1}{\eCase{l~x \dra e_2}}], \EXN) \mapsto
                  (E[\eRestore{\EXN}{e_1}], \EXN')$.
By given, $\wfc{E[\eHandle{e_1}{\eCase{l~x \dra e_2}}]}{\EXN}$. 
Then, by \defref{wf}, we know that $\varnothing;\Gamma_0 \ts \eHandle{e_1}{\eCase{l~x \dra e_2}} : \tau;\rho$ and $\ts (E,\EXN) : \tau;\rho$ (\cc{3}).
By inv of \rname{handle},
$\varnothing;\Gamma_0 \ts e_1 : \tau;l:\tau',\rho$ (\cc{4}) and
$\varnothing;\Gamma_0,x:\tau' \ts e_2 : \tau;\rho$ (\cc{5}).
TS: $\wfc{E[\eRestore{\EXN}{e_1}]}{\EXN'}$.
Then, it is sufficient to show that (STS): $\ts (E[\eRestore{\EXN}{[]}], \EXN') : \tau;l:\tau',\rho$ because of \cc{4}.
Then, with \cc{3}, STS: $\tsp \EXN' : l:\tau',\rho$.
By exception context lemma, \cc{3} also shows that $\tsp \EXN : \rho$.
Because $\EXN' = \EXN \otimes \eRec{l=E[\eLet{x}{\eRestore{\EXN}{[]}}{e_2}]}$,
we only need to show that 
$\tsp \eRec{l=E[\eLet{x}{\eRestore{\EXN}{[]}}{e_2}]} : ~ l:\tau'$
which is true by restore lemma with \cc{3} and \cc{5}. \\
\end{itemize}
\end{proof} 

To prove progress, we need the unique decomposition lemma:

\begin{lemma}[Unique decomposition]
\label{lem:unique}
Let $e$ be a closed term but not a value.
Then, there exist unique $E$ and 
redex $r$ such that $e \equiv E[r]$.
\label{lemma:decomposition}
\end{lemma}

\begin{proof}
By definition of $E$.
\end{proof}

Given this lemma, we can show progress:

\begin{lemma}[Progress]
If a configuration $(e, \EXN)$ is well-formed,
either it is a final configuration $(v,\{ \})$ or else
there exists a single-step transition to another configuration,
i.e, $(E[e'], \EXN) \mapsto (E''[e''], \EXN')$ where $e \equiv E[e']$.
\end{lemma}

\begin{proof}
For value terms, they are immediately final configurations by definition.
For non-value terms, there exist unique $E$ and $e'$ such that $e \equiv E[e']$
by \lemref{unique}. Then, the proof proceeds by case analysis on $e'$.
\end{proof}

The main result is the type soundness (i.e., safety)
 of the \EL{} programs:

\begin{theorem}[Type soundness]
\label{thm:safety}
If a configuration is well-formed,
either it is a final configuration or eles there exists
a single-step transition to another well-formed configuration.
\end{theorem}

\begin{proof}
Type soundness follows from the preservation and progress lemmas.
\end{proof}

\begin{corollary}[Type safe exception handling]
Well-typed $\EL{}$ programs do not have uncaught exceptions.
\end{corollary}

\begin{proof}
By \thmref{safety}.
\end{proof}




\section{The Internal Language (\IL)}
\label{sec:il}

\EL{} expressions can be translated into expressions of
a variant of \systemf{} with records and named functions.
We call this language \IL{}. Recall that the semantics for \EL{} 
shown in \figref{el-opsem} uses
non-determinism in its \rname{handle all} rule.  The need for this
arises because with \eHandle{e_1}{x \dra e_2} a new exception context
with one field for every exception that $e_1$ might raise must be
built.  This set of exceptions
is not always fixed and does not only depend on $e_1$ itself:
exceptions can be passed in, either directly as first-class values or
perhaps by a way of functional parameters to higher-order functions.
Therefore, to remove the non-determinism a combination of static
analysis and runtime techniques is needed.

In essence, we need access to the type of $e_1$, and we must be able
to utilize this type when building a new exception context.  To make
this idea precise, we provide an elaboration semantics for \EL{}. We
define an explicitly typed internal language \IL{} and augment the
\EL{} typing judgments with a translation component. \IL{} is a
variant of \systemf enriched with extensible records as well as a
special type-sensitive \textbf{reify} construct which provides the
``canonical'' translation from functions on sums to records of
functions.  Using \textbf{reify} we are able to give a deterministic
account of ``catch-all'' exception handlers.

Unlike \EL{}, \IL{} does not have dedicated mechanisms for raising
and handling exceptions.  Therefore, we will use {\em continuation
  passing style} and represent exception contexts explicitly as
extensible records of continuations.  In \EL{}, exceptions are
simply members of a sum type, and the translation treats them as such:
they are translated via {\em dual transformation}
into polymorphic functions on records of
functions.  Therefore, they are applicable to both exception contexts
(i.e., records of continuations) and to first-class cases (i.e.,
records of ordinary functions).

\subsection{Syntax and semantics}
%
%

\begin{figure}
\[
\begin{array}{@{}lrcl@{}}
\mathsf{Terms} &
\eb & \bnfdef & n \bnfalt x \bnfalt \eAbs{x:\taub}{\eb} 
                \bnfalt \eTabs{\alpha:\kappa}{\eb} \bnfalt 
                \eb_1~\eb_2 \bnfalt \eTapp{\eb}{\thetab} \bnfalt
                \eLet{x:\taub}{\eb_1}{\eb_2} \bnfalt \\
&   &         & \eLetrec{f:\taub}{\eAbs{x:\taub_2}{\eb_1}}{\eb_2} \bnfalt 
                \eLetrec{f:\taub}{\eTabs{\alpha:\kappa}{\eb_1}}{\eb_2} \bnfalt \\
&   &         & \eRecN{i=1}{n}{l_i = \eb_i} \bnfalt
		\eRecExt{\eb_1}{\eRec{l=\eb_2}} \bnfalt 
		\eRecSub{\eb}{l} \bnfalt \eb.l \bnfalt \reify{\rhob}{\taub}{\eb}\\

\mathsf{Values} &
\vb & \bnfdef & n \bnfalt
               \eAbs{x:\taub}{\eb} \bnfalt
	       \eTabs{\alpha:\kappa}{\eb} \bnfalt 
	       \eRecN{i=1}{n}{l_i = \vb_i} \\[1mm]

\mathsf{Types} &
\taub & \bnfdef & \alpha \bnfalt \Int \bnfalt
                  \taub_1 \ra \taub_2 \bnfalt
		  \tRec{\rhob} \bnfalt
		  \forall\alpha:\kappa.\taub \bnfalt
                  \tBind{\alpha}{\taub}\\
&
\rhob & \bnfdef & \alpha \bnfalt \tRowEmp \bnfalt
                  l:\taub, \rhob \bnfalt
                  \alpha \rowArrow \taub \\
&
\thetab & \bnfdef & \taub \bnfalt \rhob

\end{array}
\]
\caption{Internal language (\IL) syntax.}
\label{fig:il-syntax}
\end{figure}


\figref{il-syntax} shows the syntax for \IL{}.  We use
meta-variables such as $\eb$, $\taub$, and $\rhob$ for terms and types
of \IL{} to visually distinguish them from their \EL{} counterparts
$e$, $\tau$, and $\rho$.  
The term language consists of constants ($n$), variables ($x$), term-
and type abstractions ($\eAbs{x:\taub}{\eb}$ and
$\eTabs{\alpha:\kappa}{\eb}$), term- and type applications
($\eb_1~\eb_2$ and $\eb[\thetab]$), recursive bindings for abstractions
(\textbf{letrec}), \textbf{let}-bindings, records---including
constructs for creation $\eRec{l=\eb}$, extension $\otimes$, field
deletion $\oslash$, and projection $\eb.l$---as well as the aforementioned
\textbf{reify} operation which turns functions on sums into
corresponding records of functions.
\IL{} types consist of ordinary types
$\taub$ and row types $\rhob$.  Ordinary types include base types
(\Int), function types ($\taub_1 \ra \taub_2$), records
(\tRec{\rhob}), polymorphic types ($\forall \alpha:\kappa.\taub$), 
recursive types (\tBind{\alpha}{\taub}) and
(appropriately kinded) type variables $\alpha$. The set of type
variables and their kinds is shared between \EL{} and \IL{}. Row types
are either the empty row (\tRowEmp), a typed label followed by another
row type ($l:\taub,\rhob$), a row type variable ($\alpha$) or a {\em
  row arrow} applied to a row type variable and a type ($\alpha
\rowArrow \taub$).  The key difference between the row types of
\EL{} and \IL{} is the inclusion of such row arrows.
They are critical to represent sums and cases in terms of
records. As usual, well-formedness of potentially open type
terms is stated relative to a kinding environment $\bar{\Delta}$
mapping type variables to their kinds, so judgments have the form
$\bar{\Delta} \ts \taub : \kappa$.  For brevity we omit rules
because they are either standard or closely follow the ones we used for \EL{}
(see \figref{el-wf}).

A small-step operational semantics for \IL{} is shown in
\figref{il-opsem}. With the exception of \textbf{reify}, most rules
are standard. There are three definitions of substitution rules 
for free variables
(\figref{il-subs-1}) and for free type variables (\figref{il-subs-2}
and \ref{fig:il-subs-3}). 
For example, let \(\rhob=l_1:\taub_1,\ldots,l_n:\taub_n,\tRowEmp\)
\noindent and consider $(\alpha \rowArrow \taub)[\rhob/\alpha]$.  
Substitution cannot simply replace
$\alpha$ with $\rhob$, since the result would not even be
syntactically valid.  Instead, it must {\em normalize}, resulting in
$l_1:(\taub_1 \ra \taub'),\ldots,l_n:(\taub_n \ra \taub'),\tRowEmp$
where $\taub' = \taub[\rhob/\alpha]$.

\figref{il-static} shows typing rules for \IL{} which are
mostly standard with the exception of \textbf{reify}.
The rule for type application involves type substitution, 
and, as before, we must use a
row-normalizing version of substitution.  A formal definition of row
normalization as a judgment is shown in \figref{il-row-norm}.

%
%

\begin{landscape}
\begin{figure}
\[
\begin{array}{lcl}
\Eb  & \bnfdef & [] \bnfalt \Eb~\eb \bnfalt \vb~\Eb \bnfalt
                 \Eb~[\taub] \bnfalt 
                 \eLet{x:\taub}{\Eb}{\eb_2} \bnfalt 
                 \eRecExt{\Eb}{\eRec{l=\eb_2}} \bnfalt \eRecExt{\vb}{\eRec{l=\Eb}} \bnfalt \\
     &         & \eRec{\ldots,l_{i-1}=\vb_{i-1},l_i=\Eb,l_{i+1}=\eb_{i+1},\ldots} \bnfalt
                 \eRecSub{\Eb}{l} \bnfalt \Eb.l
\end{array}
\]
\caption{Evaluation contexts for \IL.}
\label{fig:il-contexts}

%
%

\[
\begin{array}{rcll}

\Eb[(\eAbs{x:\taub}{\eb})~\vb] & \mapsto 
& \Eb[\eb\,[\vb/x]] 
& \rlabel{app} \\[1mm]

\Eb[(\eTabs{\alpha:\kappa}{\eb})~[\taub]] & \mapsto 
& \Eb[\eb\,[\taub/\alpha]] 
& \rlabel{type/app} \\[1mm]

\Eb[\eLet{x:\taub}{\vb}{\eb}] & \mapsto 
& \Eb[\eb\,[\vb/x]]
& \rlabel{let} \\[1mm]

\Eb[\eLetrec{f:\taub}{\eAbs{x:\taub_2}{\eb_1}}{\eb_2}] & \mapsto 
& \Eb[\eb_2\,[(\eAbs{x:\taub_2}{\eb_1\,[(\eLetrec{f:\taub}
                                             {\eAbs{x:\taub_2}{\eb_1}}
                                             {f})/f]})/f]]
& \rlabel{rec/fun} \\[1mm]

\Eb[\eLetrec{f:\taub}{\eTabs{\alpha:\kappa}{\eb_1}}{\eb_2}] & \mapsto 
& \Eb[\eb_2\,[(\eTabs{\alpha:\kappa}{\eb_1\,[(\eLetrec{f:\taub}
                                             {\eTabs{\alpha:\kappa}{\eb_1}}
                                             {f})/f]})/f]]
& \rlabel{polyrec/fun} \\[1mm]

\Eb[\eRecExt{\eRecN{i=1}{n}{l_i=\vb_i}}{\eRec{l=\vb}}] & \mapsto 
& \Eb[\eRec{l_1=\vb_1,\ldots,l_n=\vb_n,l=\vb}] 
& \rlabel{r/ext} \\[1mm]

\Eb[\eRecSub{\eRec{\ldots,l_i=\vb_i,\ldots}}{l_i}] & \mapsto 
& \Eb[\eRec{\ldots,l_{i-1}=\vb_{i-1},l_{i+1}=\vb_{i+1},\ldots}] 
& \rlabel{r/sub} \\[1mm]

\Eb[\eRec{\ldots,l_i=\vb_i,\ldots}.l_i] & \mapsto 
& \Eb[\vb_i] 
& \rlabel{select} \\[2mm]

\Eb[\reify{l_1:\taub_1,\ldots,l_n:\taub_n,\tRowEmp}{\taub}{\vb}] & \mapsto 
& \Eb[\eRecN{i=1}{n}{ l_i = \eAbs{x_i:\taub_i}
                 {\vb~(\eTabs{\alpha:\star}
                            {\eAbs{c:\tRecN{j=1}{n}{l_j:\taub_j \ra  \alpha}}{c.l_i~x_i}})} }]
& \rlabel{reify}
\end{array}
\]
\caption{Operational semantics for \IL.}
\label{fig:il-opsem}
\end{figure}
\end{landscape}


%
%

\begin{landscape}
\begin{figure}
\[
\begin{array}{r@{~=~}l}
n~[\vb/x] & n   \\
x~[\vb/x] & \vb \\
y~[\vb/x] & y \mathbf{~if~} x \neq y \\[1mm]
(\eAbs{x:\taub}{\eb})~[\vb/x] & \eAbs{x:\taub}{\eb} \\[1mm] 
(\eAbs{y:\taub}{\eb})~[\vb/x] & \eAbs{y:\taub}{(\eb~[\vb/x])} 
                                \mathbf{~if~} x \neq y, y \notin \mathbf{FV}(\vb) \\[1mm]
(\eTabs{\alpha:\kappa}{\eb})~[\vb/x] & \eTabs{\alpha:\kappa}{(\eb~[\vb/x])} \\[1mm]
(\eb_1~\eb_2)~[\vb/x] & (\eb_1~[\vb/x])~(\eb_2~[\vb/x]) \\
(\eTapp{\eb}{\thetab})~[\vb/x] & \eTapp{(\eb~[\vb/x])}{\thetab} \\[1mm]
(\eLetrec{f:\taub}{\eAbs{x:\taub_2}{\eb_1}}{\eb_2})~[\vb/f] & 
	\eLetrec{f:\taub}{\eAbs{x:\taub_2}{\eb_1}}{\eb_2} \\[1mm]
(\eLetrec{f:\taub}{\eAbs{x:\taub_2}{\eb_1}}{\eb_2})~[\vb/x] &
        \eLetrec{f:\taub}{(\eAbs{x:\taub_2}{\eb_1})~[\vb/x]}{(\eb_2~[\vb/x])} 
\mathbf{~if~} x \neq f, f \notin \mathbf{FV}(\vb) \\[1mm]
(\eLetrec{f:\taub}{\eTabs{\alpha:\kappa}{\eb_1}}{\eb_2})~[\vb/f] & 
  \eLetrec{f:\taub}{\eTabs{\alpha:\kappa}{\eb_1}}{\eb_2} \\[1mm] 
(\eLetrec{f:\taub}{\eTabs{\alpha:\kappa}{\eb_1}}{\eb_2})~[\vb/x] &
  \eLetrec{f:\taub}{\eTabs{\alpha:\kappa}{(\eb_1~[\vb/x])}}{(\eb_2~[\vb/x])}
                  \mathbf{~if~} x \neq f, f \notin \mathbf{FV}(\vb) \\[1mm]
(\eLet{x:\taub}{\eb_1}{\eb_2})~[\vb/x] & 
\eLet{x:\taub}{(\eb_1~[\vb/x])}{\eb_2} \\[1mm]
(\eLet{y:\taub}{\eb_1}{\eb_2})~[\vb/x] &
                  \eLet{x:\taub}{(\eb_1~[\vb/x]}{(\eb_2~[\vb/x])} 
                  \mathbf{~if~} x \neq f, y \notin \mathbf{FV}(\vb) \\[1mm]
(\eRecN{i=1}{n}{l_i = \eb_i})~[\vb/x] & 
\eRecN{i=1}{n}{l_i = \eb_i~[\vb/x]} \\[1mm]
(\eRecExt{\eb_1}{\eRec{l=\eb_2}})~[\vb/x] & 
           \eRecExt{(\eb_1~[\vb/x])}{\eRec{l=\eb_2~[\vb/x]}} \\[1mm]
(\eRecSub{\eb}{l})~[\vb/x] & \eRecSub{(\eb~[\vb/x])}{l} \\
(\eb.l)~[\vb/x] & (\eb~[\vb/x]).l \\[2mm]
(\reify{\rhob}{\taub}{\eb})~[\vb/x] & \reify{\rhob}{\taub}{(\eb~[\vb/x])}
\end{array}
\]
\caption{Substituting $\vb$ for free variable $x$, $\eb~[\vb/x]$.}
\label{fig:il-subs-1}
\end{figure}
\end{landscape}
\begin{landscape}
\begin{figure}
\[
\begin{array}{r@{~=~}l}
n~[\thetab/\alpha] & n \\
x~[\thetab/\alpha] & x \\
(\eAbs{x:\taub'}{\eb})~[\thetab/\alpha] & 
        \eAbs{x:\taub'~[\thetab/\alpha]}{(\eb~[\thetab/\alpha])} \\
(\eTabs{\alpha:\kappa}{\eb})~[\thetab/\alpha] & 
        \eTabs{\alpha:\kappa}{\eb} \\
(\eTabs{\beta:\kappa}{\eb})~[\thetab/\alpha] &
        \eTabs{\beta:\kappa}{(\eb~[\thetab/\alpha])}
        \mathbf{~if~} \alpha \neq \beta, \beta \notin \mathbf{FTV}(\thetab) \\
(\eb_1~\eb_2)~[\thetab/\alpha] & 
(\eb_1~[\thetab/\alpha])~(\eb_2~[\thetab/\alpha]) \\
(\eTapp{\eb}{\thetab'})~[\thetab/\alpha] & 
        \eTapp{(\eb~[\thetab/\alpha])}{\thetab'} \\
(\eLetrec{f:\taub_1}{\eAbs{x:\taub_2}{\eb_1}}{\eb_2})~[\thetab/\alpha] &
        \eLetrec{f:\taub_1~[\thetab/\alpha]}
        {\eAbs{x:\taub_2~[\thetab/\alpha]}{(\eb_1~[\thetab/\alpha])}}
        {(\eb_2~[\thetab/\alpha])} \\
(\eLetrec{f:\taub'}{\eTabs{\alpha:\kappa}{\eb_1}}{\eb_2})~[\thetab/\alpha] &
        \eLetrec{f:\taub'~[\thetab/\alpha]}
        {\eTabs{\alpha:\kappa}{\eb_1}}{(\eb_2~[\thetab/\alpha])} \\
(\eLetrec{f:\taub'}{\eTabs{\beta:\kappa}{\eb_1}}{\eb_2})~[\thetab/\alpha] &
        \eLetrec{f:\taub~[\thetab/\alpha]}
        {\eTabs{\beta:\kappa}{(\eb_1~[\thetab/\alpha])}}
        {(\eb_2~[\thetab/\alpha])} 
        \mathbf{~if~} 
        \alpha \neq \beta, \beta \notin \mathbf{FTV}(\thetab) \\
(\eLet{x:\taub'}{\eb_1}{\eb_2})~[\thetab/\alpha] & 
        \eLet{x:\taub'~[\thetab/\alpha]}
        {(\eb_1~[\thetab/\alpha])}{(\eb_2~[\thetab/\alpha])} \\
(\eRecN{i=1}{n}{l_i = \eb_i})~[\thetab/\alpha] & 
        \eRecN{i=1}{n}{l_i = \eb_i~[\thetab/\alpha]} \\
(\eRecExt{\eb_1}{\eRec{l=\eb_2}})~[\thetab/\alpha] &
        \eRecExt{(\eb_1~[\thetab/\alpha])}
        {\eRec{l=\eb_2~[\thetab/\alpha]}} \\
(\eRecSub{\eb}{l})~[\thetab/\alpha] & 
\eRecSub{(\eb~[\thetab/\alpha])}{l} \\
(\eb.l)~[\thetab/\alpha] &
(\eb~[\thetab/\alpha]).l \\
(\reify{\rhob}{\taub'}{\eb})~[\thetab/\alpha] & 
        \reify{\rhob~[\thetab/\alpha]}{\taub~[\thetab/\alpha]}
        {(\eb~[\thetab/\alpha])}
\end{array}
\]
\caption{Substituting $\thetab$ for free type variable $\alpha$, $\eb~[\thetab/\alpha]$.}
\label{fig:il-subs-2}
\end{figure}

\begin{figure}
\[
\begin{array}{r@{~=~}l}
\alpha~[\thetab/\alpha] & \thetab \\
\beta~[\thetab/\alpha] & \beta \mathbf{~if~} \alpha \neq \beta \\
\Int~[\thetab/\alpha] & \Int \\
(\taub_1 \ra \taub_2)~[\thetab/\alpha] & 
(\taub_1~[\thetab/\alpha]) \ra (\taub_2~[\thetab/\alpha]) \\
(\forall\alpha:\kappa.\taub)~[\thetab/\alpha] &
\forall\alpha:\kappa.\taub \\
(\forall\beta:\kappa.\taub)~[\thetab/\alpha] &
       \forall\beta:\kappa.(\taub~[\thetab/\alpha]) 
       \mathbf{~if~} 
       \alpha \neq \beta, \beta \notin \mathbf{FTV}(\theta) \\
\tRec{\rhob}~[\thetab/\alpha] & 
\tRec{\rhob~[\thetab/\alpha]} \\
\tBind{\alpha}{\taub}~[\thetab/\alpha] & \tBind{\alpha}{\taub} \\
\tBind{\beta}{\taub}~[\thetab/\alpha] & \tBind{\beta}{(\taub[\thetab/\alpha])}
     \mathbf{~if~} \alpha \neq \beta \\
\tRowEmp~[\thetab/\alpha] & \tRowEmp \\
(l:\taub, \rhob)~[\thetab/\alpha] & 
   l:\taub'~[\thetab/\alpha], \rhob~[\thetab/\alpha] \\
(\beta \rowArrow \taub)~[\thetab/\alpha] &
       \beta \rowArrow (\taub'~[\thetab/\alpha]) 
       \mathbf{~if~} \alpha \neq \beta \\
(\alpha \rowArrow \taub)~[\tRowEmp/\alpha] & \tRowEmp \\
(\alpha \rowArrow \taub)~[\beta/\alpha] & 
   \beta \rowArrow (\taub'~[\beta/\alpha]) \\
(\alpha \rowArrow \taub)~[l:\taub_l,\rhob/\alpha] & 
   l:\taub_l \ra \taub', (\alpha \rowArrow \taub)~[\rhob/\alpha]
   \mathbf{~where~} \taub' = \taub[l:\taub_l,\rhob/\alpha]
\end{array}
\]
\vspace{-5mm}
\caption{Substituting $\thetab$ for free type variable $\alpha$, $\thetab'~[\thetab/\alpha]$.}
\label{fig:il-subs-3}
\end{figure}
\end{landscape}


%
%

%
%

\begin{figure*}
\[
\begin{array}{c}

\infer[\rlabel{T-int}]
{\Deltab;\Gammab  \ts  n : \Int}
{\strut}
\qquad

\infer[\rlabel{T-var}]
{\Deltab;\Gammab  \ts  x : \taub}
{
\begin{array}{c}
\Gammab(x) = \forall\alpha_1:\kappa_1\ldots\forall\alpha_n:\kappa_n.\taub' \\[1mm]
\forall i_{\in 1..n} .\, \Deltab \ts \taub_i : \kappa_i \qquad
\taub = \taub'[\taub_1 / \alpha_1,\ldots,\taub_n / \alpha_n]
\end{array}
}
\\[4mm]

\infer[\rlabel{T-abs}]
{\Deltab;\Gammab \ts \eAbs{x:\taub'}{\eb} : \taub' \ra \taub}
{\Deltab;\Gammab,x:\taub' \ts \eb : \taub}
\quad

\infer[\rlabel{T-app}]
{\Deltab;\Gammab \ts \eb_1~\eb_2 : \taub}
{
\begin{array}{c}
\Deltab;\Gammab \ts \eb_1 : \taub_2 \ra \taub \quad
\Deltab;\Gammab \ts \eb_2 : \taub_2
\end{array}
}

\\[4mm]

\infer[\rlabel{T-abs/type}]
{\Deltab;\Gammab \ts \eTabs{\alpha:\kappa}{\eb} : \forall\alpha:\kappa.\taub}
{\Deltab, \alpha:\kappa;\Gammab \ts \eb : \taub}

\quad

\infer[\rlabel{T-app/type}]
{\Deltab;\Gammab \ts \eb[\taub'] : \taub[\taub'/\alpha]}
{
\begin{array}{c}
\Deltab;\Gammab \ts \eb : \forall\alpha:\kappa.\taub \quad
\Deltab \ts \taub' : \kappa
\end{array}
}
\\[4mm]

\infer[\rlabel{T-let}]
{\Deltab;\Gammab \ts \eLet{x:\taub}{\eb_1}{\eb_2} : \taub_2}
{
\Deltab;\Gammab \ts \eb_1 : \taub \qquad
\Deltab;\Gammab,x:\taub \ts \eb_2 : \taub_2
}
\\[4mm]

\infer[\rlabel{T-letrec}]
{\Deltab;\Gammab \ts \mathbf{letrec~} f:\taub_2 \ra \taub_1 = \eAbs{x:\taub_2}{\eb_1}
                     \mathbf{~in~} \eb_2 : \taub
}
{
\Deltab;\Gammab,f:\taub_2 \ra \taub_1,x:\taub_2 \ts \eb_1 : \taub_1 \qquad
\Deltab;\Gammab,f:\taub_2 \ra \taub_1,x:\taub_2 \ts \eb_2 : \taub
}
\\[4mm]

\infer[\rlabel{T-letrec/type}]
{\Deltab;\Gammab \ts \mathbf{letrec~} f:\forall\alpha:\kappa.\taub_1 = \eTabs{\alpha:\kappa}{\eb_1}
                     \mathbf{~in~} \eb_2 : \taub
}
{
\Deltab;\Gammab,f:\forall\alpha:\kappa.\taub_1 \ts \eb_1 : \taub_1 \qquad
\Deltab;\Gammab,f:\forall\alpha:\kappa.\taub_1 \ts \eb_2 : \taub
}
\\[4mm]

\infer[\rlabel{roll}]
{
\Deltab; \Gammab \ts \eb : \tBind{\alpha}{\rhob}
}
{
\Deltab; \Gammab \ts \eb : \tSum{\rhob[\tBind{\alpha}{\rhob}/\alpha]}
}

\qquad

\infer[\rlabel{unroll}]
{
\Deltab; \Gammab \ts \eb : \tSum{\rhob[\tBind{\alpha}{\rhob}/\alpha]}
}
{
\Deltab; \Gammab \ts \eb : \tBind{\alpha}{\rhob}
}
\\[4mm]

\infer[\rlabel{T-r}]
{\Deltab;\Gammab \ts \eRecN{i=1}{n}{l_i = \eb_i} : \tRecN{i=1}{n}{l_i:\taub_i}}
{\forall i.\Deltab;\Gammab \ts \eb_i : \taub_i}
\qquad

\infer[\rlabel{T-select}]
{\Deltab; \Gammab \ts \eb.l: \taub}
{\Deltab; \Gammab \ts \eb :  \tRec{l:\taub, \rhob}}
\\[4mm]

\infer[\rlabel{T-r/ext}]
{\Deltab;\Gammab \ts \eRecExt{\eb_1}{\eRec{l = \eb_2}} : \tRec{l : \taub_2, \rhob}}
{
\begin{array}{l}
\Deltab;\Gammab \ts \eb_1 : \tRec{\rhob} \qquad
\Deltab \ts (l:\taub_2, \rhob) : \varnothing \qquad
\Deltab;\Gammab \ts \eb_2 : \taub_2 
\end{array}
}
\\[4mm]

\infer[\rlabel{T-r/sub}]
{\Deltab;\Gammab \ts \eRecSub{\eb}{l} : \tRec{\rhob}}
{\Deltab;\Gammab \ts \eb : \tRec{l : \taub, \rhob}}
\qquad

\infer[\rlabel{T-reify}]
{\Deltab; \Gammab \ts \reify{\rhob}{\taub}{\eb} : \tRec{\rhob \rowArrow \taub}}
{\Deltab; \Gammab \ts \eb : \cpsSum{\rhob} \ra \taub}

\end{array}
\]
\caption{The static semantics for \IL{}.}
\label{fig:il-static}
\end{figure*}


%
%

\begin{figure}
\begin{small}
\[
\begin{array}{c}
\infer
{\alpha;\taub \tmaps \alpha \rowArrow \taub}
{\strut}
\qquad

\infer
{\tRowEmp;\taub \tmaps \tRowEmp}
{\strut}
\qquad

\infer
{(l:\taub_1,\rhob);\taub_2 \tmaps l:\taub_1 \ra \taub_2,\rhob'}
{\rhob;\taub_2 \tmaps \rhob'}

\end{array}
\]
\end{small}
\caption{Row arrow normalization.}
\label{fig:il-row-norm}
\vspace{-3mm}
\end{figure}


\subsection{Properties of \IL}

To prove type soundness, we need some standard lemmas such as
substitution and canonical lemmas:

\begin{lemma}[Substitution]
If $\varnothing;\varnothing,x:\taub' \ts \eb:\taub$
and $\varnothing;\varnothing \ts \vb : \taub'$,
then $\varnothing;\varnothing \ts \eb[\vb/x]:\taub$.
\label{lemma:sub-1}
\end{lemma}

\begin{proof}
By induction of a derivation of $\varnothing;\varnothing,x:\taub' \ts \eb:\taub$.
\end{proof}

\begin{lemma}[Type substitution]
If $\varnothing,\alpha:\kappa;\varnothing \ts \eb:\taub$
and $\varnothing \ts \thetab : \kappa$,
then $\varnothing;\varnothing \ts \eb[\thetab/\alpha]:\taub[\thetab/\alpha]$.
\end{lemma}

\begin{proof}
By induction of a derivation of $\varnothing,\alpha:\kappa;\varnothing \ts \eb:\taub$.
Similar to the proof of lemma~\ref{lemma:sub-1}.
\end{proof}

\begin{lemma}[Canonical forms]
\label{lem:canonical}
\begin{enumerate}
 \item if $\vb$ is a value of type $\Int$, then $\vb = n$.
 \item if $\vb$ is a value of type $\taub_1 \ra \taub_2$, then $\vb = \eAbs{x:\taub_1}{\eb}$.
 \item if $\vb$ is a value of type $\forall\alpha:\kappa.\taub$, then $\vb = \eTabs{\alpha:\kappa}{\eb}$.
 \item if $\vb$ is a value of type $\tRec{\rhob}$, then $\vb = \eRecN{i=0}{n}{l_i=\vb_i}$ for some $n$.
\end{enumerate}
\end{lemma}

\begin{proof}
By induction of $\taub$ with the inversion lemma.
\end{proof}

We can prove type soundness using the standard technique of
preservation and progress:

\begin{lemma}[Preservation]
If $\varnothing;\varnothing \ts \eb:\taub$ and $\Eb[\eb] \mapsto \Eb[\eb']$,
then $\varnothing;\varnothing \ts \eb':\taub$.
\end{lemma}

\begin{proof}
The proof proceeds by case analysis according to the derivation of $\Eb[\eb] \mapsto \Eb[\eb']$.
The cases are entirely standard except for the $\mathbf{reify}$ expression.
We present only this.
\begin{itemize}
\item {\em Case} $\eb = \reify{l_1:\taub_1,\ldots,l_n:\taub_n,\tRowEmp}{\taub'}{\vb}$ and
                 $\eb' = \{ l_i = \lambda~x_i:\taub_i.
                  \vb~(\Lambda~\alpha:\star.
                            \eAbs{c:\tRecN{j=1}{n}{l_j:\taub_j \ra  \alpha}}{c.l_i~x_i})\}_{i=1}^{n}$.
By given,
$\varnothing;\varnothing \ts \mathbf{reify}[l_1:\taub_1,\ldots,l_n:\taub_n,\tRowEmp][\taub']~\vb:\tau$
where $\taub = \eRec{\rhob \rowArrow \taub'}
             = \eRec{l_1:\taub_1 \ra \taub',\ldots,l_n:\taub_n \ra \taub'}$.
By inv of \rname{T-reify}, $\varnothing;\varnothing \ts \vb : \cpsSum{\rhob} \ra \taub$ (\cc{1}).
TS: $\varnothing;\varnothing \ts \{ l_i = \lambda~x_i:\taub_i.
                  \vb~(\Lambda~\alpha:\star.
                  \eAbs{c:\tRecN{j=1}{n}{l_j:\taub_j \ra  \alpha}}{c.l_i~x_i}) \}^n_{i=1}:\tau$.
By inv of \rname{T-r} and \rname{T-abs},\\
STS: $\forall i.\varnothing;\varnothing,x_i:\taub_i \ts
\vb~(\LAM{\alpha:\star}.
     \lambda c:\{l_j:\taub_j \ra  \alpha\}_{j=1}^{n}.{c.l_i~x_i}) : \taub'$.
By inv of \rname{T-app},
STS: $\forall i.\varnothing;\varnothing,x_i:\taub_i \ts \vb : \cpsSum{\rhob} \ra \taub$
which is true by \cc{1}) and
$\forall i.\varnothing;\varnothing,x_i:\taub_i \ts \eTabs{\alpha:\star}
                  {\eAbs{c:\tRecN{j=1}{n}{l_j:\taub_j \ra  \alpha}}{c.l_i~x_i}}: 
\cpsSum{\rhob}$ (which is provable by typing rules).
\end{itemize}
\end{proof}
\begin{lemma}[Progress]
If $\varnothing;\varnothing \ts \eb:\taub$,
then either $\eb$ is a value or else there is some $\eb'$ with
$\eb \mapsto \Eb[\eb']$ where $\eb = \Eb[\bar{r}]$ and $\bar{r}$ is a redex.
\end{lemma}

\begin{proof}
By induction of a derivation of $\varnothing;\varnothing \ts \eb :\taub$.
The cases are entirely standard except for the $\mathbf{reify}$ expression.
We present only this.
\begin{itemize}
\item {\em Case} $\eb =\reify{\rhob}{\taub}{\eb_1}$. \\
By given, $\varnothing;\varnothing \ts
\reify{\rhob}{\taub}{\eb_1} : \tRec{\rhob \rowArrow \taub}$.
By inv of \rname{T-reify},
$\varnothing; \varnothing \ts \eb_1 : \cpsSum{\rhob} \ra \taub$.
Because of its type, $\eb_1$ should be a function, which is a value.
Then, done by \rname{reify}.
\end{itemize}
\end{proof}

The main result is the type soundness of the \IL{} programs:

\begin{theorem}[Type soundness]
If $\varnothing;\varnothing \ts \eb:\taub$, 
either $\eb$ is a value or else there is some $\eb'$ with
$\eb \mapsto \eb'$ where $\varnothing;\varnothing \ts \eb' : \taub$.
\end{theorem}

\begin{proof}
Type soundness follows from the preservation and progress lemmas.
\end{proof}

\subsection{From \EL{} to \IL{}}
The translation from \EL{} into \IL{} is somewhat involved 
because it performs two transformations at once: 
(1) a transformation into continuation-passing style (CPS) 
\cite{appel:92:cps}, and (2) a dual
translation that eliminates sums and cases in favor of records of
functions and polymorphic functions on such records.

There are two translation judgments: one for syntactic values, and one
for all expressions.  The judgment for a syntactic value $e$ has the
form $\Delta;\Gamma \tsv e : \tau \leadsto \eb : \taub$.  Notice the
absence of exception types.  Since $e$ is a value, its \IL{}
counterpart $\eb$ requires neither continuation nor handler.  For
non-values there is no derivation for a $\tsv$ judgment.

The \IL{} counterpart for non-values is a {\em computation}.
Computations are suspensions that await a continuation and a handler
record.  Once continuation and handlers are supplied, a computation
will run until a final answer is produced and the program terminates.
The translation of an expression $e$ to its computation counterpart
is expressed by a judgment of the form $\Delta;\Gamma \ts e :
\tau;\rho \leadsto \cb : \comp{\taub}{\rhob}$ where $\cb$ is the \IL{}
term representing the computation denoted by $e$.  The type of $\cb$ is
always $\comp{\taub}{\rhob}$ where $\taub$ and $\rhob$ are the \IL{}
counterparts of $\tau$ and $\rho$.

\paragraph{Notation:} To talk about continuations, handlers, and
computations, it is convenient to introduce some notational shorthands
(see \figref{il-type-macros}).  We write $\Ans$ for the type of the
final answer, 
\cont{\taub} for the type of continuations accepting values of type
$\taub$, \hdlr{\rhob} for the type of exception handlers, i.e.,
records of continuations whose argument types are described by
$\rhob$, and \comp{\taub}{\rhob} for the type of computations awaiting
a \cont{\taub} and a \hdlr{\rhob}.  The CPS-converted \IL{} equivalent
of an \EL{} function type is $\cpsRa{\taub_1}{\taub_2}{\rhob}$. It
describes functions from $\taub_1$ to $\comp{\taub_2}{\rhob}$.
Similarly, the type $\etCase{\tSum{\rhob}}{\taub}{\rhob'}$ is the \IL{}
encoding of a first-class cases type, i.e., a record of functions that
produce computations of type $\comp{\taub}{\rhob'}$.  Finally,
$\cpsSum{\rhob}$ is the dual encoding of a sum: the polymorphic type
of functions from records of functions to their common co-domain.
%
%

\begin{figure}
\[
\begin{array}{rcl@{\qquad}rcl}
\Ans                  & \equiv & \Int &
\comp{\taub}{\rhob}   & \equiv & \cont{\taub} \ra \hdlr{\rhob} \ra \Ans \\
\rhob \rowArrow \taub & \equiv & \rhob' \quad \tx{such~that~} \rhob;\taub \tmaps \rhob' &
\cpsRa{\taub}{\rhob}{\taub'} 
                      & \equiv & \taub \ra \comp{\taub'}{\rhob} \\
\cont{\taub}          & \equiv & \taub \ra \Ans  &
\cpsCa{\tSum{\rhob'}}{\rhob}{\taub} 
                      & \equiv & \tRec{\rhob' \rowArrow \comp{\taub}{\rhob}} \\
\hdlr{\rhob}          & \equiv & \tRec{\rhob \rowArrow \Ans} &
\cpsSum{\rhob}        & \equiv & \forall\alpha:\star.\tRec{\rhob \rowArrow \alpha} \ra \alpha
\end{array}
\]
\caption{Type synonyms for \IL{} types.}
\label{fig:il-type-macros}
\end{figure}


Notice that most of the type synonyms in \figref{il-type-macros} make
use of the notation $\rhob \rowArrow \taub$.  It stands for the unique
row type $\rhob'$ for which the {\em row normalization} judgment
$\rhob;\taub \tmaps \rhob'$ holds (see~\figref{il-row-norm}).  Our
presentation relies on the convention that any direct or indirect use
of the $\rowArrow$ shorthand in a rule introduces an implicit row
normalization judgment to the premises of that rule.

To improve the readability of the rules, we omit many ``obvious''
types from \IL{} terms.  For example, we write $\lambda k \lambda h.
\eb: \comp{\taub}{\rhob}$ without the types for $k$ and $h$, since
these types clearly can only be $\cont{\taub}$ and $\hdlr{\rhob}$,
respectively.

\paragraph{Type translation:} \figref{il-types-x} shows the
translation of \EL{} types to \IL{} types.  The use of type synonyms
makes the presentation look straightforward. (But beware of implicit
normalization judgments!)

\paragraph{Value translation:} \figref{il-x-val} shows the translation
of syntactic values: constants, variables, functions, and cases.
Constants are trivial while variables may produce type applications if
their types are polymorphic.

The transformation of functions depends on whether the body itself is
a syntactic value or not.  If the body $e$ of function $f$ is a value,
then it is transformed {\em as a value}, i.e., using the $\tsv$
judgment, into an \IL{} term $\eb$.  Then a recursively polymorphic
CPS function is constructed.  When instantiated and applied, it simply
passes $\eb$ to its continuation $k'$.  Its exception handler $h'$ is
never used.  Since the constructed function is polymorphic, it must be
instantiated at $\rhob$ to form the final result.
If the body $e$ is a non-value, then rule \rname{fun/non-val} applies
and $e$ is turned into a computation $\cb$ that becomes the body of the
constructed \IL{} function.

Cases are treated as a sequence of individual non-value functions that
are not recursive.  Each of these functions is translated and placed
into the result record at the appropriate label.

\paragraph{Basic computations:} \figref{il-x-basic} shows the
translation of basic terms: injection into sums, applications, and
\textit{let}-bindings.  Also shown is rule \rname{value} for lifting
syntactic values into the domain of computations.  From $\eb$ (the
result of translating value $e$) it constructs a computation term
that passes $\eb$ to its continuation $k$.  The computation's
exception handler $h$ is never used, which is justification for
leaving the exception type of syntactic values unspecified.

The computation representing $l~e$, i.e., the creation of a sum value,
first runs sub-computation $\cb$ corresponding to $e$ to obtain the
intended ``payload'' $x$.  The result that is sent to the continuation
is a polymorphic function which receives a record $r$ of other
functions, selects $l$ from $r$, and invokes the result with the $x$
(the payload) as its argument.  This is simply the dual encoding of
sums as functions taking records as arguments.

Application is simple: after running two sub-computations
$\cb_1$ and $\cb_2$ to obtain the callee $x_1$ and its intended argument
$x_2$, the callee is invoked with $x_2$ to obtain the third and final
computation.  All three computations are invoked with the same
handler argument.

Non-value \textit{let}-bindings simply chain two computations together
without altering any handlers.  The translation of a polymorphic
\textit{let}-bindings invokes the value translation judgment on the
definien expression $e_1$ to obtain $\eb_1$ which is then turned into
a polymorphic value via type abstraction.  The constructed value is
available to the sub-computation $\cb_2$ representing the body $e_2$.

We omitted the rules for type equality, since they are somewhat
tedious but straightforward.

\paragraph{Computations involving records, cases and exceptions:}
The translations for records, cases and exception-related
expressions are shown in \figref{il-x-rec}, \ref{fig:il-x-case}
and \ref{fig:il-x-exn}, respectively.
A \rname{match} computation instantiates its sum argument
(bound to $x_1$) at computation type and applies it to the record of
functions $x_2$ representing the cases.  The \rname{raise}
computation, on the other hand, instantiates the sum at type \Ans and
applies it to $h$, i.e., the current record of exception handlers.  It
does not use its regular continuation $k$, justifying the typing rule
that leaves the result type unconstrained.

A case extension computation
extends a record of functions representing cases, while the
\rname{handle} computation extends the record of
(continuation-)functions representing handlers.  The rules for
\rname{unhandle} and \rname{rehandle} are similar to that for
\rname{handle}: in the former case a field is dropped from the
handler record, while in the latter a field is replaced.  Similar
operations exist for cases, but for brevity we have omitted them from
the discussion.

The \rname{handle-all} rule is the only rule introducing
\rname{reify} into its output term.  It is used to build a new
exception-handler record from $\rhob$, which is the exception type of
$e_1$.  Each field $l_i$ of this record receives the payload of
exception $l_i$, injects it into $\cpsSum{\rhob}$, and passes the
result (as a binding to $x$) to the computation specified by $e_2$.

\begin{figure}[t]
\[
\begin{array}{@{}c@{}}

\infer{\alpha \leadsto \alpha}{\strut}

\qquad

\infer{\Int \leadsto \Int}{\strut}

\qquad

\infer
{\tRowEmp \leadsto \tRowEmp}{\strut}

\qquad

\infer
{
l:\tau, \rho \leadsto l:\taub, \rhob
}
{
\tau \leadsto \taub &  \rho \leadsto \rhob \\
}

\qquad

\infer{\tSum{\rho} \leadsto \cpsSum{\rhob}}
      {\rho \leadsto \rhob}

\\[1.5mm]

\infer{\tau_1 \era{\rho} \tau_2 \leadsto \cpsRa{\taub_1}{\rhob}{\taub_2}}
      {\tau_1 \leadsto \taub_1 & \tau_2 \leadsto \taub_2 & \rho \leadsto \rhob}

\qquad

\infer
{
\etCase{\tSum{\rho'}}{\tau}{\rho} \leadsto \cpsCa{\tSum{\rhob'}}{\rhob}{\taub}
}
{
\rho' \leadsto \rhob' & \tau \leadsto \taub & \rho \leadsto \rhob
}

\end{array}
\]
\caption{Translation of EL types to IL types.}
\label{fig:il-types-x}
\vspace{-3mm}
\end{figure}


%
%

%
%

\begin{figure}
\[
\begin{array}{c}

\infer[\rlabel{var}]
{
\Delta;\Gamma  \tsv  x : \tau \leadsto x[\thetab_1]\ldots[\thetab_n] : \taub
}
{
\begin{array}{c}
\Gamma(x) = \forall\alpha_1:\kappa_1\ldots\forall\alpha_n:\kappa_n.\tau' \quad 
\forall i_{\in 1..n} .\, \Delta \ts \theta_i : \kappa_i \\[1mm]
\tau = \tau'[\theta_1 / \alpha_1,\ldots,\theta_n / \alpha_n] \quad
\tau \leadsto \taub \quad 
\forall i_{\in 1..n} .\, \theta_i \leadsto \thetab_i 
\end{array}
}

\quad

\infer[\rlabel{int}]
{
\Delta;\Gamma  \tsv  n : \Int \leadsto n : \Int
}
{\strut}
\\[6mm]

\infer[\rlabel{fun/val}]
{
\begin{array}{rl}
\Delta;\Gamma \tsv & \eFun{f}{x}{e} : \tau_2 \era{\rho} \tau \\
\leadsto &
 \mathbf{letrec}~f:\forall\alpha:\varnothing.\cpsRa{\taub_2}{\alpha}{\taub}
   = \Lambda \alpha \lambda x \lambda k' \lambda h' . k' \eb  ~\mathbf{in}~
    f[\rhob] : \cpsRa{\taub_2}{\rhob}{\taub}
\end{array}
}
{
\begin{array}{c}
\Delta;\Gamma, f : \forall\alpha:\varnothing.\tau_2 \era{\alpha} \tau,  x:\tau_2  \tsv  e : \tau \leadsto \eb : \taub \\
\istyp{\Delta}{\tau_2} \qquad
\tau_2 \leadsto \taub_2 \qquad
\isrow{\Delta}{\rho} \qquad
\rho \leadsto \rhob
\end{array}
}
\\[6mm]

\infer[\rlabel{fun/non-val}]
{
\Delta;\Gamma \tsv \eFun{f}{x}{e} : \tau_2 \era{\rho} \tau 
\leadsto  
 \mathbf{letrec}~f:\cpsRa{\taub_2}{\rhob}{\taub} 
   = \eAbs{x}{\cb}~\mathbf{in}~
   f : \cpsRa{\taub_2}{\rhob}{\taub}
}
{
\begin{array}{c}
\Delta;\Gamma, f : \tau_2 \era{\rho} \tau,  x:\tau_2  \ts  e : \tau ; \rho 
\leadsto \cb : \comp{\taub}{\rhob} \\
\istyp{\Delta}{\tau_2} \quad
\tau_2  \leadsto \taub_2 \quad
\isrow{\Delta}{\rho}  
\end{array}
}
\\[6mm]

\infer[\rlabel{c}]
{
\Delta;\Gamma \tsv \eCaseN{i=1}{n}{l_i~x_i \dra e_i} : \etCase{\tSumN{i=1}{n}{l_i:\tau_i}}{\tau}{\rho} 
\leadsto 
 \eRecN{i=1}{n}{l_i=\eAbs{x_i}{\cb_i}} : \cpsCa{\tSumN{i=1}{n}{l_i:\taub_i}}{\rhob}{\taub}
}
{
\begin{array}{c}
\forall i_{\in 1..n}.\Delta;\Gamma, x_i: \tau_i \ts e_i : \tau ; \rho \leadsto \cb_i
: \comp{\taub}{\rhob}\\
\isrow{\Delta}{(l_1:\tau_1, \ldots, l_n:\tau_n,\tRowEmp)} \qquad
\forall i_{\in 1..n}.\,\tau_i \leadsto \taub_i
\end{array}
}
\end{array}
\]
\caption{The translation from \EL{} to \IL{} for syntactic values.}
\label{fig:il-x-val}
\end{figure}

%
%

\begin{figure}
\[
\begin{array}{c}

\infer[\rlabel{value}]
{
\Delta;\Gamma \ts e : \tau ; \rho \leadsto \lambda k \lambda h. k~\eb
: \comp{\taub}{\rhob}
}
{
\Delta;\Gamma \tsv e : \tau \leadsto \eb : \taub \qquad
\isrow{\Delta}{\rho} \qquad
\rho \leadsto \rhob
}


\\[6mm]

\infer[\rlabel{dcon}]
{
\begin{array}{rl}
\Delta; \Gamma  \ts & l \ e : \tSum{l:\tau, \rho} ; \rho' \\
\leadsto &
 \lambda k\lambda h.
    \cb(\lambda x.k(\Lambda\alpha\lambda r.(r.l~x)))~h 
    : \comp{\cpsSum{l:\taub,\rhob}}{\rhob'}
\end{array}
}
{
\begin{array}{c}
\Delta; \Gamma \ts e : \tau ; \rho' \leadsto \cb:\comp{\taub}{\rhob'} \qquad
\isrow{\Delta}{(l:\tau, \rho)} \qquad \rho \leadsto \rhob
\end{array}
}
\\[6mm]

\infer[\rlabel{app}]
{
\begin{array}{c}
\Delta; \Gamma  \ts  e_1 \ e_2 : \tau ; \rho 
\leadsto 
 \lambda k\lambda h.
 \cb_1(\eAbs{x_1}{\cb_2(\eAbs{x_2}{x_1~x_2~k~h})~h})~h 
 : \comp{\taub}{\rhob}
\end{array}
}
{
\begin{array}{c}
\Delta; \Gamma  \ts  e_1 : \tau_2 \era{\rho} \tau ; \rho \leadsto 
\cb_1 : \comp{\cpsRa{\taub_2}{\rhob}{\taub}}{\rhob} \\
\Delta; \Gamma  \ts  e_2 : \tau_2 ; \rho \leadsto 
\cb_2 : \comp{\taub_2}{\rhob}
\end{array}
}

\\[6mm]

\infer[\rlabel{let/val}]
{
\begin{array}{rl}
\Delta; \Gamma \ts & \eLet{x}{e_1}{e_2}: \tau_2 ; \rho \\
\leadsto &
 \lambda k\lambda h.
 \mathbf{let}~x:\forall\alpha_1:\kappa_1\ldots\forall\alpha_n:\kappa_n.\taub_1 =
 \Lambda\alpha_1\ldots\Lambda\alpha_n.\eb_1 \\
 & ~~~~~~~~\mathbf{in}~\cb_2~k~h : \comp{\taub_2}{\rhob}
\end{array}
}
{
\begin{array}{c}
\{\alpha_1,\ldots,\alpha_n\} = \FTV(\tau_1) \setminus \FTV(\Gamma) \\
\Delta,\alpha_1:\kappa_1,\ldots,\alpha_n:\kappa_n;\Gamma
   \tsv e_1 : \tau_1 \leadsto \eb_1 : \taub_1 \\
\Delta;\Gamma,x:\forall\alpha_1:\kappa_1\ldots\forall\alpha_n:\kappa_n.\tau_1
   \ts e_2 : \tau_2 ; \rho \leadsto \cb_2 : \comp{\taub_2}{\rhob}
\end{array}
}
\\[8mm]

\infer[\rlabel{let/non-val}]
{
\begin{array}{c}
\Delta; \Gamma \ts \eLet{x}{e_1}{e_2}: \tau_2 ; \rho 
\leadsto 
 \lambda k\lambda h. \cb_1(\eAbs{x}{\cb_2~k~h})~h : \comp{\taub_2}{\rhob}
\end{array}
}
{
\begin{array}{c}
\Delta; \Gamma \ts e_1 : \tau_1 ; \rho \leadsto \cb_1 : \comp{\taub_1}{\rhob} \\ 
\Delta; \Gamma, x : \tau_1 \ts e_2 : \tau_2 ; \rho \leadsto \cb_2 :
\comp{\taub_2}{\rhob}
\end{array}
}
\\[6mm]

\infer[\rlabel{roll}]
{
\Delta; \Gamma \ts e : \tBind{\alpha}{\tSum{\rho}} ; \rho'
\leadsto \cb:\comp{\tBind{\alpha}{\cpsSum{\rhob'}}}{\rhob'}
}
{
\Delta; \Gamma \ts e : \tSum{\rho[\tBind{\alpha}{\tSum{\rho}}/\alpha]} ; \rho'
\leadsto \cb:\comp{\cpsSum{\rhob[\tBind{\alpha}{\cpsSum{\rhob}}/\alpha]}}{\rhob'}
}
\\[4mm]

\infer[\rlabel{unroll}]
{
\Delta; \Gamma \ts e : \tSum{\rho[\tBind{\alpha}{\tSum{\rho}}/\alpha]} ; \rho'
\leadsto \cb:\comp{\cpsSum{\rhob[\tBind{\alpha}{\cpsSum{\rhob}}/\alpha]}}{\rhob'}
}
{
\Delta; \Gamma \ts e : \tBind{\alpha}{\tSum{\rho}} ; \rho'
\leadsto \cb:\comp{\tBind{\alpha}{\cpsSum{\rhob'}}}{\rhob'}
}

\end{array}
\]
\caption{The translation from \EL{} to \IL{} for basic computations.}
\label{fig:il-x-basic}
\end{figure}

%
%

\begin{figure}
\begin{mathpar}
%
%
\irr{\rlabel{r}}
{
\forall i_{\in 1..n} . \Delta;\Gamma \ts e_i : \tau_i ; \rho \leadsto \cb_i : \comp{\taub_i}{\rhob} \+
\Delta \ts (l_1:\tau_1, \ldots, l_n : \tau_n) : \varnothing \+
(l_1:\tau_1, \ldots, l_n : \tau_n) \leadsto (l_1:\taub_1, \ldots, l_n : \taub_n)
}
{
\begin{array}{rl}
\Delta;\Gamma \ts & \eRecN{i=1}{n}{l_i = e_i} : 
                    \tRecN{i=1}{n}{l_i:\tau_i} ; \rho \\
         \leadsto & \lambda k\lambda h.
                    k~(\eRecN{i=1}{n}{l_i=\cb_i})~h
                    :\comp{\eRecN{i=1}{n}{l_i:\taub_i}}{\rhob}
\end{array}
}
\and
\irr{\rlabel{r/ext}}
{
\Delta;\Gamma \ts e_1 : \tRec{\rho} ; \rho' \leadsto \cb_1:\comp{\tRec{\rhob}}{\rhob'} \+
\Delta;\Gamma \ts e_2 : \tau_2 ; \rho' \leadsto \cb_2 : \comp{\taub_2}{\rhob'} \+
\Delta \ts (l:\tau_2, \rho) : \varnothing \+
(l:\tau_2, \rho) \leadsto (l:\taub_2, \rhob)
}
{
\begin{array}{l}
\Delta;\Gamma \ts \eRecExt{e_1}{\eRec{l = e_2}} : \tRec{l : \tau_2, \rho} ; \rho' \\
\leadsto 
 \lambda k\lambda h.
  \cb_1~(\eAbs{v_1:\tRec{\rhob}}
  {\cb_2~(\eAbs{v_2:\taub_2}
  {k~(\eRecExt{v_1}{\eRec{l = v_2}})})~h})~h 
  : \comp{\tRec{l:\taub_2,\rhob}}{\rhob'}
\end{array}
}
\and
\irr{\rlabel{r/sub}}
{
\Delta;\Gamma \ts e : \tRec{l : \tau, \rho} ; \rho' \leadsto 
\cb:\comp{\eRec{l:\taub,\rhob}}{\rhob'}
}
{
\Delta;\Gamma \ts  \eRecSub{e}{l} : \tRec{\rho} ; \rho'
\leadsto           \lambda k\lambda h.
                    \cb~(\eAbs{v:\eRec{l:\taub,\rhob}}{k~(\eRecSub{v}{l})})~h :
                    \comp{\tRec{\rhob}}{\rhob'}
}
\and
\irr{\rlabel{select}}
{
\Delta; \Gamma \ts e :  \tRec{l:\tau, \rho} ; \rho' \leadsto \cb:\comp{\eRec{l:\taub,\rhob}}{\rhob'}
}
{
\Delta; \Gamma \ts e.l: \tau ; \rho'
\leadsto           \lambda k\lambda h.
                     \cb~(~\eAbs{r:\eRec{l:\taub,\rhob}}{(r.l)~k~h})~h : 
                     \comp{\taub}{\rhob'}
}
\end{mathpar}
\caption{The translation from \EL{} to \IL{} for computations involving records.}
\label{fig:il-x-rec}
\end{figure}

%
%

\begin{figure}
\begin{mathpar}
\irr{\rlabel{c/ext}}
{
\begin{array}{c}
\Delta;\Gamma \ts e_1 : \etCase{\tSum{\rho_1}}{\tau}{\rho} ; \rho'
   \leadsto \cb_1 : \comp{\cpsCa{\tSum{\rhob_1}}{\rhob}{\taub}}{\rhob'} \qquad
\isrow{\Delta}{(l:\tau_1,\rho_1)} \\
\Delta;\Gamma, x:\tau_1 \ts e_2 : \tau ; \rho \leadsto \cb_2: \comp{\taub}{\rhob} \qquad
(l:\tau_1,\rho_1) \leadsto (l:\taub_1,\rhob_1) 
\end{array}
}
{
\begin{array}{rl}
\Delta;\Gamma \ts & \eCaseExt{e_1}{\eCase{l~x \dra e_2}} :
   \etCase{\tSum{l:\tau_1, \rho_1}}{\tau}{\rho} ; \rho' \\
\leadsto &
 \lambda k\lambda h.
 \cb_1 (\lambda x_1 .
 k (\eRecExt{x_1}{\eRec{l=\eAbs{x}{\cb_2}}})) h 
 : \comp{{\cpsCa{\tSum{l:\taub_1,\rhob_1}}{\rhob}{\taub}}}{\rhob'} 
\end{array}
}
\and
\irr{\rlabel{c/sub}}
{
\Delta;\Gamma \ts e_1 : \etCase{\tSum{l:\tau',\rho_1}}{\tau}{\rho} ; \rho'
   \leadsto \cb_1 : \comp{\cpsCa{\tSum{l:\taub',\rhob_1}}{\rhob}{\taub}}{\rhob'} 
}
{
\begin{array}{rl}
\Delta;\Gamma \ts & \eCaseSub{e_1}{l} :
   \etCase{\tSum{\rho_1}}{\tau}{\rho} ; \rho' \\
\leadsto &
 \lambda k\lambda h.
 \cb_1 (\lambda x_1 .
 k (\eRecSub{x_1}{l})) h 
 : \comp{{\cpsCa{\tSum{\rhob_1}}{\rhob}{\taub}}}{\rhob'} 
\end{array}
}
\and
\irr{\rlabel{match}}
{
\begin{array}{c}
\Delta; \Gamma \ts e_1 : \tSum{\rho} ; \rho' \leadsto \cb_1:\comp{\cpsSum{\rhob}}{\rhob'} \\
\Delta; \Gamma \ts e_2 : \etCase{\tSum{\rho}}{\tau}{\rho'} ; \rho'
 \leadsto \cb_2:\comp{\cpsCa{\tSum{\rhob}}{\rhob'}{\taub}}{\rhob'}
\end{array}
}
{
\begin{array}{rl}
\Delta; \Gamma \ts & \eMat{e_1}{e_2}: \tau ; \rho' \\
\leadsto &
 \lambda k\lambda h.
 \cb_1 (\eAbs{x_1}{\cb_2 (\eAbs{x_2}{x_1[\comp{\taub}{\rhob'}]~x_2~k~h})~h})~h 
 : \comp{\taub}{\rhob'}
\end{array}
}
\end{mathpar}
\caption{The translation from \EL{} to \IL{} for computations involving cases.}
\label{fig:il-x-case}
\end{figure}

%
%

\begin{figure}
\[
\begin{array}{c}

\infer[\rlabel{raise}]
{
\begin{array}{c}
\Delta; \Gamma \ts \eRaise{e} : \tau ; \rho 
\leadsto 
 \lambda k\lambda h.
 \cb (\eAbs{x}{x [\Ans]~h})~h : \comp{\taub}{\rhob}
\end{array}
}
{
\begin{array}{c}
\Delta; \Gamma \ts e : \tSum{\rho} ; \rho \leadsto \cb:\comp{\cpsSum{\rhob}}{\rhob} \qquad
\Delta \ts \tau : \star \qquad
 \tau \leadsto \taub
\end{array}
}

\\[6mm]

\infer[\rlabel{handle}]
{
\begin{array}{rl}
\Delta; \Gamma \ts & \eHandle{e_1}{\eCase{l~x \dra e_2}} : \tau ; \rho \\
\leadsto &
 \lambda k
 \lambda h.
 \cb_1~k~(\eRecExt{h}{\eRec{l=\eAbs{x}{\cb_2~k~h}}}) : \comp{\taub}{\rhob}
\end{array}
}
{
\begin{array}{c}
\Delta; \Gamma \ts e_1 : \tau ;~l:\tau', \rho \leadsto \cb_1:\comp{\taub}{(l:\taub',\rhob)} \\
\Delta; \Gamma, x:\tau' \ts e_2 : \tau ; \rho \leadsto \cb_2:\comp{\taub}{\rhob}
\end{array}
}

\\[6mm]

\infer[\rlabel{unhandle}]
{
\Delta; \Gamma \ts  \eUnhandle{e}{l} : \tau;~l:\tau',\rho 
\leadsto 
  \lambda k \lambda h . \cb~k~(\eRecSub{h}{l}) : \comp{\taub}{l:\taub',\rhob}
}
{
\begin{array}{c}
\Delta; \Gamma \ts e : \tau;~\rho \leadsto \cb : \comp{\taub}{\rhob} \quad
\isrow{\Delta}{(l:\tau',\rho)} \quad
\tau' \leadsto \taub' \quad 
\rho \leadsto \rhob
\end{array}
}

\\[6mm]

\infer[\rlabel{rehandle}]
{
\begin{array}{rl}
\Delta; \Gamma \ts & \eRehandle{e_1}{\eCase{l~x \dra e_2}} : \tau ; l:\tau'', \rho \\
\leadsto &
 \lambda k
 \lambda h.
 \cb_1~k~(\eRecExt{(\eRecSub{h}{l})}{\eRec{l=\eAbs{x}{\cb_2~k~h}}})  
 : \comp{\taub}{(l:\taub'',\rhob)}
\end{array}
}
{
\begin{array}{c}
\Delta; \Gamma \ts e_1 : \tau ;~l:\tau', \rho \leadsto \cb_1:\comp{\taub}{(l:\taub',\rhob)} \\
\Delta; \Gamma, x:\tau' \ts e_2 : \tau ;~l:\tau'', \rho \leadsto \cb_2:\comp{\taub}{(l:\taub'',\rhob)}
\end{array}
}

\\[6mm]

\infer[\rlabel{handle-all}]
{
\begin{array}{rl}
\Delta; \Gamma \ts & \eHandle{e_1}{\eCase{x \dra e_2}} : \tau ; \rho \\
\leadsto &
 \lambda k
 \lambda h.
 \cb_1~k~(\reify{\rhob'}{\Ans}(\eAbs{x}{\cb_2~k~h})) 
 : \comp{\taub}{\rhob} \\[4mm]
\end{array}
}
{
\begin{array}{c}
\Delta; \Gamma \ts e_1 : \tau ; \rho' \leadsto \cb_1 : \comp{\taub}{\rhob'} \\
\Delta; \Gamma, x:\tSum{\rho'} \ts e_2 : \tau ; \rho \leadsto \cb_2 : \comp{\taub}{\rhob}
\end{array}
}

\\[6mm]

\framebox{
\infer[\rlabel{program}]
{
\Gamma_0 \ts \prog{e} \leadsto \cb~(\eAbs{x}{x})~\eRec{} : \Ans
}
{
\varnothing;\Gamma_0 \ts e : \Int ; \tRowEmp \leadsto \cb : \comp{\Int}{\tRowEmp}
}
}

\end{array}
\]
\caption{The translation from \EL{} to \IL{} for computations involving exceptions.}
\label{fig:il-x-exn}
\end{figure}


\subsection*{Properties of $\leadsto$}

An important property of the translation is that it translates 
well-formed \EL{} expressions to well-formed \IL{} expressions.
Before we proceed to establishing the correctness of $\leadsto$,
we set up a few helper lemmas:

\begin{lemma}[Type synonyms]
\begin{enumerate}
 \item If $\Deltab;\Gammab \ts \lambda k:\cont{\taub}.\lambda h:\hdlr{\rhob}.\eb : \comp{\taub}{\rhob}$,
       then $\Deltab;\Gammab,k:\cont{\taub},h:\hdlr{\rhob} \ts \eb : \Ans$.
 \item If $\Deltab;\Gammab, h:\hdlr{\rhob} \ts \cb : \comp{\taub}{\rhob}$ and
       $\Deltab;\Gammab, h:\hdlr{\rhob} \ts \cb~\eb~h : \Ans$,
       then $\Deltab;\Gammab, h:\hdlr{\rhob} \ts \eb: \cont{\taub}$.
 \item If $\Deltab;\Gammab, k:\cont{\taub} \ts \cb : \comp{\taub}{\rhob}$ and
       $\Deltab;\Gammab, k:\cont{\taub} \ts \cb~k~\eb : \Ans$,
       then $\Deltab;\Gammab, k:\cont{\taub} \ts \eb: \hdlr{\rhob}$.
\end{enumerate}
\label{lemma:synonyms}
\end{lemma}

\begin{proof}
By defintion of $\comp{\taub}{\rhob}$ which is $\cont{\taub} \ra \hdlr{\rhob} \ra \Ans$ and
by the typing rule of \rname{T-abs} and \rname{T-app}.
\end{proof}

\begin{lemma}[Weakening-$\Deltab;\Gammab$]
If $\Deltab;\Gammab \ts \cb : \comp{\taub}{\rhob}$,
then $\Deltab';\Gammab' \ts \cb : \comp{\taub}{\rhob}$
for all $\Gammab'$ and $\Deltab'$ such that $\Gammab'\supseteq\Gammab$ and $\Deltab'\supseteq\Deltab$.
\label{lemma:weaking}
\end{lemma}

\begin{proof}
By induction of a derivation of $\Deltab;\Gammab \ts \cb : \comp{\taub}{\rhob}$.
\end{proof}

\begin{definition}[Translation of environments]
\[
\begin{array}{rcl}
C(\varnothing) & = & \varnothing \\
C(\Gamma, x \mapsto \sigma) & = & C(\Gamma),x \mapsto C(\sigma) \\
C(\tau) & = & \taub \mathrm{~~~where~} \tau \leadsto \taub \\
C(\forall \alpha:\kappa.\sigma) & = & \forall \alpha:\kappa.C(\sigma) \\
C(\Delta, \alpha \mapsto \kappa) & = & C(\Delta), \alpha \mapsto \kappa \\
\end{array}
\]
\label{def:trans-environ}
\end{definition}

\begin{lemma}[Translation of $\Gamma$]
If $\Delta \ts \tau : \kappa$, then $C(\Delta) \ts \taub : \kappa$.
\label{lemma:trans-gamma}
\end{lemma}

\begin{proof}
By induction of a derivation of $\Delta \ts \tau : \kappa$.
\end{proof}

\begin{lemma}[Substitution]
If $\tau = \tau'[\tau_1/\alpha_1,\ldots,\tau_n/\alpha_n]$ and $\tau \leadsto \taub$,
then $\taub = \taub'[\taub_1/\alpha_1,\ldots,\taub_n/\alpha_n]$
where $\tau' \leadsto \taub'$ and $\forall n_{\in 1..n}.\tau_n \leadsto \taub_n'$.
\label{lemma:substitution}
\end{lemma}

\begin{proof}
By induction of $\tau$.
\end{proof}

These lemmas allow us to prove correctness of $\leadsto$:

\begin{lemma}[Correctness of translation~$\leadsto$]
If $\Delta;\Gamma \ts e:\tau;\rho \leadsto \cb :\comp{\taub}{\rhob}$ and $\Gammab \supseteq C(\Gamma)$
and $\Deltab \supseteq C(\Delta)$,
then $\Deltab;\Gammab \ts \cb : \comp{\taub}{\rhob}$.
\end{lemma}

\begin{proof}
By induction of a derivation of $\Delta;\Gamma \ts e:\tau;\rho \leadsto \cb :\comp{\taub}{\rhob}$.
At each step of induction, we assume that the desired property holds for all subderivations and
proceed by case on the possible shape of $e$ to show that
$\Deltab;\Gammab \ts \cb:\comp{\taub}{\rhob}$.
By Lemma~\ref{lemma:synonyms}, it is sufficient to show that (STS)
$\Deltab;\Gammab,k:\cont{\taub},h:\hdlr{\rhob} \ts \eb:\Ans$ where $\cb = \kh{\taub}{\rhob}.\eb$.
Then, proofs are straightforward. We present the case \rname{handle/all}
for example.
\begin{itemize}
\item {\em Case} $e = \eHandle{e_1}{\eCase{x \dra e_2}}$ and
$\eb =   \cb_1~k~(\reify{\rhob'}{\Ans}(\eAbs{x}{\cb_2~k~h}))$. \\
STS: $\Deltab;\Gammab,k:\cont{\taub},h:\hdlr{\rhob} \ts
\cb_1~k~(\reify{\rhob'}{\Ans}(\lambda x:\cpsSum{\rhob'}.\cb_2~k~h))$.
By IH for $e_1$ and lemma~\ref{lemma:synonyms},
STS: $\Deltab;\Gammab,k:\_,h:\_ \ts \reify{\rhob'}{\Ans}(\eAbs{x:\cpsSum{\rhob'}}{\cb_2~k~h}) :
\hdlr{\rhob'}$ (which is true by \rname{T-reify}).
\end{itemize}
\end{proof}

\section{Untyped $\lambda$-Calculus with records (\lambdarec)}
\label{sec:lrec}

\IL{} expressions are translated into expressions of a variant
of an untyped language, called \lambdarec, which is closer
to machine code. 
Its essence is that records are
represented as vectors with slots that are addressed 
numerically. Therefore, the labels in every row are mapped to indices
that form an initial segment of the natural numbers.  Individual
labels are assigned to slots in increasing order, relying on an
arbitrary but fixed total order on the set of labels.   

The \lambdarec language extends the untyped
$\lambda$-calculus with ($n$-ary) tuples and named functions;
\figref{lambda::syntax} shows the abstract syntax for \lambdarec.  The
terms of the language, denoted by $\eeu$, consist of
numbers $n$, variables $x$, the operations plus and minus,
$\eLength{\eeu}$ for determining the number of fields in a tuple $\eeu$,
named functions, function application, and introduction and
eliminations forms for tuples.  The introduction form for tuples,
$\elRecN{i=1}{n}{s_i}$, specifies a sequence of slices from which the
tuple is being constructed.  The elimination form for tuples is
selection (projection), written $\eeu_1.\eeu_2$, that projects out the field
with index $\eeu_2$ from the tuple $\eeu_1$.  The terms include a let
expression (as syntactic sugar for application) and a simple conditional
expression $\eIfzero{\eeu}{\eeu}{\eeu}$.  A {\em slice},
denoted by $s$, is either a term, or a triple of terms
$(\eeu_1,\eeu_2,\eeu_3)$, where $\eeu_1$ yields a record 
while $\eeu_2$ and $\eeu_3$ must evaluate to numbers. 
A slice $(\eeu_1, \eeu_2, \eeu_3)$ specifies consecutive
fields of the record $\eeu_1$ between the indices of $\eeu_2$ (including)
and $\eeu_3$ (excluding).

\figref{lrec-opsem} shows the dynamic semantics for
\lambdarec.  We enforce an order on evaluation by assuming that the
premises are evaluated from left to right and top to bottom (in that
order).  The semantics is largely standard. The only interesting
judgments concern evaluation of slices and construction of tuples.
Slices evaluate to a sequence of values selected by the specified
indices (if any).  Tuple selection projects out the specified field
with the specified index from the tuple.  Since tuples can be
implemented as arrays, selection can be implemented in constant time.
Thus, if records can be transformed into tuples and record selection
can be transformed into tuple selection, record operations can be
implemented in constant time.  The computation of the indices is the
key component of the translation from \IL{} to \lambdarec.

%
%

\begin{landscape}
\begin{figure}
\[
\begin{array}{lrrl}
\mathsf{Terms} &
\eeu    & \bnfdef  & n \bnfalt x \bnfalt \eeu_1 + \eeu_2 \bnfalt 
                  \eeu_1 - \eeu_2 \bnfalt \eLength{\eeu} \bnfalt 
                  \eAbs{x}{\eeu} \bnfalt \eeu_1~\eeu_2 \bnfalt 
                  \elRecN{i=1}{n}{\ssu_i} \bnfalt \eeu.\eeu \bnfalt \\
&       &          & \eLet{x}{\eeu_1}{\eeu_2} \bnfalt 
                   \eLetrec{f}{\eAbs{x}{\eeu_1}}{\eeu_2} 
                   \bnfalt \eIfzero{\eeu_1}{\eeu_2}{\eeu_3}
                   \\[1mm]
\mathsf{Slices} &
\ssu       & \bnfdef  & \eeu \bnfalt (\eeu, \eeu, \eeu)  \\
\mathsf{Values} &
\vvu    & \bnfdef  & n \bnfalt \elRecN{i=1}{n}{\vvu_i} \bnfalt 
                     \eAbs{x}{\eeu} \\
\end{array}
\]
\caption{The syntax for the \lambdarec language.}
\label{fig:lambda::syntax}
\end{figure}

\begin{figure}
\[
\begin{array}{lcl}
\Eu  & \bnfdef    & [] \bnfalt \Eu~\eeu \bnfalt \vvu~\Eu \bnfalt
                    \Eu + \eeu \bnfalt \vvu + \Eu \bnfalt \Eu - t \bnfalt
                    \vvu - \Eu \bnfalt \eLength{E} \bnfalt
                    \eLet{x}{\Eu}{\eeu} \bnfalt 
                    \eIfzero{\Eu}{\eeu}{\eeu} \bnfalt
                    \\
     &            & 
                    \Eu.\eeu \bnfalt \vvu.\Eu \bnfalt
                    \elRec{\ldots,\vvu_{i-1},
                                  \Eu_{s},\ssu_{i+1},\ldots} \\
\Eu_{s} & \bnfdef & [] \bnfalt \Eu \bnfalt (\Eu,\eeu,\eeu) \bnfalt 
                    (\vvu,\Eu,\eeu) \bnfalt (\vvu,\vvu,\Eu) \\ 
\end{array}
\]
\caption{Evaluation contexts for \lambdarec.}
\label{fig:lrec-contexts}

%
%

\[
\begin{array}{rcll}

\Eu[(\eAbs{x}{\eeu})~\vvu] & \mapsto 
& \Eu[\eeu\,[\vvu/x]] 
& \rlabel{app} \\[1mm]

\Eu[n_1 + n_2] & \mapsto 
& \Eu[n] ~~~ \mathsf{where} ~~~ n = n_1 + n_2 
& \rlabel{plus} \\[1mm]

\Eu[n_1 - n_2] & \mapsto 
& \Eu[n] ~~~ \mathsf{where} ~~~ n = n_1 - n_2 
& \rlabel{minus} \\[1mm]

\Eu[\eLength{\elRec{\vvu_1,\ldots,\vvu_n}}] & \mapsto 
& \Eu[n] 
& \rlabel{len} \\[1mm]

\Eu[\eLet{x}{\vvu}{\eeu}] & \mapsto 
& \Eu[\eeu\,[\vvu/x]]
& \rlabel{let} \\[1mm]

\Eu[\eLetrec{f}{\eAbs{x}{\eeu_1}}{\eeu_2}] & \mapsto 
& \Eu[\eeu_2[(\eAbs{x}{\eeu_1[(\eLetrec{f}
                              {\eAbs{x}{\eeu_1}}
                              {f})/f]})/f]]
& \rlabel{rec/fun} \\[1mm]

\Eu[\eIfzero{0}{\eeu_1}{\eeu_2}] & \mapsto 
& \Eu[\eeu_1]
& \rlabel{ifzero/true} \\[1mm]

\Eu[\eIfzero{n}{\eeu_1}{\eeu_2}] & \mapsto 
& \Eu[\eeu_2] ~~~ \mathsf{where} ~~~ n \neq 0
& \rlabel{ifzero/false} \\[1mm]

\Eu[\elRec{\vvu_1,\ldots,\vvu_i,\ldots,\vvu_n}.i] 
& \mapsto 
& \vvu_i
& \rlabel{select} \\[1mm]


\Eu_{s}[\vvu] & \mapsto 
& \vvu
& \rlabel{slice/singleton} \\[1mm]

\Eu_{s}[(\elRec{\vvu_1,\ldots,\vvu_i,\ldots,\vvu_j,\ldots,\vvu_{n}},i,j)] 
& \mapsto 
& \vvu_i,\ldots,\vvu_{j-1}
& \rlabel{slice/sequence}
\end{array}
\]
\caption{Operational semantics for \lambdarec.}
\label{fig:lrec-opsem}
\end{figure}
\end{landscape}


\subsection{From \IL{} to \lambdarec}

\figref{lambda::systemf-lambda} shows the translation from \IL{}
into the \lambdarec language.  The translation
takes place under an {\em index context}, denoted by
$\Sigma$ that maps row variables to sets consisting of label and term
pairs:
\[
\begin{array}{lcl}
\Sigma & \bnfdef & \varnothing \bnfalt 
                   \Sigma,\beta \mapsto \tRecN{i=1}{n}{(l_i,\eeu_i)}
\end{array}
\]

\noindent Then, for a row variable $\beta$, 
$\Sigma(\beta) = \{(l_1, \eeu_1), \ldots, (l_n, \eeu_n)\}$
where $\eeu_i$ is the term that will aid in computing the index for $l_i$ 
in a record.  Additionally, 
we define two auxiliary functions $\projT{\Sigma}{\beta}{l}$ for
the index (term) of $l$ for $\beta$ and $\projL{\Sigma}{\beta}$
for projecting out the labels from a row variable $\beta$.

\[
\begin{array}{lcl}
\projT{\Sigma}{\beta}{l} & = & \eeu \mathsf{~~~if~} (l,\eeu) \in \Sigma(\beta)\\
\projL{\Sigma}{\beta}    & = & \{l~|~(l,\eeu) \in \Sigma(\beta)\} \\
\end{array}
\]

The translation of numbers, variables, functions, applications, and
let expressions are straightforward.  A record is translated into a
tuple of slices, each of which is obtained by translating the label
expressions.  The slices are sorted based on the corresponding labels.
Since sorting can re-arrange the ordering of the fields, the
transformation first evaluates the fields in their original order by
binding them to variables and then constructs the tuple using those
variables.

A record selection is translated by computing the index for the label
being projected based on the type of the record.  To compute indices
for record labels, the translation relies on two functions
$\mathsf{pos}$ and $\mathsf{labels}$. Given a
set of labels $L$ and a label $l$, define the {\em position} of
$l$ in $L$, denoted $\pos{l}{L}$, as the number of labels of
$L$ that are less than $l$ in the total order defined on labels:
\[
\begin{array}{lcl}
\pos{l}{L} & = & |\{l'~|~l' \in L \land l' <_l l\}|
\end{array}
\]
\noindent where $|\{l_1,\ldots,l_n\}| = n$ and 
$<_l$ denotes the ordering relation on labels.  For a
given record type $\tRec{\rhob}$, 
define $\labelsOf{\tRec{\rhob}}$ to be the pair
consisting of the set of labels and the remainder row,
which is either empty or a row variable. More precisely:
\[
\begin{array}{lcl}
\labelsOf{\{l_1:\taub_1, \ldots, l_k:\taub_k, \cdot\}} &=& (\{l_1, \ldots, l_k\}, \cdot)\\
\labelsOf{\{l_1:\taub_1, \ldots, l_k:\taub_k, \beta\}} &=& (\{l_1, \ldots, l_k\}, \beta)\\
\labelsOf{\{l_1:\taub_1, \ldots, l_k:\taub_k, \beta \rowArrow \taub \}} &=& (\{l_1, \ldots, l_k\}, \beta)
\end{array}
\]
Notice that we treat $\beta \rowArrow \tau$ just like plain $\beta$,
taking advantage of the fact that $(\beta \rowArrow \tau) \lacks l$ if
and only if $\beta \lacks l$.

Let $\rhob$ be some row type.
We can compute the {\em index} of a label $l$ in $\rhob$, denoted
$\indexOf{\Sigma}{l}{\labelsOf{\tRec{\rhob}}}$, 
depending on $\labelsOf{\tRec{\rhob}}$, as follows:
\[
\begin{array}{lcl}
\indexOf{\Sigma}{l}{(L,\cdot)} &=& \pos{l}{L} \\
\indexOf{\Sigma}{l}{(L,\beta)} &=& \projT{\Sigma}{\beta}{l} - 
                                   \pos{l}{\projL{\Sigma}{\beta} \setminus L}
\end{array}
\]

\begin{figure}
\centering
\includegraphics[scale=0.45]{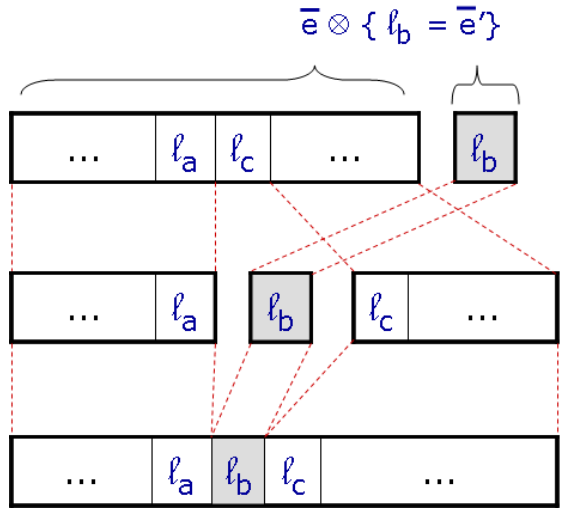}
\caption{Record extension.}
\label{fig:recext}
\end{figure}

For example, the record extension $\eRecExt{\eb_1}{\eRec{l=\eb_2}}$ 
is translated by
first finding the index of $l$ in the tuple corresponding to $e_1$,
then splitting the tuple into two slices at that index, and finally
creating a tuple that consists of the these two slices along with a
slice consisting of the new field as \figref{recext} illustrates.  
Similarly, record subtraction
splits the tuple for the record immediately before and immediately
after the label being subtracted into two slices and creates a tuple
from these slices.  

Type abstractions are translated into functions by
creating an argument $x_i^j$ for each label $l_i^j$ in the kind
$\kappa_i$ of the $\beta_i$.  Note that abstractions of ordinary type
variables ($\alpha_i$'s) are simply dropped.  
{\em Let}-bindings for type abstraction (for the purpose of 
representing polymorphic recursion) are also straightforward.
Type applications are transformed into function applications 
by generating ``evidence'' for
each substituted row-type variable.  As with type abstractions,
substitutions into ordinary type variables are dropped.  Evidence
generation requires computing the indices of each label $l_i^j \in
\kappa_i$ in any record type that extends $\{\rho_i\}$ by adding
fields for every such $l_i^j$. 

The situation is somewhat more complicated in
the case of \textbf{reify}. As we have explained earlier,
\textbf{reify} is special because its dynamic semantics are inherently
type-sensitive and cannot be explained via type erasure.  At runtime
\textbf{reify} needs to know the indices of each label in its row type
argument.  
But since all indices are allocated to an initial segment
of the naturals, it suffices to know the {\em length} of the row.
Therefore, our solution is to pass an additional ``length
index'' argument for every row type variable that is bound by a type
abstraction.

To do so, we represent the length of a row by a ``pseudo-label'' 
\texttt{\$len} in an index context ($\Sigma$):  
\[
\begin{array}{lcl}
\Sigma & \bnfdef & \ldots \bnfalt 
                   \Sigma,\beta \mapsto 
                          \tRec{(l_1,\eeu_1),\ldots,(l_n,\eeu_n),(\$len,\eeu)}
\end{array}
\]
Then, we can define a helper function $\mathsf{lengthOf}$ 
to determine the length of a row:
\[
\begin{array}{lcl}
\lengthOf{\Sigma}{\labelsOf{\taub}} &=& \indexOf{\Sigma}{\$len}{\labelsOf{\taub}} \\
\end{array}
\]
Assuming that \texttt{\$len} is greater than any other label
in the total order on labels, 
we can use $\mathsf{indexOf}$ to compute the length of a row.

\begin{figure}
\begin{mathpar}
\irr{\rlabel{int}}
{ }
{\Deltab;\Gammab;\Sigma \ts n : \Int \fmaps n }
\and
\irr{\rlabel{var}}
{
\Deltab;\Gammab \ts x : \taub
}
{\Deltab;\Gammab;\Sigma \ts x : \taub \fmaps x}
\and
\irr{\rlabel{fun}}
{
\Deltab;\Gammab,x:\taub';\Sigma \ts \eb : \taub \fmaps \eeu \+
}
{
\Deltab;\Gammab;\Sigma \ts \eAbs{x:\taub'}{\eb} : \taub' \ra \taub \fmaps 
\eAbs{x}{\eeu}
}
\and
\irr{\rlabel{app}}
{
\Deltab;\Gammab;\Sigma \ts \eb_1 : \taub_2 \ra \taub \fmaps \eeu_1 \+ 
\Deltab;\Gammab;\Sigma \ts \eb_2 : \taub_2 \fmaps \eeu_2
}
{
\Deltab;\Gammab;\Sigma \ts \eb_1~\eb_2 : \taub \fmaps \eeu_1~\eeu_2
}
\and
\irr{\rlabel{let}}
{
\Deltab;\Gammab;\Sigma \ts \eb_1 : \taub \fmaps \eeu_1 \+
\Deltab;\Gammab,x:\taub;\Sigma \ts \eb_2 : \taub_2 \fmaps \eeu_2
}
{
\Deltab;\Gammab;\Sigma \ts \eLet{x:\taub}{\eb_1}{\eb_2} : \taub_2 \fmaps 
\eLet{x}{\eeu_1}{\eeu_2}
}
\and
\irr{\rlabel{letrec}}
{
\Deltab;\Gammab,f:\taub_2 \ra \taub_1,x:\taub_2;\Sigma \ts \eb_1 : \taub_1 \fmaps \eeu_1 \+
\Deltab;\Gammab,f:\taub_2 \ra \taub_1,x:\taub_2;\Sigma \ts \eb_2 : \taub \fmaps \eeu_2 
}
{\Deltab;\Gammab;\Sigma \ts \eLetrec{f:\taub_2 \ra \taub_1}{\eAbs{x:\taub_2}{\eb_1}}{\eb_2} : \taub \fmaps
\eLetrec{f}{\eAbs{x}{\eeu_1}}{\eeu_2}
}
\and
\irr{\rlabel{ty/letrec}}
{
\Deltab,\alpha:\kappa;\Gammab,f:\forall\alpha:\kappa.\taub_1;\Sigma,\alpha:\{(l_1,x_1), \ldots, (l_n,x_n), (\$len,x)\} \ts \eb_1 : \taub_1 \fmaps \eeu_1 \+
\Deltab,\alpha:\kappa;\Gammab,f:\forall\alpha:\kappa.\taub_1;\Sigma \ts \eb_2 : \taub_1 \fmaps \eeu_2 \+
\kappa = \{ l_1, \ldots, l_n \}
}
{
\begin{array}{lcl}
\Deltab;\Gammab;\Sigma & \ts & \eLetrec{f:\forall\alpha:\kappa.\taub_1}{\eTabs{\alpha:\kappa}{\eb_1}}{\eb_2} : \taub \\
                       & \fmaps & \eLetrec{f}{\lambda x_1 \ldots \lambda x_n.\eeu_1}{\eeu_2}
\end{array}
}
%
\and
\irr{\rlabel{ty/abs}}
{
\Deltab,\alpha:\kappa;\Gammab;\Sigma,\alpha:\{(l_1,x_1), \ldots, (l_n,x_n), (\$len,x)\} 
\ts 
\eb :\taub \fmaps \eeu \+
\kappa = \{ l_1, \ldots, l_n \}
}
{
\Deltab;\Gammab;\Sigma  \ts  \Lambda \alpha:\kappa.\eb : \forall \alpha:\kappa.\taub \fmaps  
\lambda x_1 \ldots \lambda x_n.\lambda x~\eeu
}
\and
\irr{\rlabel{ty/app}}
{
\Deltab;\Gammab;\Sigma \ts \eb : \forall \alpha:\kappa.\taub \fmaps \eeu \+
\Deltab \ts \taub' : \kappa \+
(L,\rhob) = \labelsOf{\taub'} \+
\kappa  =   \{ l_1, \ldots, l_n \} \+
\forall i \in \{1,\ldots,n\} . 
\eeu_i =  \indexOf{\Sigma}{l_i}{(L \cup \kappa, \rhob)} \+ 
\eeu' = \lengthOf{\Sigma}{\labelsOf{\taub'}}
}
{
\Deltab;\Gammab;\Sigma \ts \eb[\taub'] : \taub[\taub'/\alpha]
 \fmaps  \eeu~\eeu_1 \ldots \eeu_n~\eeu'
}
\end{mathpar}
\caption{The translation from \IL{} into \lambdarec 
for basic computations.}
\label{fig:lambda::systemf-lambda}
\end{figure}

\begin{figure}
\begin{mathpar}
\irr{\rlabel{select}}
{
\Deltab;\Gammab;\Sigma \ts \eb : \tRec{l:\taub,\rhob} \fmaps \eeu \+
\eeu' = \indexOf{\Sigma}{l}{\labelsOf{\tRec{l:\taub,\rhob}}}
}
{
\Deltab;\Gammab;\Sigma \ts \eb.l : \taub \fmaps \eeu.\eeu'
}
\and
\irr{\rlabel{r}}
{
\forall i_{\in 1..n},j_{\in 1..n}. i<j \Rightarrow l_{\#(i)} <_l l_{\#(j)} \+
\{l_{\#(1)}, \ldots, l_{\#(n)} \} = \{l_1, \ldots, l_n\} \+
\forall i. \left( \Deltab;\Gammab;\Sigma \ts \eb_i : \taub_i \fmaps \eeu_i \right)
}
{
\begin{array}{lcl}
\Deltab;\Gammab;\Sigma & \ts    & \eRecN{i=1}{n}{l_i = \eb_i} : 
                                  \tRecN{i=1}{n}{l_i:\taub_i} \\
                       & \fmaps & \eLet{x_1}{\eeu_1}{\ldots\eLet{x_n}{\eeu_n}{\elRecN{i=1}{n}{x_{\#(i)}}}} 
\end{array}
}
\and
\irr{\rlabel{r/ext}}
{
\Deltab;\Gammab;\Sigma \ts \eb_1 : \tRec{\rhob} \fmaps \eeu_1 \+ 
\Deltab;\Gammab;\Sigma \ts \eb_2 : \taub_2 \fmaps \eeu_2 \+
\eeu_0 = \indexOf{\Sigma}{l}{\labelsOf{\tRec{\rhob}}} \+
}
{
\Deltab;\Gammab;\Sigma \ts \eRecExt{\eb_1}{\eRec{l=\eb_2}} : \tRec{l:\taub_2,\rhob} \fmaps 
\eLet{x}{\eeu_1}{\elRec{(x, 0, \eeu_0) ,t_2,(x, \eeu_0, \eLength{x})}}
}
\and
\irr{\rlabel{r/sub}}
{
\Deltab;\Gammab;\Sigma \ts \eb : \tRec{l:\taub,\rhob} \fmaps \eeu \+
\eeu_0 = \indexOf{\Sigma}{l}{\labelsOf{\tRec{l:\taub,\rhob}}} \+
}
{
\Deltab;\Gammab;\Sigma  \ts     \eRecSub{\eb}{l} : \tRec{\rhob}
        \fmaps  \eLet{x}{\eeu}{\elRec{(x, 0, \eeu_0), (x, \eeu_0+1, \eLength{x})}}
}
\and
\irr{\rlabel{T-reify}}
{
\Deltab; \Gammab; \Sigma \ts \eb : \cpsSum{\rhob} \ra \taub \fmaps \eeu \+
\eeu' = \lengthOf{\Sigma}{\labelsOf{\tRec{\rhob}}}
}
{
\begin{array}{lcl}
\Deltab;\Gammab;\Sigma & \ts    & \reify{\rhob}{\taub}{\eb} : 
                                  \tRec{\rhob \rowArrow \taub} \\
                       & \fmaps & \mathbf{letrec~}{f}~{=}~
                                  \lambda x_{\eeu}.\lambda x_{\eeu'}.\lambda n.\lambda \vvu.\\
                       &        & \hspace{2cm}\mathbf{ifzero~}(x_{\eeu'}, \\
                       &        & \hspace{3.5cm}\vvu,\\
                       &        & \hspace{3.5cm}f~x_{\eeu}~(x_{\eeu'}-1)~(n+1)~\elRec{v,(\eAbs{x_n}{\eeu~(\eAbs{c}{c.n~x_n})})}) \\ 
                       &        & \mathbf{~~~~~in~}{f~\eeu~\eeu'~1~\elRec{}} \\
\end{array}
}
\end{mathpar}
\caption{The translation from \IL{} into \lambdarec
for computations involving records.}
\label{fig:lambda::systemf-lambda-rec}
\end{figure}

\subsection*{Properties of $\fmaps$}

A desirable property of the translation $\fmaps$ is 
that it preserves the semantics of \IL{}. Let $P_1$ be
a program in \IL{} and $P_2$ a program in \lambdarec{}
obtained by applying $\fmaps$. We wish to show that
if $P_1$ evaluates to $n$, then $P_2$ also evaluates to $n$
assuming that both languages use the same number values.
The approach we will use is similar to Leroy's
proofs by simulation \cite{Leroy2006}.
First, we construct a relation $\eb \sim \eeu$.


\begin{definition}[$\eb \sim \eeu$]
\begin{mathpar}
\irr{~}
{ }
{n \sim n}
\and
\irr{~}
{\Deltab;\Gammab;\Sigma \ts \eb:\taub \fmaps \eeu}
{\eb \sim \eeu}
\and
\irr{~}
{\Deltab;\Gammab,x:\taub;\Sigma \ts \eb:\taub \fmaps \eeu}
{\eAbs{x:\taub}{\eb} \sim \eAbs{x}{\eeu}}
\and
\irr{~}
{\vb_i \sim \vvu_{\#(i)}}
{\eRecN{i=1}{n}{l_i=\vb_i} \sim \elRecN{i=1}{n}{\vvu_{\#(i)}}}
\end{mathpar}


%
\label{def:sim}
\end{definition}

Then, we show that this relation is preserved during evaluation
of $P_1$ and $P_2$. However, the number of evaluation steps
may not equal to each other. In particular, the number of evaluation step of
\lambdarec{} is always larger than that of \IL{}
since the translation may introduce more transitions in \lambdarec{}.
For example, the index passing mechanism adds more computations 
(\rname{ty/abs} and \rname{ty/app}) and
translating from records to slices adds additional \textbf{let}
expressions~(\rname{r}). Therefore, we use $\eeu \mapsto^{+} \eeu'$ 
instead of $\eeu \mapsto \eeu'$.

Before we proceed to establishing the main theorem,
we set up a few helper lemmas:

\begin{lemma}[Substitution]
\label{lem:subst}
\begin{mathpar}
\irr{}
{
\Deltab;\Gammab,x:\taub';\Sigma \ts \eb:\taub \fmaps \eeu \+
\Deltab;\Gammab;\Sigma \ts \vb:\taub' \fmaps \vvu
} 
{
\eb[\vb/x] \sim \eeu[\vvu/x] 
}
\end{mathpar}
\end{lemma}

\begin{proof}
By induction on $\fmaps$.
\end{proof}

\begin{lemma}[Type substitution]
\label{lem:tysubst}
\begin{mathpar}
\irr{}
{
\Deltab,\alpha:\kappa;\Gammab;\Sigma,\alpha:\{(l_1,x_1), \ldots, (l_n,x_n)\} \ts \eb :\taub \fmaps \eeu \+
\Deltab \ts \taub' : \kappa \+
(L,\rhob) = \labelsOf{\taub'} \+
\kappa  =   \{ l_1, \ldots, l_n \} \+\+
\forall i_{\in 1..n}. 
\eeu_i \mapsto \vvu_i \mathsf{~~where~~} \eeu_i =  \indexOf{\Sigma}{l_i}{L \cup \kappa, \rhob}
} 
{ 
\eb[\taub'/\alpha] \sim \eeu[\vvu_1/x_1,\ldots,\vvu_n/x_n]
}
\end{mathpar}
\end{lemma}

\begin{proof}
By induction on $\fmaps$.
\end{proof}

\begin{lemma}
\label{lem:sim}
If $\eb \sim \eeu$ and $\eb \mapsto \eb'$, 
then $\exists~\eeu'$ such that $\eeu \mapsto^{+} \eeu'$ and $\eb' \sim \eeu'$.
\end{lemma}
\begin{proof}
By induction of a deriation of $\eb \sim \eeu$ (i.e.,
$\Deltab;\Gammab;\Sigma \ts \eb \fmaps \eeu$).
At each step of induction, we assume that the desired
property holds for all subderivations and proceed by case on
the possible shape of $\eb$:
\begin{itemize}
\item {\em Case}~\rname{int, var, fun}:  Already values. Not applicable.
\item {\em Case}~\rname{app}: 
  $\Deltab;\Gammab;\Sigma \ts \eb_1~\eb_2 :\taub \fmaps \eeu_1~\eeu_2$. 
  There are three subcases on whether $\eb_1$ and $\eb_2$ are values or not:
  \begin{itemize}
  \itemsubcase{} 
     Neither. Then, by given, $\eb_1~\eb_2 \mapsto \eb_1'~\eb_2$.
     By $\mapsto$ of \IL{}, we know that $\eb_1 \mapsto \eb_1'$ (\cc{1}).
     By inv of \rname{app}, we also know that $\eb_1 \sim \eeu_1$ (\cc{2}).
     By IH with \cc{1} and \cc{2}, there exists $\eeu_1'$ such that
     $\eeu_1 \mapsto \eeu_1'$ and $\eb_1' \sim \eeu_1'$.
     By \rname{app}, therefore, there exists $\eeu_1'~\eeu_2$ such that
     $\eb_1'~\eb_2 \sim \eeu_1'~\eeu_2$ and 
     $\eeu_1~\eeu_2 \mapsto \eeu_1'~\eeu_2$.
  \itemsubcase{} 
     Only $\eb_1$ is a value. Similar.
  \itemsubcase{} 
     Both are values. Then, by given, 
     $(\lambda x:\taub'.\eb_1')~\vb_2 \mapsto \eb_1'~[\vb_2/x]$.
     By inv of \rname{app} and \rname{fun}, we know that
     $\lambda x:\taub'.\eb_1' \sim \lambda x.\eeu_1'$ and
     furthermore, $\Deltab;\Gammab,x:\taub';\Sigma\ts\eb_1':\taub \fmaps \eeu_1'$ 
     (\cc{1}). At the same time, $\vb_2 \sim \eeu_2$.
     There are two cases on whether $\eeu_2$ is a value or not.
     If $\eeu_2$ is not a value, then it should have a form of a \textbf{let}
     expression which eventually becomes a value (i.e., slices)
     in a few steps. Therefore, we can safely assume that $\eeu_2$ is 
     a value ($\vvu_2$). Then 
     $(\lambda x.\eeu_1')~\vvu_2 \mapsto \eeu_1'~[\vvu_2/x]$
     and also by \lemref{subst} with \cc{1} and $\vvu_2$,
     $\eb_1'~[\vb_2/x] \sim \eeu_1'~[\vvu_2/x]$.
  \end{itemize}
\item {\em Case}~\rname{let}:
  $\Deltab;\Gammab;\Sigma \ts \eLet{x:\taub}{\eb_1}{\eb_2} : \taub_2 \fmaps 
  \eLet{x}{\eeu_1}{\eeu_2}$.
  There are two subcases on whether $\eb_1$ and $\eb_2$ are values or not.
  Then, similar to the case \rname{app}.
\item {\em Case}~\rname{letrec}:
  $\Deltab;\Gammab;\Sigma \ts \eLetrec{f:\taub_2 \ra \taub_1}{\eAbs{x:\taub_2}{\eb_1}}{\eb_2} : \taub \fmaps
  \eLetrec{f}{\eAbs{x}{\eeu_1}}{\eeu_2}$.
  By inv of \rname{letrec}, we have $\eb_1 \sim \eeu_1$ and
  $\eb_2 \sim \eeu_2$ under $\Gammab,f:\taub_2\ts\taub,x:\taub_2$.
  Then, by \lemref{subst} we can easily show that 
  $\eb_2~[\vb/f] \sim \eeu_2~[\vvu/f]$ 
  where $\vb = \lambda x.(\eb_1~[\eLetrec{f}{\lambda{x}.{\eb_1}}{f}])$ 
    and $\vvu = \lambda x.(\eeu_1~[\eLetrec{f}{\lambda{x}.{\eeu_1}}{f}])$ 
    and $\vb \sim \vvu$.
  
\item {\em Case}~\rname{ty/letrec}: Similar to the case \rname{letrec}.
\item {\em Case}~\rname{ty/abs}: Not applicable.
\item {\em Case}~\rname{ty/app}: 
  $\Deltab;\Gammab;\Sigma \ts \eb[\taub'] : \taub[\taub'/\alpha]
   \fmaps  \eeu~\eeu_1 \ldots \eeu_n$.
  There are two subcases on whether $\eb$ is a value or not:
  \begin{itemize}
  \itemsubcase{} $\eb$ is not a value. Then, by given, we have
     $\eb~[\taub'] \mapsto \eb'~[\taub']$ which implies
     $\eb \mapsto \eb'$ (\cc{1}). Then, by IH with \cc{1} and
     $\eb \sim \eeu$, there exists $\eeu'$ which satisfies
     $\eeu \mapsto \eeu'$ and $\eb' \sim \eeu'$.
     Therefore, by $\mapsto$ of \lambdarec{},
     $\eeu~\eeu_1 \ldots \eeu_n \mapsto \eeu'~\eeu_1 \ldots \eeu_n$
     and $\eb'~[\taub'] \sim \eeu'~\eeu_1 \ldots \eeu_n$.
  \itemsubcase{} $\eb$ is a value. Then, by \lemref{canonical}
     (the canonical lemma),
     it is $\Lambda \alpha:\kappa.\eb'$.
     Then, by \rname{ty/abs}, 
     $\Lambda \alpha:\kappa.\eb' \sim \lambda x_1 \ldots \lambda x_n.~\eeu'$.
     By inv of \rname{ty/abs} and \lemref{tysubst}, we can see that
     $\eb'[\taub'/\alpha] \sim \eeu'[\vvu_1/x_1,\ldots,\vvu_n/x_n]$.
  \end{itemize}

\item {\em Case}~\rname{select}: 
  $\Deltab;\Gammab;\Sigma \ts \eb.l : \taub \fmaps \eeu.\eeu_l$.
  There are two subcases on whether $\eb$ is a value or not:
  \begin{itemize}
  \itemsubcase{} $\eb$ is not a value. By given, we have
     $\eb~ \mapsto \eb'$. We can easily get 
     $\eb'.l \sim \eeu'.\eeu_l$.
  \itemsubcase{} $\eb$ is a value. Then, by \lemref{canonical}
     and \rname{select},
     $\eRec{\ldots,l_l=\vvu_l,\ldots}.l \sim \eeu.\eeu_l$
     where $\eeu = \eLet{x_1}{\vvu_1}{\ldots\eLet{x_n}{\vvu_n}{\elRecN{i=1}{n}{x_{\#(i)}}}}$
     and   $\eeu_l = \indexOf{\Sigma}{l}{\labelsOf{\tRec{l:\taub,\rhob}}}$.
     By \rname{select}, $\eRec{\ldots,l_l=\vb_l,\ldots}.l \mapsto \vb_l$.
     Similarly, $\eeu.\eeu_l \mapsto^{+} \elRecN{i=1}{n}{\vvu_{\#(i)}}.j \mapsto \vvu_{\#(l)}$.
     We can easily show the exsitence of $\vvu_{\#(l)}$ such that
     $\vb_l \sim \vvu_{\#(l)}$ and $\eeu.\eeu_l \mapsto^{+} \vvu_{\#(l)}$.
  \end{itemize}
  
\item {\em Case}~\rname{r}: 
  $\Deltab;\Gammab;\Sigma \ts \eRecN{i=1}{n}{l_i = \eb_i} : 
                              \tRecN{i=1}{n}{l_i:\taub_i} 
   \fmaps \eLet{x_1}{\eeu_1}{\ldots\eLet{x_n}{\eeu_n}{}}{}$
  $\elRecN{i=1}{n}{x_{\#(i)}}$. By inv of \rname{R},
  $\eb_i \sim \eeu_i$ for $1 \leq i \leq n$. By given,
  $\eb_i \mapsto \eb'_i$ and by IH, there exists $\eeu'_i$
  which makes the remains straightforward.
  
\item {\em Case}~\rname{r/ext}:
  $\Deltab;\Gammab;\Sigma \ts \eRecExt{\eb_1}{\eRec{l=\eb_2}} : \tRec{l:\taub_2,\rhob} \fmaps 
  \eLet{x}{\eeu_1}{\elRec{(x, 0, \eeu_0) ,\eeu_2,(x, \eeu_0, n)}}$.
  There are two subcases. If either $\eb_1$ or $\eb_2$ is not a value, 
  then a proof is straightforward. If both are values,
  we assume that $\eRecExt{\eRec{l_1=\vb_1,\ldots,l_n=\vb_n}}{\eRec{l=\vb}} \mapsto 
  \eRec{l_1=\vb_1,\ldots,l_n=\vb_n,l=\vb}$.
  Similarly, $\eLet{x}{\eeu_1}{\elRec{(x, 0, \eeu_0) ,\eeu_2,(x, \eeu_0, n)}}
  \mapsto^{+} \elRecN{i=1}{n+1}{\vvu_{\#(i)}} $ where $\#(i)$ denotes
  slice sorting. Then, by \defref{sim} and by IH, 
  $\eRec{l_1=\vb_1,\ldots,l_n=\vb_n,l=\vb} \sim
  \elRecN{i=1}{n+1}{\vvu_{\#(i)}}$.

\item {\em Case}~\rname{r/sub}: Similar to the case \rname{r/ext}.

\item {\em Case}~\rname{t-reify}:
   $\Deltab; \Gammab; \Sigma \ts \reify{\rhob}{\taub}{\eb} : \tRec{\rhob \rowArrow \taub}
   \fmaps \eLetrec{f}{\ldots}{f~\eeu~\eeu'~1~\eRec{}}$.
   If $\eb$ is not a value, a proof is straightforward. If it is a value,
   by \rname{reify}, 
   $\reify{\ldots,l_n:\taub_n,\tRowEmp}{\taub}{\vb} \mapsto 
    \eRecN{i=1}{n}{ l_i = \eAbs{x_i:\taub_i}
                  {\vb~(\eTabs{\alpha:\star}
                             {\eAbs{c:\tRecN{j=1}{n}{l_j:\taub_j \ra  \alpha}}{c.l_i~x_i}})} }$.
   By $\mapsto$ of \lambdarec, 
   $\mathbf{letrec~}f~=~\lambda x_{\eeu}.\lambda x_{\eeu'}.\lambda n.\lambda \vvu.\mathbf{ifzero~(}x_{\eeu'},~\vvu,~f~x_{\eeu}~(x_{\eeu'}-1)~(n+1)\\ 
\elRec{\vvu,(\eAbs{x_n}{x_{\eeu}~(\eAbs{c}{c.n~x_n})}
)})\mathbf{~in~}
{f~\vvu~n~1~\elRec{}} \mapsto^n
    \elRecN{i=1}{n}{\eAbs{x_i}{\vvu~(\eAbs{c}{c.i~x_i})}}$.
   By \\ the fact of $\vb \sim \vvu$,
   $\eRecN{i=1}{n}{ l_i = \eAbs{x_i:\taub_i}
                  {\vb~(\eTabs{\alpha:\star}
                             {\eAbs{c:\tRecN{j=1}{n}{l_j:\taub_j \ra  \alpha}}{c.l_i~x_i}})}} \sim \elRecN{i=1}{n}{\eAbs{x_i}{\vvu~(\eAbs{c}{c.i~x_i})}}$.

\end{itemize}
\end{proof}





\vspace{-0.3cm}
\begin{theorem}
Let $P_1$ be an \IL{} program of type \Int and $P_2$ a \lambdarec{} program 
obtained by applying $\fmaps$. Then, whenever $P_1$ evaluates to $n$,
$P_2$ evaluates to $n$.
\end{theorem}

\begin{proof}
$\varnothing;\varnothing;\varnothing\ts\eb:\Int \fmaps \eeu$ and
$\eb \mapsto^{*} n$ immediately imply that $\eeu \mapsto^{*} n$
by \defref{sim} and \lemref{sim}.
\end{proof}

\section{Implementation}
\label{sec:implementation}
We have implemented a prototype compiler for the \mlpolyr{} language
in Standard ML. It retains all of the features that we have
disscussed, including row polymorphism for records and sums,
polymorphic sums, extensible first-class cases as well as type-safe
exception handlers.  The compiler produces machine code for the
PowerPC architecture that can run on Apple Macintosh computers. It
also supports x86 backend based on C-{}-\cite{Jones_cmm}. 

\subsection{Compiler Phases}

The compiler is structured in a fairly traditional way and consists of
the following phases:

\begin{itemize}
\item{\bf lexer} lexical analysis, tokenization
\item{\bf parser} LALR(1) parser, generating abstract syntax trees (AST)
\item{\bf elaborator} perform type reconstruction and generation of
  annotated abstract syntax (Absyn)
\item{\bf translate} generate index-passing \lambdarec code
\item{\bf anf-convert} convert \lambdarec code into A-normal
  form~\cite{felleisen93:essence}
\item{\bf anf-optimize} perform various optimization including
  flattening, uncurrying, constant folding, simple constant- and value
  propagation, elimination of useless bindings, short-circuit
  selection from known tuples, inline tiny functions, some arithmetic
  expression simplification
\item{\bf closure} convert to first-order code by closure conversion
\item{\bf clusters} separate closure-converted blocks into clusters of
  blocks; each cluster roughly corresponds to a single C function but
  may have multiple entry points
\item{\bf value-numbering} perform simple common subexpression (CSE) within
  basic blocks
\item{\bf treeify} re-grow larger expression trees to make tree-tiling
  instruction selection more useful
\item{\bf traceschedule} arrange basic blocks to minimize unconditional jumps
\item{\bf cg} perform instruction selection by tree-tiling (maximum-munch algorithm), graph-coloring register allocation; emit assembly code
\end{itemize}

Each phase is implemented in a separate module and a main driver calls
them sequentially as illustrated in \figref{compiler}.

\begin{figure}
\begin{lstlisting}[style=mlpolyr]
(* val main : string * string list -> OS.Process.status *)
fun main (self, args) =
    let val file = Command.parse args
        val ast = Parse.parse file
        val absyn = Elaborate.elaborate ast
        val lambda = Translate.translate absyn
        val anf = LambdaToANF.convert lambda
        val anf_op = Optimize.optimize anf
        val closed = Closure.convert anf
        val {entrylabel, clusters} = Clusters.clusterify closed
        val clusters_cse = ValueNumbering.cse clusters
        val bbt_clusters = Treeify.treeify clusters_cse
        val traces = TraceSchedule.schedule bbt_clusters
        val _ = CodeGen.codegen (traces, entrylabel, file)
    in OS.Process.success
    end
\end{lstlisting}
\caption{A main driver for the \mlpolyr{} compiler.}
\label{fig:compiler}
\end{figure}

\subsection{Runtime system}
The runtime system, written in C, implements a simple two-space
copying garbage collector~\cite{Pierce02} and provides basic
facilities for input and output.

For the tracing garbage collector to be able to reliably distinguish
between pointers and integers, we employ the usual tagging trick.
Integers are 31-bit 2's-complement numbers.  An integer value $i$ is
represented internally as a 2's-complement 32-bit quantity of value
$2i$. This makes all integers even, with their least significant bits
cleared.  Heap pointers, on the other hand, are represented as odd
32-bit values.  In effect, instead of pointing to the beginning of a
word-aligned heap object, they point to the object's second byte.
Generated load- and store-instructions account for this skew by using
an accordingly adjusted displacement value.  With this representation
trick, the most common arithmetic operations (addition and
subtraction) can be implemented as single instructions as usual; they
do not need to manipulate tag bits.  The same is true for most loads
and stores.

%
%

\begin{figure}
\begin{lstlisting}[style=mlpolyr]
val String : { cmdline_args : string list,
               cmdline_pgm  : string,
               compare      : string * string -> int,
               concat       : string list -> string,
               fromInt      : int -> string,
               inputLine    : () -> string,
               output       : string -> (),
               size         : string -> int,
               sub          : string * int -> int,
               substring    : string * int * int -> string,
               toInt        : string -> int }
\end{lstlisting}
\caption{\mlpolyr{} supports minimal built-in functions which perform
  simple I/O tasks and string manipulations.}
\label{fig:string}
\end{figure}

\mlpolyr{} also supports minimal built-in functions as a record value
bound to the global variable {\tt String} as shown in
\figref{string}. This record is allocated using C code and does not
reside within the \mlpolyr{} heap. It contains routines for
manipulating string values, for converting from and to strings, and
for performing simple I/O operations.  Each routine can be accessed by
dot notation. For example, $\mathsf{String.compare}$ could be used to
compare two string values.  Their implementations are hidden inside
the \mlpolyr{} runtime system.

\chapter{Large-scale extensible programming}
\label{large-prog}

Today most programming languages support programming at the large scale
by breaking programs into pieces and developing these pieces 
separately. For example, the Standard ML module language provides
mechanisms for structuring programs into separate units called 
{\em structures}. Each structure has its own namespace and
they are hierarchically composable so that one structure can contain
other structures. The Standard ML module system also supports
module-level parameterization which makes code reuse easy.

In this section, we propose the module system for \mlpolyr{} in order
to provide an ML-like module system which provides separate
compilation and independent extension in presence of polymorphic
records, first-class cases and type safe exception handlers.  After
presenting the module language, we will discuss a way to implement it
by translating module language terms into ordinary \mlpolyr{} core
language terms and we will also disuss how to support separate
compilation. Then, we will revisit the elaborated expression problem
by Zenger and Odersky~\cite{Zenger-Odersky05} with our module-level
solution.

\section{The module system}
The syntax of our proposed module language is presented in
\figref{module-syntax}.  We use $X$ and $T$ as meta-variables for
module names and template names, respectively.  The core language
($e$) is extended to support the dot notation ($X.x$) for accessing a
component (named $x$) in a module ($X$).  A module itself consists of
a sequence of value components ($\{\{C_1 \ldots C_n\}\}$). A value
component is defined as a value declaration ($\eVal{x}{e_m}$). A
component in the module also can be added ($M \boxplus \{\{C\}\}$) or
removed ($M \boxminus x$). A module can also be optained by applying a
template to modules ($T~ (M_1,\ldots,M_n)$).  A program is a sequence
of declarations which can be either definitions of modules or those of
templates.  A template can take other modules as arguments.

%
%

\begin{figure}
\[
\begin{array}{lrrl}
\mathsf{Terms}       & e_m & \bnfdef & e \bnfalt X.x \\
\mathsf{Modules}     & M & \bnfdef & \{\{C_1\ldots C_n\}\} \bnfalt 
                                     M \boxplus \{\{C\}\} \bnfalt 
                                     M \boxminus x \bnfalt
                                     T~(M_1,\ldots,M_n)\\
\mathsf{Components}  & C & \bnfdef & \eVal{x}{e_m} \\
\mathsf{Declarations} & D & \bnfdef & \eMod{X}{M} \bnfalt 
                                      \eTemp{T}{X_1,\ldots,X_n}{M} \\
\mathsf{Program}     & P & \bnfdef & D_1 \ldots D_n 
\end{array}
\]
\caption{The syntax for the module language.}
\label{fig:module-syntax}
\end{figure}


We treat modules as packages that contain only value components, so
module language does not have type components unlike the SML module
language.  For example, we can define a module $\mathsf{Queue}$ which
contains basic operations such as $\mathsf{insert}$ and
$\mathsf{delete}$:

\begin{lstlisting}[style=mlpolyr]
module Queue = {{
   val empty = []
   fun insert (q, x) = List.rev (x::(List.rev q))
   fun delete q = case q of 
	       [] => raise `Empty ()
	     | h::tl => (tl, h)
}}
\end{lstlisting}

\noindent Each component in the module can be accessed 
by the usual dot notation: e.g., 
$\mathsf{Queue.empty}$ or $\mathsf{Queue.insert (q, 5)}$.
Then, we can add more operations by extending the basic $\mathsf{Queue}$
into $\mathsf{EQueue}$:

\begin{lstlisting}[style=mlpolyr]
module EQueue = Queue with {{
   fun size q = List.length q
   fun insertLog (q, x) = (log "insert"; Queue.insert (q, x))
}}
\end{lstlisting}
where the clause $\mathsf{with}$ is a syntactic sugar
for $M \boxplus \{\{D\}\}$.

We may consider a priority queue which retrieves the element with
the highest priority. In our implementation, we only have to
modify the function $\mathsf{insert}$ in a way that 
a sorted list is built on an entry time:
\begin{lstlisting}[style=mlpolyr]
module IntPriorityQueue = Queue where {{
   fun insert (q, x) = case q of
			   [] => [x]
			 | h::tl => if (x>h) then x::q
				    else h::(insert (tl, x))
}}
\end{lstlisting}
where the clause $\mathsf{where}$ is a syntactic sugar
for $(M \boxminus l) \boxplus \{\{D\}\}$, similar to a record update operator.
However, this priority queue works only over integers. 
Alternatively, we may keep queues in an alphabetic order,
and then the code should be changed as follows:
\begin{lstlisting}[style=mlpolyr]
module StrPriorityQueue = Queue where {{
   fun insert (q, x) = case q of
			   [] => [x]
			 | h::tl => if (String.compare (x,h) > 0) 
                                    then x::q
				    else h::(insert (tl, x))
}}
\end{lstlisting}
We can make code more reusable by generalizing this code so that it
can work over any types. Similar to functors in the Standard ML module
system, we provide a parameterized mechanism called a {\em template}
which takes other modules as arguments.  For example, we can
parameterize a comparison function, so that a priority queue can work
over any type depending on its argument:
\begin{lstlisting}[style=mlpolyr]
template PriorityQueue (Order) = Queue where {{
   fun insert (q, x) = case q of
			   [] => [x]
			 | h::tl => if (Order.lt(x,h)) then x::q
				    else h::(insert (tl, x))
}}
\end{lstlisting}
Unlike functors, we do not pose any type constraints except that
the module $\mathsf{Order}$ should have a component named $\mathsf{lt}$.
By applying this template to any modules 
that have the component $\mathsf{lt}$, 
a new priority queue can be instantiated:

\begin{lstlisting}[style=mlpolyr]
module IntPrioiryQueue = PriotityQueue (IntOrder)
module StrPrioiryQueue = PriotityQueue (StrOrder)
\end{lstlisting}
where $\mathsf{IntOrder}$ and $\mathsf{StrOrder}$ can be implemented
as follows:
\begin{lstlisting}[style=mlpolyr]
module IntOrder = {{
       fun lt (x, y) = x > y
}}

module StrOrder = {{
       fun lt (x, y) = String.compare (x, y) > 0
}}
\end{lstlisting}

\section{An implementation of the module language}

Our main idea of implementing the module language is to translate the
module language constructs into ordinary \mlpolyr{} core language
ones. 
In particular, we can take advantage of the fact that each operator on
module expressions has a corresponding record operator as illustrated
in Table~\ref{table:rec-mod}.

\begin{table}
\[
\begin{array}{|@{~~}l@{~~}||@{~~}l@{~~}|@{~~}l@{~~}|}
\hline
                      & \mathsf{record}~~e
                      & \mathsf{module}~~M \\
\hline
\mathsf{Introduction} & \eRec{l_1=e_1,\ldots,l_n=e_n} 
                      & \{\{\eVal{l_1}{e_1},\ldots,\eVal{l_n}{e_n}\}\} \\
\hline
\mathsf{Selection}    & r.l                     & M.l \\
\hline
\mathsf{Extension}    & \eRecExt{r}{\eRec{l=e}} & M ~\boxplus~ {\{\{\eVal{l}{e}\}\}} \\
\hline
\mathsf{Substraction} & \eRecSub{r}{l}          & M ~\boxminus~ {l} \\
\hline
\end{array}
\]
\caption{Symmetry between record and module operations.}
\label{table:rec-mod}
\end{table}

For example, the module $\mathsf{Queue}$ can be translated 
into a form of records:
\begin{lstlisting}[style=mlpolyr]
   val Queue =             
       let val empty = []
           fun insert (q, x) = ...
           fun delete () = ...
       in  { empty = empty,
	     insert = insert,
             delete = delete
           }
       end
\end{lstlisting}
where all components are exposed as record fields. 
In case of the module $\mathsf{EQueue}$,
we need polymorphic and extensible records
which \EL{} provides:
\begin{lstlisting}[style=mlpolyr]
   val EQueue =             
       let fun size q = ...
           fun insertLog (q, x) = ...
       in  { size = size,
	     insertLog = insertLog,
             ... = Queue
           }
       end
\end{lstlisting}

Similarly, we can translate the module $\mathsf{IntPriorityQueue}$
into the record with replacement of a field $\mathsf{insert}$:
\begin{lstlisting}[style=mlpolyr]
   val IntPriorityQueue =             
       let fun insert' (q, x) = ...
           val {insert, ... = rest} = Queue
       in  {insert = insert',
	    ... = rest
           }
       end
\end{lstlisting}

A template becomes a function taking arguments and producing
a module (i.e., a record). For example, the template
$\mathsf{PriorityQueue}$ is translated as follows:

\begin{lstlisting}[style=mlpolyr]
   val PriorityQueue = fn Order =>
       let fun insert' (q, x) = ...if (Order.lt (x, h)) then ...
           val {insert, ... = rest} = Queue
       in  {insert = insert',
	    ... = rest
           }
       end
\end{lstlisting}


In sum, \figref{trans-module} shows the translation
rules from module expressions ($M$) into \EL{} expressions ($e$). 

%
%

\begin{figure}
\framebox{$M \leadsto e$}\\
\begin{mathpar}
\irr{(Module)}
{
l_{x_1}\ldots l_{x_n} \mathsf{~fresh~labels} \+
\forall i_{\in 1..n}. e_{m_i} \leadsto e'_i
}
{
\begin{array}{lcl}
  \{\{\eVal{x_1}{e_{m_1}},\ldots,\eVal{x_n}{e_{m_n}}\}\} 
& \leadsto 
& \mathsf{let} ~~ \eVal{x_1}{e_1'} \\
&         & ~~~~~ \ldots \\
&         & ~~~~~ \eVal{x_n}{e_n'} \\
&         & \mathsf{in~~} \eRec{l_{x_1} = x_1,\ldots,l_{x_n}=x_n}\\
&         & \mathsf{end} \\[4mm]
\end{array}
}
\and
\irr{(Extension)}
{
M \leadsto e \+ e_m \leadsto e' \+ l_x \mathsf{~fresh~label}
}
{
\begin{array}{lcl}
  M~\boxplus~\{\{\eVal{x}{e_m}\}\} 
& \leadsto 
& \mathsf{let} ~~ \eVal{x}{e'} \\
& & \mathsf{in}  ~~ e\otimes\eRec{l_x = x}\\
& & \mathsf{end} \\
\end{array} 
}
\and
\irr{(Subtraction)}
{
M \leadsto e \+ l_x \mathsf{~fresh~label~for~} M.x
}
{
\begin{array}{lcl}
M~\boxminus~x & \leadsto & e\oslash l_x \\[4mm]
\end{array}
}
\and
\irr{(Application)}
{
\forall i_{\in 1..n}.M_i \leadsto e_i
}
{
\begin{array}{lcl}
T~(M_1,\ldots,M_n) & \leadsto & T~(e_1,\ldots,e_n)
\end{array}
}
\end{mathpar}

\framebox{$e_m \leadsto e$}\\
\begin{mathpar}
\irr{(Path)}
{
l_x \mathsf{~fresh~label}
}
{
X.x \leadsto X.l_x 
}
\and
\irr{(Non/path)}
{
}
{
e \leadsto e
}
\end{mathpar}

\framebox{$D \leadsto e$}\\
\begin{mathpar}
\irr{(Module-declaration)}
{
M \leadsto e
}
{
\begin{array}{lcl}
\eMod{X}{M} & \leadsto & \eVal{X}{e}
\end{array}
}
\and
\irr{(Template~declaration)}
{
M \leadsto e
}
{
\begin{array}{cl}
         & \eTemp{X}{X_1,\ldots,X_n}{M} \\
\leadsto & \eVal{X}{\mathbf{fn}~(X_1,\ldots,X_n)~=>~e}
\end{array}
}
\end{mathpar}
\caption{The translation from the module language into the core
language.}
\label{fig:trans-module}
\end{figure}

\section{Separate compilation}
\label{sec:separate}
Separate compilation has been considered as one of key factors
for the development of extensible software~\cite{Zenger-Odersky05}.
Without the support of separate compilation,
any extensions to the base system may require re-typechecking or
re-compilation of the existing ones.

Suppose we have the following program fragment:
\begin{lstlisting}[style=mlpolyr]
module EQueue = Queue with {{
   fun size q = List.length q
   fun insertLog (q, x) = (log "insert"; Queue.insert (q, x))
}}
\end{lstlisting}
It would be surprising if we had to compile the module
$\mathsf{Queue}$ whenever we compile the module $\mathsf{EQueue}$, but
many extensibility mechanisms require such redos.  For instance, in
AspectJ, aspects can clearly modularize all extensions in separate
aspect code~\cite{aspectj}.  However, their composition does not
provide separate compilation, so it is necessary for base code to be
either re-typechecked or re-compiled (or both) for every
composition. If we can compile $\mathsf{EQueue}$ without compiling the
module $\mathsf{Queue}$, we would say that they can be compiled
separately.

Generally, separate compilation can be implemented 
in two ways~\cite{elsman:08:tr}.
Suppose we want to compile a program fragment $P$ 
which depends on a module $M$:
\begin{itemize}
\item {\em Incremental compilation} does not require explicit type 
information on $M$, but requires $M$ to be compiled prior to $P$.
\item {\em (True) separate compilation} requires explicit type information 
on $M$, but does not require the prior compilation of $M$.
\end{itemize}

Because all types are fully inferred, the core language does not
require type annotations. Taking the incremental compilation approach,
we may omit type annotation even for modules.  Some may argue that it
would be desirable to explicitly write the intended type, especially
for the sake of consistency and documentation purposes.  However, it
does not seem practical for a user to spell out all types in
\mlpolyr{} where a type may contain row types and kind information.
For example, suppose higher-order functions such as $\mathsf{map}$:
\begin{lstlisting}[style=mlpolyr]
  fun map f [] = []
    | map f (x::xs) = f x :: map f xs
\end{lstlisting}
Here, $\mathsf{map}$ does not raise exceptions but its arguments
might.  With this in mind, $\mathsf{map}$'s type should be as follows
(using Haskell-style notation for lists types [$\tau$]):

\parbox{5in}{\begin{tabbing}
{\bf val} map : $\forall \alpha : \star.  \forall \beta : \star .
                 \forall \gamma : \varnothing . \forall \delta : \varnothing .
                 (\alpha \era{\gamma} \beta) \era{\delta}
                 ([\alpha] \era{\gamma} [\beta])$
\end{tabbing}}

In order to avoid the need for this prohibitively excessive programmer
annotations,
the approach we use is to allow the type checker to infer module
signatures and to record them, so that we can use this information
later when we typecheck or compile a program which depends on this
type information. Therefore, our compiler now produces intermediate
information including typing (e.g., $\mathsf{foo.t}$) and machine code
(e.g., $\mathsf{foo.l}$ written in \lambdarec{}) as the following
sequences:
\[
\stackrel{\mathit{foo.mlpr~~~}=~~~}
         {\stackrel{~}
                   {\stackrel{\mathit{~}}
                             {\mathit{~}}}}
\stackrel{\mathsf{EL~~~}}
         {\stackrel{~}
                   {\stackrel{\mathit{~}}
                             {\mathit{~}}}}
\stackrel{\stackrel{\mathit{foo.t}}{\Uparrow}}
         {\stackrel{\vector(1,0){80}}
                   {\stackrel{\mathit{~}}
                             {\mathsf{Type~checking}}}}
\stackrel{\mathsf{~~~IL~~~}}
         {\stackrel{~}
                   {\stackrel{\mathit{~}}
                             {\mathit{~}}}}
\stackrel{\stackrel{\mathit{foo.l}}{\Uparrow}}
         {\stackrel{\vector(1,0){60}}
                   {\stackrel{\mathit{~}}
                             {\mathsf{Compilation}}}}
\stackrel{\mathsf{~~~LRec~~~}}
         {\stackrel{~}
                   {\stackrel{\mathit{~}}
                             {\mathit{~}}}}
\stackrel{\stackrel{\mathit{~}}{~}}
         {\stackrel{\vector(1,0){60}}
                   {\stackrel{\mathit{~}}
                             {\mathsf{Evaluation}}}}
\stackrel{\mathsf{~~~Value~~~}}
         {\stackrel{~}
                   {\stackrel{\mathit{~}}
                             {\mathit{~}}}}
\]
Then, this information will be used during type checking and evaluating
$\mathsf{bar.mlpr}$ which depends on the module defined in 
$\mathsf{foo.mlpr}$: 
\[
\stackrel{\mathit{bar.mlpr~~~}=~~~}
         {\stackrel{~}
                   {\stackrel{\mathit{~}}
                             {\mathit{~}}}}
\stackrel{\mathsf{EL~~~}}
         {\stackrel{~}
                   {\stackrel{\mathit{~}}
                             {\mathit{~}}}}
\stackrel{\stackrel{\mathit{bar.t}}{\Uparrow}~~~
          \stackrel{\mathit{foo.t}}{\Downarrow}}
         {\stackrel{\vector(1,0){80}}
                   {\stackrel{\mathit{~}}
                             {\mathsf{Type~checking}}}}
\stackrel{\mathsf{~~~IL~~~}}
         {\stackrel{~}
                   {\stackrel{\mathit{~}}
                             {\mathit{~}}}}
\stackrel{\stackrel{\mathit{bar.l}}{\Uparrow}}
         {\stackrel{\vector(1,0){60}}
                   {\stackrel{\mathit{~}}
                             {\mathsf{Compilation}}}}
\stackrel{\mathsf{~~~LRec~~~}}
         {\stackrel{~}
                   {\stackrel{\mathit{~}}
                             {\mathit{~}}}}
\stackrel{\stackrel{\mathit{foo.l}}{\Downarrow}}
         {\stackrel{\vector(1,0){60}}
                   {\stackrel{\mathit{~}}
                             {\mathsf{Evaluation}}}}
\stackrel{\mathsf{~~~Value~~~}}
         {\stackrel{~}
                   {\stackrel{\mathit{~}}
                             {\mathit{~}}}}
\]

This setup is virtually straightforward, with a few notable exceptions:

\begin{itemize}
\item Even though our module language does not have type components,
  our type inference creates unification variables and some of them
  may escape without generalization.  Here, the subtlety lies in
  whether the type checker allows them to escape to the module
  level. Dreyer and Blume explore this subtlety and note that many
  different policies exist regarding how to handle non-generalized
  unification variables~\cite{Dreyer06principaltype}.  According to
  their work, the SML/NJ compiler disallows unification variables to
  escape. Even though it has the benefit of being consistent and
  predictable, it can be too restrictive in some cases. Suppose we
  have the following code in SML:

\begin{lstlisting}[style=mlpolyr]
structure A = 
struct
   val id0 = fn x => x
   val id  = id0 id0
end

val _ = A.id ``hello''
\end{lstlisting}

While the SML/NJ compiler rejects this code but the MLton compiler
accepts it in a more liberal way but it still requires access to the
whole program. Since we do not have type components, we can take such
a liberal way relatively easily. We allow non-generalized unification
variables to escape up to the module level in a similar way to MLton,
but we can also manage to support separate compilation. Let us see
such examples:
\begin{lstlisting}[style=mlpolyr]
module ID0 = {{
   val id0 = fn x => x
   val id  = id0 id0
}}
\end{lstlisting}
where $\mathsf{id0}$ has a polymorphic type of
$\forall\alpha.\alpha\ra\alpha$ but $\mathsf{id}$ has a monomorphic
type of $\beta\ra\beta$. Note that $\beta$ is not a polymorphic
variable because $\mathsf{id0}~\mathsf{id0}$ is not a syntactic value and
the value restriction forces it to be monomorphic~\cite{Pierce02}.
Therefore, the following code will not pass the type checker
since monomorphic type variable $\beta$ can not be instantiated
into both $\Int$ and $\mathsf{string}$ at the same time:
\begin{lstlisting}[style=mlpolyr]
val _ = (ID0.id 5, ID0.id ``hello'') (* ill-typed *)
\end{lstlisting}

However, the situation can change when separate compilation is
considered. Suppose we have modules $\mathsf{A}$, $\mathsf{B}$ and
$\mathsf{C}$ as follows:

\begin{lstlisting}[style=mlpolyr]
module A = {{
   val a = ID0.id 5
}}

module B = {{ 
   val b = ID0.id ``hello''
}}

module C = {{
   val _ = (A.a, B.b) (* ill-typed *)
}}
\end{lstlisting}
Even MLton would reject $\mathsf{A}$ and $\mathsf{B}$ when they are
compiled together. As long as we separately compile $\mathsf{A}$ and
$\mathsf{B}$, on the contrary, there is no reason to disallow them to
pass the type checker. They can be used independently. However, they
can not be linked together because it implies that an unification
variable is instantiated inconsistently across the module
boundary. Therefore, the type checker should disallow module
$\mathsf{C}$ even after $\mathsf{A}$ and $\mathsf{B}$ are separately
compiled.  In order to detect this inconsistency across the module
boundary, we may need to track all instances of unification variables
and check their consistency during linking time.  So far, our \EL{}
does not have any imperative features so we do not need such a
checking mechanism during the link time. However, we will need one in
case that we add mutable references since it is possible to assign two
different types into one reference cell and the usual typing rule for
the polymorphic let-binding may be unsound.

\item Higher-order modules cause another such complication. Consider the
following code:
\begin{lstlisting}[style=mlpolyr]
template ID () = {{
   val id0 = fn x => x
   val id  = id0 id0
}}

module D = ID ()
module E = ID ()

val _ = (D.id 5, E.id ``hello'') (* value in question *)
\end{lstlisting}

Since we translate a template into an abstraction, we generate new
fresh type variables whenever we see unbounded unification variables
along with templates. Under this scheme, the above value in question
becomes accepted since $\mathsf{D.id}$ now has a type of $\alpha \ra
\alpha$ and $\mathsf{E.id}$ has a type of $\beta \ra \beta$ (assuming
that $\alpha$ and $\beta$ are fresh type variables).  Then, when they
are applied to $5$ and ``hello'', respectively (Line 9), $\alpha$ and
$\beta$ will be instantiated to $\Int$ and $\mathsf{string}$,
independently.
However, it might be surprising to see the type checker rejecting the
following code:
\begin{lstlisting}[style=mlpolyr]
val _ = (D.id0 5, D.id0 ``hello'') (* ill-typed *)
val _ = (E.id0 5, E.id0 ``hello'') (* ill-typed *)
\end{lstlisting}
We may expect to translate the template $\mathsf{ID}$ into a core term
with a type of $() \ra \{\mathsf{id0}:\forall
\alpha.\alpha\ra\alpha,\mathsf{id}:\beta\ra\beta\}$. Since our core
language does not support rank-1 polymorphism as in
SML\#~\cite{Ohori99}, the translated type will actually be $\forall
\alpha.() \ra
\{\mathsf{id0}:\alpha\ra\alpha,\mathsf{id}:\beta\ra\beta\}$.
Therefore, after instantiation, a type of $\mathsf{id0}$ becomes
$\alpha\ra\alpha$ where $\alpha$ is not a polymorphic variable any
more but just a placeholder for type instantiation. Thus, $\alpha$ can
not be instantiated into both $\Int$ and $\mathsf{string}$. This
limitation can be overcome by adopting rank-1 polymorphism in our core
language or by improving our module language up to the level of the ML
module language.

\item In our core language, we have the nice property that well-typed
  programs do not have uncaught exceptions.  Similarly, uncaught
  exceptions cannot escape up to the module level without being
  caught.  For example, the following example will be ill-typed:
\begin{lstlisting}[style=mlpolyr]
module Ex = {{
   ...
   val _ = raise `Fail ()  (* ill-typed *)
   ...
}}
\end{lstlisting}
However, the exception may be caught across the module boundary. 
Let us see the module $\mathsf{List}$:
\begin{lstlisting}[style=mlpolyr]
module List = {{
   ...
   fun hd l = case l of
	        [] => raise `Empty ()
	      | h::tl => h
   ...
}}
\end{lstlisting}
Any exception would not be raised until when $\mathsf{hd}$ is applied,
and the type of $\mathsf{hd}$ captures this fact:
$\forall\alpha.\forall\rho:\{\mathsf{Empty}\}.~\alpha~\mathsf{list}~
\stackrel{\underrightarrow{\mathsf{Empty}: (); \rho}}{~}\alpha$.
Then, the exception $\mathsf{Empty}$ is required to be caught
when an argument is supplied:
\begin{lstlisting}[style=mlpolyr]
val h = List.hd [1,2,3] (* ill-typed *)
\end{lstlisting}
To guarantee exception safety, the proper handler must be prepared
at a caller's site:
\begin{lstlisting}[style=mlpolyr]
val h = try x = List.hd [1,2,3]
        in  x
        handling `Empty () => 0
        end
\end{lstlisting}
\end{itemize} 

\section{Case study: the SAL interpreter example revisited}
\label{sub:exp-mod}
In the previous chapter (\secref{exp-prob}), 
we have implemented the base SAL interpreter
and its extensions mainly by using extensible cases.
In this section, we revisit the same example with the support
of modules.

\subsection*{Base interpreter}

We reorganize the previous implementation, 
making use of our module language. 
Figure~\ref{fig:code-base-mod} shows the module version of 
a base interpreter for SAL. 
First, we structure programs into separate units.
For example, the module $\mathsf{Envt}$ consists of
a collection of functions for dealing with environments:
$\mathsf{bind}$ and $\mathsf{empty}$:
\begin{lstlisting}[style=mlpolyr]
module Envt = {{
    fun bind (a, x, e) y = 
        if String.compare (x, y) == 0 then a else e y
    fun empty x = 
        raise `Fail (String.concat ["unbound variable: ", x, "\n"])
}}
\end{lstlisting}

The modules $\mathsf{Checker}$, $\mathsf{BigStep}$
and $\mathsf{Interp}$ are organized in a similar manner.
Notice that each module has its own namespace,
so that we do not have to make up new names such as
$\mathsf{check\_case}$ or $\mathsf{eval\_case}$
(as in \secref{exp-prob}). 

\begin{figure}
\begin{lstlisting}[style=mlpolyr]
(* module for the static semantics *)
module Checker = {{
    fun bases (check, env) =
        cases `VAR x => env x
            | `NUM n => ()
            | `PLUS (e1, e2) =>
              (check (env, e1);
               check (env, e2))
            | `LET (x, e1, e2) =>
              (check (env, e1);
               check (Envt.bind ((), x, env), e2))

    fun check e =
        let fun run (env, e) = match e with bases (run, env)
        in (run (Envt.empty, e); e)
        end
}}

(* module for the evaluation semantics *)
module BigStep = {{
    fun bases (eval, env) =
        cases `VAR x => env x
            | `NUM n => n
            | `PLUS (e1, e2) => eval (env, e1) + eval (env, e2)
            | `LET (x, e1, e2) => 
               eval (Envt.bind (eval (env, e1), x, env), e2)

    fun eval e =
        let fun run (env, e) = match e with bases (run, env)
        in  run (Envt.empty, e)
        end
}}

(* module for the interpreter *)
module Interp = {{
    fun interp e =
        try r = BigStep.eval (Checker.check e)
        in r
        handling `Fail msg => (String.output msg ; -1)
        end
}}
\end{lstlisting}
\caption{The module version of a base interpreter.}
\label{fig:code-base-mod}
\end{figure}

\subsection*{Extensions}

As the language grows, the corresponding rules such as static
semantics ($\mathsf{check}$) and dynamic semantics ($\mathsf{eval}$)
are changed.  \figref{ext-int-mod} shows modules for an extended
checker $\mathsf{EChecker}$ and an extended evaluator
$\mathsf{EBigStep}$.  Note that we can now use more uniform naming
(i.e., $\mathsf{check}$ instead of $\mathsf{echeck}$) due to the
availability of separate namespaces.

\begin{figure}
\begin{lstlisting}[style=mlpolyr]
(* module for the extended static semantics *)
module EChecker = {{
   fun bases (check, env) =
       cases `If0 (e1, e2, e3) => 
             (check (e1, env); check (e2, env); check (e3, env))
       default: Checker.bases (check, env)
     
   fun check e =
       let fun run (env, e) = match e with bases (run, env)
       in (run (Envt.empty, e); e)
       end
}}

(* module for the extended evaluation semantics *)
module EBigStep = {{
    fun bases (eval, env) =
        cases `IF0 (e1, e2, e3) =>
           if eval (env, e1) == 0 then eval (env, e2)
           else eval (env, e3)
        default: BigStep.bases (eval, env)

    fun eval e =
        let fun run (env, e) = match e with bases (run, env)
        in  run (Envt.empty, e)
        end
}}

(* module for the extended interpreter *)
module EInterp = {{
    fun interp e =
        try r = EBigStep.eval (EChecker.check e)
        in r
        handling `Fail msg => (String.output msg ; -1)
        end
}}
\end{lstlisting}
\caption{Implementation for an extended interpreter.}
\label{fig:ext-int-mod}
\end{figure}

\subsection*{Independent extensions}

Moreover, we can utilize templates, i.e., ``module functions'' which
take concrete modules as arguments. The result is a composite module:
\begin{lstlisting}[style=mlpolyr]
template InterpFun (C, E) = {{
    fun interp e =
        try r = E.eval (C.check e)
        in r
        handling `Fail msg => (String.output msg ; -1)
        end
}}
\end{lstlisting}
Then, we can instantiate different interpreters depending on their
parameters:

\begin{lstlisting}[style=mlpolyr]
module I  = InterpFun (Check, BigStep) 
module I' = InterpFun (ECheck, EBigStep)
\end{lstlisting}

In this way, it becomes possible to combine independently developed
extensions (e.g., $\mathsf{ECheck}$ and $\mathsf{EBigStep}$) 
so that they can be used jointly.

\chapter{Beyond the very large: feature-oriented programming}
\label{pl-prog}
\section{Introduction}

Previous work on extensible compilers has proposed
new techniques on how to easily add extensions to existing
programming languages and their compilers.
For example, JaCo is an extensible compiler for Java based on extensible 
algebraic types~\cite{Zenger-Odersky01,Zenger-Odersky05}. 
The Polyglot framework implements an extensible compiler where even changes of 
compilation phases and manipulation of internal abstract syntax trees are 
possible~\cite{nystrom03polyglot}.
Aspect-oriented concepts (i.e., cross-cutting concerns) are also applied 
to extensible compiler construction~\cite{Xiaoqing+:05}.
While all this work successfully demonstrates 
that a base compiler can be extended easily, 
most of these existing solutions do not attempt to pay special
attention to the {\em set of extensions} they produce.
Sometimes all the extensions can be integrated together to become
a new version of the system, in which case these existing solutions work well.

However, there are many cases where software changes cannot be merged
back so that different versions evolve and begin to coexist
independently. Moreover, there are even situations where such
divergence is planned from the beginning. A marketing plan may
introduce a product lineup with multiple editions. Aa mentioned
in~\chref{intro}, Windows Vista which ships in six editions is such an
example. Unless we carefully manage each change in different editions,
multiple versions that originate from one source start to coexist
separately. They quickly become so incompatible that they require
separate maintenance, even though much of their code is
duplicated. This quickly leads to a maintenance nightmare. In such a
case, the role of programming languages becomes limited and, instead,
we need a way to manage variability in the product lineup.

One possible way of addressing these issues is to adopt the product
line engineering paradigm.  Product line engineering is an emerging paradigm of
developing a family of products ~\cite{Kyo+:02,Kwanwoo+:02,SEI}.  It defines
a software product line to be a set of software
systems that share a common set of features with variations.
Therefore, it is expected to be developed from a common set of
software components (called {\it core assets}) on the same software
architecture.  The paradigm encourages developers to focus on
developing a set of products, rather than on developing one particular
product. Products are built from core assets rather than from scratch,
so mechanisms for managing variability are essential.

In many cases, however, product line methods do not
impose any specific synthesis mechanisms on product line implementation,
so implementation details are left to developers. 
As a consequence, feature-oriented programming (FOP) emerges
as an attempt to realize this paradigm at the code level.
For example, AHEAD, FeatureC++ and FFJ support the composition
of features in various ways~\cite{Batory2004,Apel05featurec++:on,Apel08}.

Although FOP has become popular in product line engineering,
comparative studies of the corresponding mechanisms for product line
implementation have rarely been conducted.  Lopez-Herrejon et
al. compared five technologies in order to evaluate feature
modularization \cite{Lopez05} but their experiment was
conducted entirely at the code level, which lead them to conclude that
a technology-independent model would be needed in order to reason
about product lines.

In this section, we first propose a two-way extensible interpreter as
a canonical example for product line engineering.  Our intention with
this example is to provide a framework for comparison of language
support for product line implementation.  Then, we identify some
issues that an implementation technique is expected to resolve,
illustrate how the MLPolyR language can be used to implement a two-way
extensible interpreter, and evaluate how effective our solution is.

\section{A two-way extensible interpreter as a generator}

We have seen how the MLPolyR language implements a two-way extensible
interpreter in various ways.  Similarly, many programming language
solutions have already been developed to solve the dilemma caused
by simultaneous two-way extensibility.  For example, Zenger and
Odersky presents a hybrid language specifically designed to solve this
issue~\citep{Zenger-Odersky05}.

Most of these existing solutions, however, do not consider the {\em
  set of extensions} they produce.  For example, assume one wants to
build an interpreter $\mathsf{I}$, which is the composition of the
combinators $\mathsf{eval}$ (realizing the evaluation semantics) and
$\mathsf{check}$ (realizing the static semantics) where $\mathsf{o}$
means function composition:
\[
\mathsf{I} = \mathsf{eval}~~\mathsf{o}~~\mathsf{check}
\]
The evaluation stage could also be implemented 
by the machine semantics $\mathsf{eval}_m$, 
instead of the evaluation semantics $\mathsf{eval}$:
\[
\mathsf{I_m} = \mathsf{eval}_{\mathsf{m}}~~ \mathsf{o}~~\mathsf{check}
\]
Optionally, the combinator $\mathsf{opt}$ which performs constant folding 
may be inserted to build an optimized interpreter $\mathsf{I_{opt}}$:
\[
\mathsf{I_{opt}} = \mathsf{eval}~~ \mathsf{o} ~~ \mathsf{opt}~~ \mathsf{o} ~~\mathsf{check}
\]
As the base language grows to support a conditional term, 
$\mathsf{eval}$, $\mathsf{opt}$ and 
$\mathsf{check}$ also evolve to constitute a new interpreter 
$\mathsf{I'_{opt}}$:
\[
\mathsf{I'_{opt}} = \mathsf{eval}'~~ \mathsf{o}~~\mathsf{opt}'~~ \mathsf{o}~~ \mathsf{check}'
\]
Since these interpreters have a lot in common, 
we should try to understand them as a {\em family} of interpreters. 
Therefore, the two-way extensible interpreter turns out to be 
a {\em generator} of a program family of SAL interpreters. 
While this two dimensional extension problem has been generally
studied within the context of how to easily extend base code
in a type safe manner, we focus on the generativity aspect of such solutions.
Moreover, our extensible interpreter example enables us
to emphasize the overall structure of the system, the so-called
{\em software architecture} \citep{Garlan94}.
Hence, we can analyze variations in terms of architectural and
component-level variations, rather than in terms of operations
or data which are rather vague and general.
Architectural variation captures 
inclusion or exclusion of certain functionality. 
For example, the extended interpreter includes an optimization phase 
while the base interpreter does not. 
Component-level variations capture that which may have 
multiple alternative implementations. For example, 
every interpreter has its own evaluator which implements 
either the evaluation semantics or the machine semantics.

\section{Feature-oriented product line engineering}

Since we set up a two-way extensible interpreter to generate a family
of products, it is natural to apply product line engineering for
better support of their development.  Among various product line
approaches, we adopt FORM product line engineering for the following
reasons:
\begin{itemize}
\item The method relies on a feature-based model which provides
  adequate means for reasoning about product lines~\citep{Kyo+:02}.
\item The method supports architecture design which plays an important
  role in bridging the gap between the concepts at the requirement
  level and their realization at the code level by deciding how
  variations are modularized by means of architectural
  components~\citep{Noda2008}.
\item The method consists of well-defined development process which
  enables us to easily identify implementation dependent phases.
\end{itemize}

\begin{figure}
\centering
\includegraphics[scale=0.6]{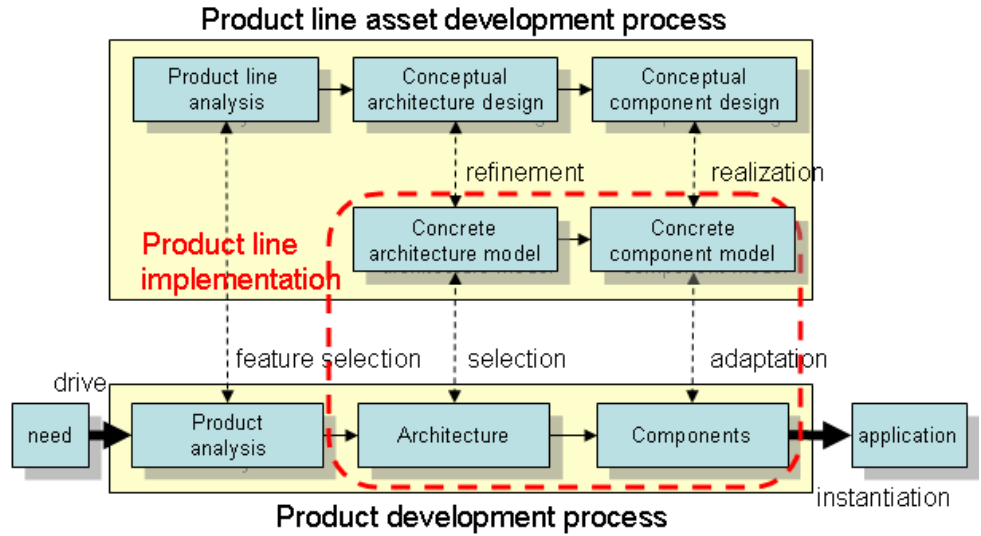}
\caption{Development process (adopted from FORM~\citep{Kyo+:02}).}
\label{fig:devprocess}
\end{figure} 

To let us focus on product line implementation as opposed 
to implementation independent processes, we highlight the former 
as shown in~\figref{devprocess}. The area surrounded 
by dashed lines is the subject of our comparative study.
In this section, we will give an overview of overall engineering
activities for a family of the SAL interpreters.
Then, in the following section, we will show how to refine conceptual models
into concrete models with the mechanisms that the MLPolyR 
language provides.

\subsection{Product line analysis}

\begin{figure}
\centering
\includegraphics[scale=0.8]{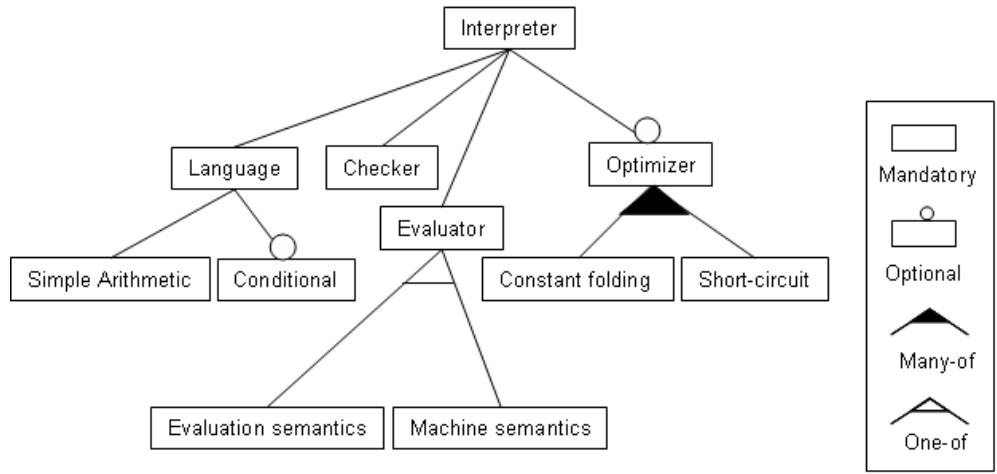}
\caption{Feature model for the SAL interpreter.}
\label{fig:feature-model}
\end{figure}

We perform commonality and variability analysis for the family
of the SAL interpreters. We can easily consider features 
in the base interpreter as commonalities and exclusive features 
only in some extensions as variations. Then, we determine 
what causes these variations. For example, we can clearly tell
that the choice of a set of language constructors differentiates 
interpreters. Similarly, the choice of evaluation strategies 
makes an impact. Optimization could optionally be performed. 
We refer to these factors that differentiate products
as {\em features}~\citep{Kyo+:02,Kang98}. 
\figref{feature-model} shows the feature model 
according to our product line analysis. 

\subsection{Product line architecture design}
\label{sec:pl-arch}
\begin{figure}
\centering
\includegraphics[scale=1]{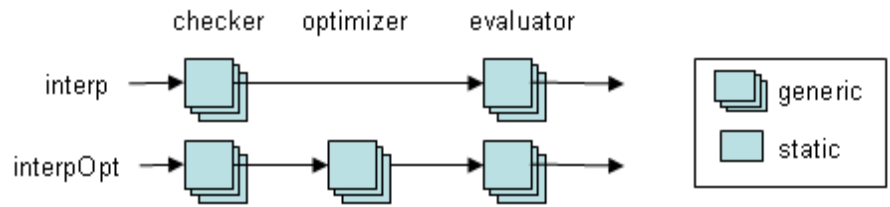}
\caption{Reference architectures.}
\label{fig:refarch}
\end{figure}

Architecture design involves identifying conceptual components 
and specifying their configuration. Based on the product line analysis, 
we define two reference architectures by mapping each combinator 
to a distinct component in~\figref{refarch}. A component can be 
either generic or static. A {\em generic} component encapsulates variations 
when a certain aspect of this component varies in different products.
The $\mathsf{evaluator}$ component is a typical example. 
A {\em static} component performs usual common functionality 
across family members. 

During this phase, we have to not only identify components 
but also define interfaces between components: 
\[
\begin{array}{lcccc}
\mathsf{checker}   & : & \mathsf{term} & \ra & \mathsf{term}\\
\mathsf{optimizer} & : & \mathsf{term} & \ra & \mathsf{term}\\
\mathsf{evaluator} & : & \mathsf{term} & \ra & \mathsf{value}
\end{array}
\]
As usual, the arrow symbol $\ra$ is used to specify a function
type. In our example, components act like pipes in a pipe-and-filter
architecture style, so all interface information is captured by the
type.  By using the above components, we can specify the overall
structure of various interpreters:
\[
\begin{array}{lcl}
\mathsf{interp}    & = & \mathsf{evaluator}~~ \mathsf{o}~~ \mathsf{checker}\\
\mathsf{interpOpt} & = & \mathsf{evaluator}~~ \mathsf{o}~~\mathsf{optimizer}~~ \mathsf{o}~~ \mathsf{checker}
\end{array}
\]

\subsection{Product line component design}
\label{sec:pl-comp}
Next, we identify conceptual components which are 
constituents of a conceptual architecture. 
A conceptual component can have multiple implementations. 
For example, there are many versions of the $\mathsf{evaluator}$ 
component depending on the evaluation strategy:
\[
\begin{array}{lcccc}
\mathsf{eval}   & : & \mathsf{term} & \ra & \mathsf{value} \\
\mathsf{eval}_{\mathsf{m}} & : & \mathsf{term} & \ra & \mathsf{value}
\end{array}
\]
At the same time, the language $\mathsf{term}$ can be extended 
to become $\mathsf{term}'$ which is an extension of $\mathsf{term}$
(for example to support conditionals):
\[
\begin{array}{lcccc}	
\mathsf{eval}'   & : & \mathsf{term}' & \ra & \mathsf{value} \\
\mathsf{eval}'_{\mathsf{m}} & : & \mathsf{term}' & \ra & \mathsf{value}
\end{array}
\]
Similarly, $\mathsf{check}$ and $\mathsf{check}'$ can be specified as follows:
\[
\begin{array}{lcccc}
\mathsf{check}  & : & \mathsf{term}  & \ra & \mathsf{term} \\
\mathsf{check}' & : & \mathsf{term}' & \ra & \mathsf{term}'
\end{array}
\]
For the $\mathsf{optimizer}$ component, there are many possible
variations due to inclusion or exclusion of various individual
optimization steps (here: constant folding and short-circuiting) and
due to the variations in the underlying term language (here: basic and
extended):

\[
\begin{array}{lcccc}
\mathsf{opt}_{\mathsf{cons}} & : & \mathsf{term} & \ra & \mathsf{term} \\
\mathsf{opt}'_{\mathsf{cons}} & : & \mathsf{term}' & \ra & \mathsf{term}' \\
\mathsf{opt}'_{\mathsf{short}} & : & \mathsf{term}' & \ra & \mathsf{term}' \\
\mathsf{opt}'_{\mathsf{cons+short}} & : & \mathsf{term}' & \ra & \mathsf{term}'
\end{array}
\]

\subsection{Product analysis}
\label{sec:product}
Product engineering starts with analyzing the requirements provided by
the user and finds a corresponding set of required features from the
feature model.  Assuming we are to build four kinds of interpreters,
we have to have four different feature selections:
\[
\begin{array}{lcl}
\mathsf{FS}(\mathsf{I})     & = & \{\mathsf{Evaluation~semantics}\} \\
\mathsf{FS}(\mathsf{I}_{\mathsf{m}})   & = & \{\mathsf{Machine~semantics}\} \\
\mathsf{FS}(\mathsf{I}_{\mathsf{opt}})  & = & \{\mathsf{Machine~semantics, Optimizer, Constant~folding}\} \\
\mathsf{FS}(\mathsf{I}'_{\mathsf{opt}}) & = & \{\mathsf{Conditional, Evaluation~semantics, Optimizer, Constant~folding, Short-circuit}\}
\end{array}
\]
Here, the function $\mathsf{FS}$ maps a feature product to its
corresponding set of its required features.  (For brevity only
non-mandatory features are shown.)

\begin{figure}
\centering
\includegraphics[scale=0.7]{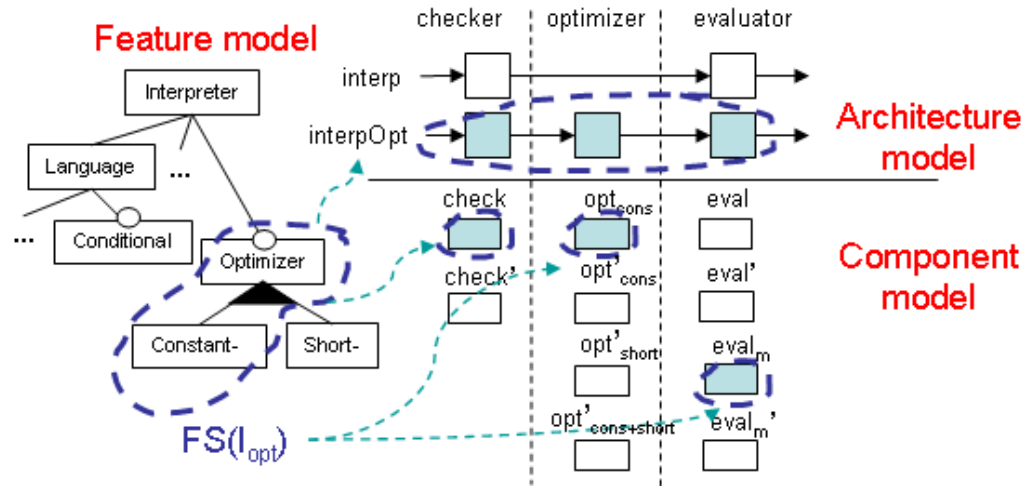}
\caption{Product engineering.}
\label{fig:prodeng}
\end{figure}

During product engineering, these selected feature sets give
advice on the selection among both reference architectures and components. 
\figref{prodeng} shows the overall product engineering process 
where the reference architecture $\mathsf{interpOpt}$ gets selected, 
guided by the presence of the $\mathsf{Optimizer}$ feature. 
Feature sets also show which components need to be selected and 
how they would be instantiated at the component level. 
For example, the presence of the $\mathsf{Constant~folding}$ feature 
guides us to choose the component $\mathsf{optimizer}$ 
with the implementation $\mathsf{opt_{cons}}$. 
Similarly, the presence of the $\mathsf{Machine~semantics}$ feature
picks the implementation $\mathsf{eval_m}$ instead of $\mathsf{eval}$. 
The target product would be instantiated 
by assembling such selections.

\section{Issues in product line implementation}
\label{sec:issues}
During the product line asset development process, we obtain reference
models which represent architectural and component-level
variations. Such variations should be realized at the code level. The
first step is to refine conceptual architectures into concrete
architectures which describe how to configure conceptual components.
Then, product line component design involves realization of conceptual
components using the proper product feature delivery methods.  This
section discusses some issues that surface during product line
implementation.

\subsection*{Product line architecture implementation}

In order to specify concrete reference architectures, 
we have to not only identify 
conceptual components but also define interfaces between components. 
Moreover, since there may be multiple reference architectures, 
it would be convenient to have mechanisms for abstracting 
architectural variations, capturing the inclusion or exclusion 
of certain components. Therefore, any adequate implementation technique 
should be able to provide mechanisms for:
\begin{itemize}
\item Declaration of required conceptual components 
($\mathsf{checker}$, $\mathsf{optimizer}$ and $\mathsf{evaluator}$) 
and their interfaces,
\item Specification of the base reference architecture $\mathsf{interp}$ 
and its optimized counterparts $\mathsf{interpOpt}$ 
by using such conceptual components. 
\end{itemize}

\subsection*{Product line component implementation.}

This phase involves realization of conceptual components. 
The main challenge of this phase is in how to implement generic components 
that encapsulate component-level variations.  
Such variations could be in the form of either code extension or 
code substitution. Any solution to the traditional expression 
problem can be a mechanism to implement code extension. 
For our running example, the following pairs correspond to code extension:
\begin{itemize}
\item $\mathsf{check}$ and $\mathsf{check}'$
\item $\mathsf{eval}$ and $\mathsf{eval}'$ 
\item $\mathsf{eval}_{\mathsf{m}}$ and $\mathsf{eval}'_{\mathsf{m}}$
\item $\mathsf{opt_{cons}}$ and $\mathsf{opt'_{cons}}$
\item $\mathsf{opt}_{\mathsf{cons}}$ and $\mathsf{opt}'_{\mathsf{cons+short}}$
\item $\mathsf{opt}'_{\mathsf{short}}$ and $\mathsf{opt}'_{\mathsf{cons+short}}$
\end{itemize}

Code substitution provides another form of variation 
at the component level when two different implementations 
provide interchangeable functionality. For example, 
$\mathsf{eval}$ and $\mathsf{eval_m}$ both implement the $\mathsf{evaluator}$
component, but neither is an extension of the other.
Language abstraction mechanisms are expected 
to handle this case elegantly. For our running example, 
the corresponding scenarios are as follows:
\begin{itemize}
\item $\mathsf{eval}$ and $\mathsf{eval}_{\mathsf{m}}$
\item $\mathsf{eval}'$ and $\mathsf{eval}'_{\mathsf{m}}$
\end{itemize}

\subsection*{Product engineering}

Based on the product analysis, a feature product is instantiated 
by assembling product line core assets. For our running example, 
the evaluated techniques should be able to instantiate 
four interpreters ($\mathsf{I}, \mathsf{I_m}, \mathsf{I_{opt}}, \mathsf{I'_{opt}}$) 
based on the selected feature set.

\section{Language supports for product line implementation}

In this section, we illustrate how the \mlpolyr{} language
can be used to implement a two-way extensible interpreter.
First, we show how each issue identified in the previous chapter
will be resolved by various mechanisms provided by \mlpolyr{}.
A comparison with other product line implementation
techniques follows.

\subsection*{Product line architecture implementation}

Each component in a reference architecture is mapped to an \mlpolyr{}
module. As specified in \secref{issues}, 
we first define types (or signatures) of the interested components
based on the outcome of product line architecture design.
(\secref{pl-arch}):
\[
\begin{array}{lclccll}
\mathsf{Checker}   & : & \{\{ & \mathsf{check} & : & 
                              \mathsf{term} \ra \mathsf{term}, & \ldots ~\}\}\\
\mathsf{Optimizer} & : & \{\{ & \mathsf{opt}   & : &
                              \mathsf{term} \ra \mathsf{term}, & \ldots ~\}\}\\
\mathsf{Evaluator} & : & \{\{ & \mathsf{eval}  & : & 
                              \mathsf{term} \ra \mathsf{int}, & \ldots ~\}\}
\end{array}
\]
where $\ldots$ indicates that there may be more parts in a component,
but they are not our concerns.  In practice, we do not have to write
such interfaces explicitly since the type checker infers the principal
types.  Then, by using these conceptual modules ($\mathsf{Checker}$,
$\mathsf{Optimizer}$ and $\mathsf{Evaluator}$), we can define two
reference architectures:
\begin{lstlisting}[style=mlpolyr]
module Interp = {{
    val interp = fn e => Evaluator.eval(Check.check e)
}}

module InterpOpt = {{
    val interp = fn e => Evaluator.eval(Optimizer.opt(Checker.check e))
}}
\end{lstlisting}

Alternatively, like functors in SML, 
we can use a parameterization technique called a {\em template} 
which takes concrete modules as arguments and 
instantiates a composite module:
\begin{lstlisting}[style=mlpolyr]
template InterpFun (C, E) = {{
    val interp = fn e => E.eval (C.check e)
}}

template InterpOptFun (C, O, E) = {{
    val interp = fn e => E.eval (O.opt (C.check e))
}}
\end{lstlisting}
where $\mathsf{C}$, $\mathsf{O}$ and $\mathsf{E}$ represent
$\mathsf{Checker}$, $\mathsf{Optimizer}$ and $\mathsf{Evaluator}$
respectively. Their signatures are captured as constraints by the type
checker. For example, the type checker infers the constraint that the
module $\mathsf{C}$ should have a component named $\mathsf{check}$
which has a type of $\alpha \ra \beta$ and $\beta$ should be either an
argument type of the module $\mathsf{E}$ (Line 1) or that of
$\mathsf{O}$ (Line 5).

The second approach with templates supports more code reuse because a
reference architecture becomes polymorphic, i.e., parameterized not
only over the values but also over the types of its components. As
long as components satisfy constraints that the type checker computes,
any components can be plugged into a reference architecture.  For
example, for the argument $\mathsf{C}$, either the base module
$\mathsf{Check}$ and its extension $\mathsf{EChecker}$ can applied to
the template $\mathsf{InterpFun}$.

\subsection*{Product line component implementation}

Modules in \mlpolyr{} implement components. 
In order to manage component-level variations,
we have to deal with both code extension and code substitution
as discussed in \secref{issues}.
For example, we will see multiple implementations 
of the component $\mathsf{Evaluator}$:
\[
\begin{array}{lclccll}
\mathsf{BigStep}   & : & \{\{ & \mathsf{eval} & : & 
                              \mathsf{term} \ra \mathsf{int}, & \ldots ~\}\}\\
\mathsf{Machine} & : & \{\{ & \mathsf{eval}   & : &
                              \mathsf{term} \ra \mathsf{int}, & \ldots ~\}\}\\
\mathsf{EBigStep} & : & \{\{ & \mathsf{eval}  & : & 
                              \mathsf{term}' \ra \mathsf{int}, & \ldots ~\}\}\\
\mathsf{EMachine} & : & \{\{ & \mathsf{eval}   & : &
                              \mathsf{term}' \ra \mathsf{int}, & \ldots ~\}\}

\end{array}
\]
where $\mathsf{term}$ represents a type of the base constructors and 
$\mathsf{term}'$ that of the extension. 
$\mathsf{BigStep}$ and $\mathsf{EBigStep}$ implement
the evaluation semantics and its extension 
while $\mathsf{Machine}$ and $\mathsf{EMachine}$ implement
the machine semantics and its extension. 
Note that the pair of $\mathsf{BigStep}$ and $\mathsf{EBigStep}$
and also the pair of $\mathsf{Machine}$ and $\mathsf{EMachine}$
correspond to code extension while the pair of $\mathsf{BigStep}$ and
$\mathsf{Machine}$ corresponds to code substitution.

Code extension is supported by first-class extensible cases
as we already studied in \secref{exp-prob}.
\figref{code-mlpolyr} shows how such extensions are made.
In an extension, only a new case is handled
(Line 19-21) and the default explicitly refers to the original set
of other cases represented by $\mathsf{BigStep.bases}$ (Line 22).

\begin{figure}
\begin{lstlisting}[style=mlpolyr]
(* module for the evaluation semantics *)
module BigStep = {{
    fun bases (eval, env) =
        cases `VAR x => env x
            | `NUM n => n
            | `PLUS (e1, e2) => eval (env, e1) + eval (env, e2)
            | `LET (x, e1, e2) => 
               eval (Envt.bind (eval (env, e1), x, env), e2)

    fun eval e =
        let fun run (env, e) = match e with bases (run, env)
        in  run (Envt.empty, e)
        end
}}

(* module for the extended evaluation semantics *)
module EBigStep = {{
    fun bases (eval, env) =
        cases `IF0 (e1, e2, e3) =>
           if eval (env, e1) == 0 then eval (env, e2)
           else eval (env, e3)
        default: BigStep.bases (eval, env)

    fun eval e =
        let fun run (env, e) = match e with bases (run, env)
        in  run (Envt.empty, e)
        end
}}
\end{lstlisting}
\caption{The module $\mathsf{BigStep}$ realizes the evaluation
semantics ($\mathsf{eval}$), and the module $\mathsf{EBigStep}$ 
realizes the extended evaluation semantics ($\mathsf{eval'}$)
by defining only a new case $\mathsf{`IF0}$.
In an extension, only a new case is handled
(Line 19-21) and the default explicitly refers to the original set
of other cases represented by $\mathsf{BigStep.bases}$ (Line 22).
Then, $\mathsf{EBigStep.bases}$ can handle five cases including $\mathsf{IF0}$.
We can obtain a new evaluator $\mathsf{EBigStep.eval}$ by closing
the recursion through applying $\mathsf{bases}$ to evaluator itself
(Line 25). Note that a helper function $\mathsf{run}$ is actually 
applied instead of $\mathsf{eval}$ in order to pass an initial environment 
in Line 26.}
\label{fig:code-mlpolyr}
\end{figure}

\begin{figure}
\begin{lstlisting}[style=mlpolyr]
(* module for the machine semantics *)
module Machine = {{
    fun ecases (K, env, estate, vstate) =
        cases `VAR x => env x
            | `NUM n => vstate (n, K)
            | `PLUS (e1, e2) => estate (`PLUSl (e2, env)::K, env, e1)
	    | `LET (x, e1, e2) => estate (`LETl (x, e2, env)::K, env, e1)
    and vcases (v, K, estate, vstate) =
        cases `PLUSl (e, env) => estate((`PLUSr v)::K, env, e) 
	    | `LETl (x, e, env) => estate (K, Envt.bind (v, x, env), e)
	    | `PLUSr v' => vstate (v'+v, K)
 
    fun estate (K, env, e) = match e with ecases (K, env, estate, vstate)
    and vstate (v, K) =
        case K of
	    [] => v
	  | h::tl => match h with vcases (v, tl, estate, vstate)

    fun eval e = estate ([], Envt.empty, e)
}}

(* module for the extended machine semantics *)
module EMachine = {{
    fun ecases (K, env, estate, vstate) =
        cases `IF0 (e1, e2, e3) => 
               estate (`IF0l (e2, e3, env)::K, env, e1)
        default: Machine.ecases (K, env, estate, vstate)
    and vcases (v, K, estate, vstate) =
        cases `IF0l (e2, e3, env) => 
	      if v == 0 then estate (K, env, e2) else estate (K, env, e3)
        default: Machine.vcases (v, K, estate, vstate)
 
    fun estate (K, env, e) = match e with ecases (K, env, estate, vstate)
    and vstate (v, K) =
        case K of
	    [] => v
	  | h::tl => match h with vcases (v, tl, estate, vstate)

    fun eval e = estate ([], Envt.empty, e)
}}
\end{lstlisting}
\caption{The module $\mathsf{Machine}$ realizes the machine
semantics ($\mathsf{eval_m}$), and the module $\mathsf{EMachine}$ 
realizes the extended machine semantics ($\mathsf{eval'_m}$)
by defining only new cases $\mathsf{`IF0}$ and $\mathsf{`IF0l}$.}
\label{fig:machine-code}
\end{figure}

Code substitution as another form of variation
at the component level does not cause any trouble. 
For example, \figref{machine-code} shows the module $\mathsf{Machine}$ 
which implements the machine semantics (i.e., $\mathsf{eval_m}$).
Like $\mathsf{BigStep}$ and $\mathsf{EBigStep}$,
$\mathsf{EMachine}$ extends $\mathsf{Machine}$ through two extensible 
cases (Line 27 and 31).
In our example two different implementations ($\mathsf{BigStep}$ and 
$\mathsf{Machine}$) provide interchangeable functionality, but
neither is an extension of the other, so they are implemented
independently.

Analogously, we can implement the remaining two conceptual components
$\mathsf{Checker}$ and $\mathsf{Optimizer}$. 
For $\mathsf{Checker}$ we have,
\[
\begin{array}{lclccll}
\mathsf{Check}   & : & \{\{ & \mathsf{check} & : & 
                              \mathsf{term} \ra \mathsf{term}, & \ldots ~\}\}\\
\mathsf{ECheck} & : & \{\{ & \mathsf{check}   & : &
                              \mathsf{term}' \ra \mathsf{term}', & \ldots ~\}\}
\end{array}
\]
where each implements the concrete component $\mathsf{check}$ and 
$\mathsf{check'}$, respectively. For the component $\mathsf{Optimizer}$,
\[
\begin{array}{lclccll}
\mathsf{COptimizer}   & : & \{\{ & \mathsf{opt} & : & 
                              \mathsf{term} \ra \mathsf{term}, & \ldots ~\}\}\\
\mathsf{ECOptimizer}   & : & \{\{ & \mathsf{opt} & : & 
                              \mathsf{term}' \ra \mathsf{term}', & \ldots ~\}\}\\
\mathsf{ESOptimizer}   & : & \{\{ & \mathsf{opt} & : & 
                              \mathsf{term}' \ra \mathsf{term}', & \ldots ~\}\}\\
\mathsf{ECSOptimizer}   & : & \{\{ & \mathsf{opt} & : & 
                              \mathsf{term}' \ra \mathsf{term}', & \ldots ~\}\}
\end{array}
\]
where each implements the concrete component $\mathsf{opt_{cons}}$,
$\mathsf{opt'_{cons}}$, $\mathsf{opt'_{short}}$ and $\mathsf{opt'_{cons+short}}$,
respectively.

\subsection*{Product engineering}

In \secref{product}, we define four interpreters 
($\mathsf{I}$, $\mathsf{I_m}$, $\mathsf{I_{opt}}$, and $\mathsf{I'_{opt}}$)
differentiated by the feature selection. Each will be instantiated
by selecting a proper architecture (either $\mathsf{InterpFun}$ 
and $\mathsf{InterOptFun}$) and choosing its components 
(either $\mathsf{BigStep}$ or $\mathsf{Machine}$, etc)
with implicit advice from the selected feature set. For example:
\begin{itemize}
\item When the feature set is $\mathsf{FS(I)}$,
the reference architecture $\mathsf{InterpFun}$ gets selected
since the $\mathsf{Optimizer}$ feature is not in the set.
Then, the proper components are  selected and instantiated.
For example, the presence of the $\mathsf{Evaluation~semantics}$ feature 
guides us to choose the component $\mathsf{BigStep}$
instead of $\mathsf{Machine}$. Therefore, we instantiate the interpreter
$\mathsf{I}$ as follows:
\[
\begin{array}{@{}l@{~~}c@{~~}ll}
\mathsf{module~I}               & = & \mathsf{InterpFun} & (\mathsf{Checker}, \mathsf{BigStep})
\end{array}
\]   

\item When the feature set is $\mathsf{FS(I_m)}$,
the reference architecture $\mathsf{InterpFun}$ gets
chosen. Here, components $\mathsf{Machine}$ and $\mathsf{Check}$
are selected because of the presence of $\mathsf{Machine~semantics}$ feature. 
Therefore, we instantiate the interpreter
$\mathsf{I_m}$ as follows:
\[
\begin{array}{@{}l@{~~}c@{~~}ll}
\mathsf{module~I_m}               & = & \mathsf{InterpFun} & (\mathsf{Checker}, \mathsf{Machine}) 
\end{array}
\]   

\item When the feature set is $\mathsf{FS(I_{opt})}$,
the reference architecture $\mathsf{InterpOptFun}$ is chosen
since the $\mathsf{Optimizer}$ feature is in the set.
Then, again, the proper components get selected and instantiated.
Here, the presence of the $\mathsf{Constant~folding}$ feature 
guides us to choose the component $\mathsf{COptimizer}$
and the presence of the $\mathsf{Machine~semantics}$ feature
leads us to instantiate the component $\mathsf{Machine}$.
Therefore, we instantiate the interpreter
$\mathsf{I_{opt}}$ as follows:
\[
\begin{array}{@{}l@{~~}c@{~~}ll}
\mathsf{module~I_{opt}}               & = & \mathsf{InterpOptFun} & (\mathsf{Checker}, \\
                                  &   &                    & ~~\mathsf{COptimizer},\\
                                  &   &                    & ~~\mathsf{Machine})
\end{array}
\]   

\item When the feature set is $\mathsf{FS(I'_{opt})}$,
the reference architecture $\mathsf{InterpOptFun}$ is chosen.
As far as the components are concerned,
the presence of the $\mathsf{Conditional}$ and 
$\mathsf{Evaluation~semantics}$ features guide
us to choose the component $\mathsf{EBigStep}$.
Similarly, the presence of the $\mathsf{Optimizer}$, 
$\mathsf{Conditional}$, $\mathsf{Constant~folding}$ and 
$\mathsf{Short-circuit}$ forces the use of component $\mathsf{ECSOptimizer}$.
Therefore, we instantiate the interpreter
$\mathsf{I'_{opt}}$ as follows:
\[
\begin{array}{@{}l@{~~}c@{~~}ll}
\mathsf{module~I'_{opt}}               & = & \mathsf{InterpOptFun} & (\mathsf{EChecker}, \\
                                  &   &                    & ~~\mathsf{ECSOptimizer},\\
                                  &   &                    & ~~\mathsf{EBigStep})
\end{array}
\]
\end{itemize}

   



\section{Evaluation}

Although they are not intended to aim specifically for
feature-oriented programming, many language constructs
can be used to manage variability in the context of 
product line implementation. For example, various mechanisms
including classes, aspects and modules can support abstraction of
features. They also support extension mechanisms such as sub-classing, 
macro processing, aspect-weaving or parameterizing,
which can be used to modularize feature composition.
Among various techniques, there are three representative 
implementation approaches which can be found frequently 
in the product line literature \cite{Gacek2001,Kastner08}.

\subsection*{The annotative approach}
As the name suggests, the annotative approaches implement features 
using some form of annotations. Typically, preprocessors, 
e.g., macro systems, have been used in many literature examples 
as the feature product delivery method \cite{Kang98,KangKLK05}. 
For example, the macro language in FORM determines inclusion or 
exclusion of some code segments based on the feature selection:
\begin{lstlisting}[style=mlpolyr]
val interp = 
    fn e => Evaluator.eval
$IF(;:$Optimizer) [
            (Optimizer.opt
             (Check.check e))
][                         
            (Check.check e)
]
\end{lstlisting}
Depending on the presence of the $\mathsf{Optimizer}$ feature,
either block (4-5 or 7) will be selected.

Macro languages have some advantage in that 
they can be mixed easily with any target programming languages. 
However, feature specific segments are scattered across 
multiple classes, so code easily becomes complicated. 
Saleh and Gomaa propose the feature description language \cite{Saleh05}. 
Its syntax looks similar to the C/C++ preprocessor 
but it supports separation of concerns by modularizing 
feature specific code in a separate file. 
In the annotation approach, however, target compilers 
do not understand the macro language and any error appearing 
in feature code segments cannot be detected 
until all feature sets are selected and the corresponding code segments 
are compiled. 

\subsection*{The compositional approach}
For taking advantage of the current compiler technology 
including static typing and separate compilation, 
we need native language supports. Therefore, language-oriented proposals 
generally take compositional approaches by providing better support 
for feature modularity \cite{Lopez05}. FeatureC++ \cite{Apel05featurec++:on}, 
AHEAD \cite{Batory2004} and AspectJ are such language extensions. 

In this approach, features are implemented as 
distinct units and then they are combined to become a product. 
Aspect-oriented programming has become popular 
as a way of implementing the compositional approach \cite{Lee06,Cho2008}. 
The main idea is to implement variations as separate aspects 
and to obtain each product by weaving base code and aspect code.
Our extensible cases provide similar composability. Furthermore,
our module language also supports extensible modules,
which make large-scale code reusable.
Note that composition in aspects does not provide separate compilation, 
so base code requires to be either re-typed-checked or re-compiled 
or both for every composition. However, our module system supports
separate compilation.

\subsection*{The parameterization approach}
The idea of parameterized programming is to implement 
the common part once and parameterize variations 
so that different products can be instantiated 
by assigning distinct values as parameters. Functors, 
as provided by Standard ML (SML), are a typical example 
in that they can be parameterized on values, types and 
other modules~\cite{smlnj}. 
The SML module system has been demonstrated to be powerful 
enough to manage variations in the context of product lines \cite{Chae08}.
However, its type system sometimes imposes restrictions
which require code duplication between functions on data types. 
Many proposals to overcome this restriction 
have been presented. For example, MLPolyR proposes extensible cases
\cite{blume+:06:mlpolyr}, 
and OCaml proposes polymorphic variants \cite{Garrigue00}. 

Similarly, templates in C++ provide parameterization over types
and have been extensively studied in the context of programming families
~\cite{Krzyszttof}. Recently, an improvement
that would provide better support of generic programming has been proposed~\cite{Reis06}. 
Originally, Java and C\# did not support parameterized types 
but now both support similar concepts~\cite{Torgersen04}.

Sometimes the parameterization approach is criticized for its difficulty 
in identifying variation points and defining parameters \cite{Gacek2001}. 
However, systematic reasoning (e.g., product line analysis done 
by product line architects) can ease such burden 
by providing essential information for product line implementation
~\cite{Chae08}.

\chapter{Conclusion}
\label{conclude}

Software evolves by means of change. Changes may be implemented
either sequentially or in parallel. Sequential changes form a series
of software releases. Some changes carried out in parallel may also be
merged back together. In this situation, we are interested in
extension mechanisms which provide a way to add extensions in a
reliable way. Some changes implemented in parallel, however, cannot be
combined together so a single software product diverges into 
different versions. In this case, multiple software versions may
evolve independently although much of their code is duplicated, which
makes it difficult to maintain them.  Under these circumstances, we
need a way of managing variability among multiple versions so that we
can easily manage the evolution of a set of products.

In this thesis, we propose type-safe extensible programming which
takes two dimensions into consideration.  In particular, our language
provides type-safe extensibility mechanisms at multiple levels of
granularity, from the fine degree (at the core expression level) to
the coarse degree (at the module level).  At the same time, in order
to manage variability, we adopt product line engineering as a
developing paradigm and then show how our extensibility mechanisms
can be used to implement a set of products:
\begin{itemize}
\item In Section \ref{ext-prog}, we propose a core language 
that supports polymorphic extensible records,
first-class cases and type safe exception handling.
With cases being first-class and extensible,
we show that our language enables a very flexible style of
composable extension;
\item In Section \ref{large-prog}, we propose a module system 
that makes extensible programming at the module level possible.
We also show how to compile each module separately
in the presence of all of the above features;
\item In Section \ref{pl-prog}, we propose a development process 
which adopts product line engineering in order to manage
variability in a family of systems. We show that our extensibility mechanisms 
can be put to good use in the context of product line implementation.
\end{itemize}

We are continuing this work in several ways.  First, we plan to
improve our type system. For example, we have constructed a prototype
compiler for \mlpolyr{} that retains all of the \mlpolyr{} features as
well as mutable record fields. Records with mutable fields have
identity, and allocation of such a record is a side-effecting operation.
However, mutable data type can weaken our polymorphic type system, in
situations where the so-called \emph{value restriction} prevents row
type variables from being generalized~\cite{pessaux+leroy:99:exn}.
Pessaux and Leroy presents such an example that shows a false
positive:
\begin{lstlisting}[style=mlpolyr]
let val r = {| i = fn x => x+1 |}
    fun f y c = if c then r!i y 
                else raise `Error ()
in r!i 0
end
\end{lstlisting}
First, $r$ has type $\eRec{|i:\Int \era{\rho} \Int|}$ where $\rho$ is
not generalized since the whole expression is not a syntactic value
(Line 1). Then, during typing $f$, a true branch with $r!i~y$ (Line 2)
is unified with a false branch with $\mathbf{raise}~`\mathsf{Error}
()$ (Line 3).  Therefore, $\rho$ becomes $\mathsf{Error()};\rho'$ and
the application $r!i~0$ falsely appears to raise $\mathsf{Error()}$
even though it does not (Line 4). Pessaux and Leroy suggests that this
false positive could be avoided with a more precise tracking of the
flow of exceptions.

Additionally, as we discussed in \secref{separate},
non-generalized unification variables in the presence of
mutable references makes our type system
unsound unless they are instantiated consistently
across the module boundary. We plan to add a consistency checking
mechanism during linking time.

Second, our module system does not require any type decoration since
the type system infers module signatures as it infers types of core
expressions. However, there will be a need for programmers to spell
out types. For example, module signatures in libraries are generally
required to be explicit. We plan to support explicit specification of
module signatures and conventional signature matching as in
SML. However, there can be situations where row types and kind
information make it difficult to specify full typing information. As
we have seen in \secref{separate}, we might ask programmers to write
the following type decoration for $\mathsf{map}$:

\parbox{5in}{\begin{tabbing}
{\bf val} map : $\forall \alpha : \star.  \forall \beta : \star .
                 \forall \gamma : \varnothing . \forall \delta : \varnothing .
                 (\alpha \era{\gamma} \beta) \era{\delta}
                 ([\alpha] \era{\gamma} [\beta])$
\end{tabbing}}

It is possible to avoid this excessive notational overhead by defining
a little language with good built-in defaults (e.g., abbreviation for
common patterns). Then, programmers would specify their intentions
using this language and these intended types can be checked against
inferred types in a style of software contract
~\cite{Findler02,Blume06:contract}. For example, we may specify
$\mathsf{map}$'s type as follows and all elided parts can be inferred
and checked by a compiler:

\parbox{5in}{\begin{tabbing}
{\bf val} map : $(\alpha \era{\gamma} \beta) \era{}
                 ([\alpha] \era{\gamma} [\beta])$
\end{tabbing}}
 
Third, we plan to integrate feature composition with our language.
Our work shows that modern programming language technology such as
extensible cases and parameterized modules is powerful enough to
manage variability identified by product line analysis.  However, in
our approach, the relations among features, architectures, and
components are implicitly expressed only during the product line
analysis. Similarly, Most feature-oriented programming languages do
not have the notion of a ``feature'' in the language syntax since
features are merely considered conceptual abstractions rather than
concrete language constructs.  Therefore, these languages cannot state
the relations between a feature and its corresponding code segments in
the program text~\cite{Apel08}.  However, other product line
model-based methods usually provide a way to express those relations
explicitly by using CASE tools.  In FORM, for example, those explicit
relations make it possible to automatically generate product code from
specifications ~\cite{Kang98}. 

In our recent work, we are proposing a macro system for \mlpolyr{},
which augments the language with an explicit notion of
features~\cite{Wonseok09::FOP}. We implemented this mechanism in order
to make it possible to write feature composition in terms of features.
Then, the compiler can integrate the corresponding code automatically
once we provide a valid feature set.  Since our expansion rules do not
support any specification of feature relationships (i.e., mutually
exclusive or required relations), however, the \mlpolyr{} compiler
cannot detect any invalid feature sets. We leave such validation to
feature modeling tools which provide various diagnoses on feature
models. Our goal is to let a front-end modeling tool generate valid
expansion rules in the \mlpolyr{} language so that an application can
automatically be assembled only by feature selection.

{
\bibliographystyle{plainnat}
\singlespacing
\pagebreak
\bibliography{main}
}

\end{document}